\documentclass[11pt]{article}
\usepackage[letterpaper, margin=1in]{geometry}
\usepackage[letterpaper]{geometry}
\usepackage[utf8]{inputenc}

\usepackage{amsmath, amssymb, amsthm, authblk, bbm, bm, xcolor}
\usepackage[ruled, lined]{algorithm2e}
\usepackage{enumerate, graphicx, subcaption, subfiles, xcolor}
\usepackage[sc]{mathpazo}
\usepackage[linktocpage=true]{hyperref}
\hypersetup{
 colorlinks,
 linkcolor=blue,          
 citecolor=red!50!black,       
 filecolor=black,   
 urlcolor=black, 
 pdftitle={},
 pdfauthor={},
 pdfcreator={},
 pdfsubject={},
 pdfkeywords={}
}
\usepackage[sort&compress,numbers]{natbib}

\newcommand*\diff{\mathop{}\!\mathrm{d}}

\numberwithin{equation}{section}
\newtheorem{theorem}{Theorem}[section]
\newtheorem{lemma}[theorem]{Lemma}

\newtheorem{corollary}[theorem]{Corollary}
\newtheorem{proposition}[theorem]{Proposition}
\newtheorem{definition}[theorem]{Definition}
\newtheorem{remark}[theorem]{Remark}
\newtheorem{claim}[theorem]{Claim}
\newtheorem{assumption}[theorem]{Assumption}

\newcommand{\pro}{\mathbb{P}}
\newcommand{\E}{\mathbb{E}}

\newcommand{\tlambda}{\tilde{\lambda}}
\newcommand{\tm}{\tilde{m}}
\newcommand{\tq}{\tilde{q}}
\newcommand{\MAE}[2]{\big\lVert #1 - #2 \big\rVert_{MAE}}

\newcommand{\blue}[1]{{#1}}

\newcommand{\keywords}[1]{\textbf{\textit{Keywords ---}} #1}

\begin{document}

%\title{Load Balancing Under Fine-grained Compatibility Constraints}
\title{Online Capacity Scaling Augmented With Unreliable\\
Machine Learning Predictions}

\author{Daan Rutten\thanks{Email: \href{mailto:drutten@gatech.edu}{drutten@gatech.edu}} }
\author{Debankur Mukherjee}
\affil{Georgia Institute of Technology}

%\author{Daan Rutten}
%\email{drutten@gatech.edu}
%\author{Debankur Mukherjee}
%\email{debankur.mukherjee@gatech.com}
%\affiliation{%
%  \institution{Georgia Institute of Technology}
%  \city{Atlanta}
%  \state{Georgia}
%}

\maketitle

\begin{abstract}
Modern data centers suffer from immense power consumption. As a result, data center operators have heavily invested in capacity scaling solutions, which dynamically deactivate servers if the demand is low and activate them again when the workload increases. We analyze a continuous-time model for capacity scaling, where the goal is to minimize the weighted sum of flow-time, switching cost, and power consumption in an online fashion. We propose a novel algorithm, \blue{called Adaptive Balanced Capacity Scaling (ABCS)}, that has access to black-box machine learning predictions. ABCS aims to adapt to the predictions and is also robust against unpredictable surges in the workload. In particular, we prove \blue{that ABCS is} $(1+\varepsilon)$-competitive if the predictions are accurate, and yet, it has a uniformly bounded competitive ratio even if the predictions are completely inaccurate. Finally, we investigate the performance of this algorithm on a real-world dataset and carry out extensive numerical experiments, which positively support the theoretical results.
\end{abstract}


\maketitle

\section{Introduction.}

\subsection{Background and motivation.}
% Overcapacity and capacity scaling as a solution
Modern data centers suffer from immense power consumption, which amounts to a massive economic and environmental impact. In 2014, data centers alone contributed to about 1.8\% of the total U.S. electricity consumption \cite{Shebabi2016} and this is projected to reach 7\% in 2030~\cite{Nature2018}. 
Consequently, data center providers are constantly striving to optimize their servers for energy efficiency, pushing the hardware's efficiency to nearly its limit. At this point, algorithmic improvements appear to be critical in order to achieve substantial further gain~\cite{Shebabi2016}. 
A common practice for data centers has been to reserve significant excess service capacity in the form of idle servers~\cite{ZoomCapacity2020},
even though a typical idle server still consumes about 44\% of its peak power consumption~\cite{Shebabi2016}. The recommendation from the U.S. Department of Energy \cite{Shebabi2016}, industry \cite{NetflixScryer2013, FacebookScaling2014, GoogleScaling2016}, and the academic research community~\cite{Albers2007, Gandhi2010, lin2012dynamic} is, therefore, to implement dynamic capacity scaling functionality based on the demand.
If the demand is low, the service capacity should be scaled down
by deactivating servers, 
while at peak times, it should be scaled up by increasing the number of active servers.
Instead of physically toggling servers on or off,
this functionality is often implemented by carefully allocating a fraction of servers to other, lower priority services and quickly bringing them back at times of high demand; see~\cite{rzadca2020autopilot, tirmazi2020borg, cortez2017resource} for a more detailed account.
This maximizes the utilization of the system and hence minimizes the power consumption. \\

% Implementation of capacity scaling
The call for algorithmic solutions to capacity scaling has inspired a vibrant line of research over the last decade~\cite{lin2012dynamic, lu2012simple, mukherjee2017optimal, mukherjee2019join, Gandhi2010, Albers2007, bansal2009speed, augustine2004optimal, gandhi2013exact, irani2002competitive}.
The problem fits into the framework of online algorithms, where the goal is to design algorithms that dynamically scale the current service capacity, based on the past and current system information. 
Here, the performance of an algorithm is captured in terms of the \emph{Competitive Ratio} (CR), which is defined as the worst possible ratio between the cost incurred by the online algorithm and that by the offline optimum algorithm.
Note that the online algorithm has information only about the past and the present, while the offline optimum has accurate information about all future input variables such as the task-arrival process in the context of the current article.
The key advantage of such strong performance guarantees lies in its robustness, that is, the algorithm safeguards against the worst-case scenario.

However, any of today's modern large-scale systems has access to massive historical data, which, combined with standard Machine Learning (ML) algorithms, can reveal definitive patterns.
In these cases, simply following the recommendations obtained from the ML predictions typically outperforms any competitive algorithm. 
Netflix is an example of a company implementing capacity scaling in practice. 
Instead of relying on competitive online algorithms, Netflix has implemented ML algorithms in their Scryer system~\cite{NetflixScryer2013}. 
They noted that their demand usually follows regular patterns, allowing them to accurately predict the demand during a day based on data from previous weeks.
Most of the time, the performance of the machine learning algorithm is therefore excellent. 
However, besides empirical verification, the performance of such ML predictions is not guaranteed. 
In fact, repeated observations show that unexpected surges in the workload are not at all uncommon~\cite{NetflixScryer2013, lassettre2003dynamic, bodik2010automating}, which cause a significant adverse impact on the system performance. \\

The contrasting approaches between academia and industry reveal a gap between what we are able to prove and what is desirable in practice. 
While online algorithms do not require any information about future arrivals, in practice, these predictions are usually available. 
At the same time, an algorithm should not blindly trust the predictions because occasionally the accuracy of the predictions can be significantly poor.
The current work aims to bridge this gap by incorporating ML predictions directly into the competitive analysis framework.
In particular, we propose a novel low-complexity algorithm for capacity scaling, \blue{called Adaptive Balanced Capacity Scaling (ABCS)}, which has access to a black-box predictor, lending predictions about future arrivals.
Critically, not only is ABCS completely unaware of the prediction's accuracy, we also restrain from making \emph{any} statistical assumptions on the accuracy.
Hence, this excludes any attempt to learn the prediction's accuracy since accurate past predictions do not necessarily warrant the quality of future predictions.
The main challenge therefore is \emph{to design near-optimal algorithms which intelligently accept and reject the recommendations given by the ML predictor without knowing or learning their accuracy}.
Note, however, that the performance \blue{of ABCS does} depend on the (unknown) error of the prediction, and it ensures, among others, two desirable properties:
(i) \emph{consistency}, i.e., if the predictions turn out to be accurate in hindsight, then ABCS automatically nearly replicates the optimal solution, and
(ii) \emph{competitiveness}, i.e., if the predictions are inaccurate in hindsight, then the performance of ABCS is at most a uniformly bounded, constant factor times the minimum cost.
The formal definitions of consistency and competitiveness are given in Section~\ref{sec:prelim}.
It is worth emphasizing that this work is not concerned with how the ML predictions are obtained and uses them as a black box.

\subsection{Our contributions.}
% Model description
We will use a canonical continuous-time dynamical system model that is used to analyze algorithms for energy efficiency; see for example~\cite{lin2012dynamic, lu2012simple, maccio2015optimal, bansal2009speed, Albers2007} for variations.
Consider a system with a large number of homogeneous servers. 
Each server is in either of two states: \emph{active} or \emph{inactive}. 
Let $m(t)$ denote the number of active servers at time~$t$.
Workload arrives into the system in continuous time and gets processed at instantaneous rate $m(t)$. 
The system has a buffer of infinite capacity, where the unprocessed workload can wait until it is executed.
We will assume that there is an unknown and arbitrary arrival rate function $\lambda(t)$ that represents the arrival process; see Section~\ref{sec:modeldescription} for further details.
We do not impose any restrictions on $\lambda(\cdot)$.
To contrast this with the often-studied case when the workload arrival is stochastic, $\lambda(\cdot)$ can be thought of as an individual sample path of the corresponding stochastic arrival process.
At any time, the system may decide to increase or decrease $m(t)$ in an online fashion.
However, it pays a switching cost each time a server is activated.
This represents the cost of terminating the lower priority service running at the inactive server and related migration costs~\cite{lin2012dynamic, lu2012simple, maccio2015optimal, rzadca2020autopilot}. 
The goal of the system is to minimize the weighted sum of the flow-time, the switching cost, and the power consumption \cite{maccio2015optimal}. 
The flow-time is defined as the total time tasks spend in the system and is a measure of the response time~\cite{bansal2009speed, Albers2007}.
We will analyze the performance of an algorithm by its competitive ratio, the worst-case ratio between the cost of the online algorithm and the minimum offline cost, over all possible arrival rate functions $\lambda(\cdot)$.
We further assume that the algorithm receives predictions about future workload through an ML oracle~\cite{mahdian2012online, lykouris2018competitive}.
More precisely, at time $t = 0$, the ML oracle predicts an arrival rate function $\tlambda(\cdot)$. 
The algorithm may use these predictions to increase or decrease the number of servers accordingly. For instance, if the oracle predicts that the demand in the next hour will increase, then the algorithm might proactively increase the number of servers. 
However, as mentioned before, it is crucial that the algorithm \blue{is} completely oblivious to the accuracy of these predictions. 
We measure the accuracy of the predictions in terms of the mean absolute error (MAE) between the predicted arrival rate function $\tlambda$ and the actual rate function $\lambda$ (see Definition~\ref{def:mae}).
% Contributions
\blue{Our contributions in the current paper are threefold:} \\

\noindent
\textbf{(1) Purely online algorithm with worst-case guarantees.} 
First, we propose a novel purely online algorithm for capacity scaling, called Balanced Capacity Scaling (BCS). 
This purely online scenario, or the scenario of traditional competitive analysis, is equivalent to having predictions with infinite error. 
There are several fundamental works that have considered the purely online scenario for capacity scaling~\cite{lu2012simple, lin2012dynamic, gandhi2013exact, mukherjee2017optimal, mukherjee2019join}.
We extend the state of the art in this area by analyzing a general model in continuous time where unprocessed workload is allowed to wait. 
In fact, we show that a class of popular algorithms that were previously proposed are not constant competitive in this more general case (see Proposition \ref{prop:timer-lower-bd}). 
We show that \blue{BCS} is $5$-competitive in the general case (Corollary~\ref{cor:5comp}) and is $2$-competitive when waiting is not allowed and workload must be processed immediately upon arrival (Theorem~\ref{th:crnowait}).
BCS is easy to implement and is memoryless, i.e., it only depends on the current state of the system and not on the past. 
Also, we prove a lower bound result that any deterministic online algorithm must have a competitive ratio of at least $2.549$ (Proposition~\ref{prop:lowerbound}), which implies that the problem is strictly harder than the classical ski-rental problem, a benchmark for online algorithms. \\

\noindent
\textbf{(2) Augmenting unreliable ML predictions.} 
When ML predictions are available, we first propose an adaptive algorithm, called Adapt to the Prediction (AP), which ensures consistency.  
That is, we prove (Theorem \ref{th:crftp}) that the competitive ratio of \blue{AP} is at most $1 + \Theta(\eta)$, where $\eta$ is a suitable measure of the prediction's accuracy and is a function of the MAE between the predicted arrival rate function $\tlambda$ and the actual rate function $\lambda$ (Definition~\ref{def:error}).
AP does not follow the predictions blindly. 
Rather, it dynamically scales the number of servers in an online fashion as the past predictions turn out to be inaccurate.
Although the performance of AP is optimal as $\eta = 0$ and it degrades gracefully with $\eta$, it is not constant competitive if predictions are completely inaccurate ($\eta = \infty$).
Thus, it does not provide any worst-case guarantees.
This is a feat shared by many recent adaptive algorithms in the literature (see Remark \ref{rem:blindlyfollow}).

Next, we combine the ideas behind \blue{BCS and AP}, to propose an algorithm that is both competitive \emph{and} consistent.
This brings us to the main contribution of the paper. 
We propose \blue{Adaptive Balanced Capacity Scaling (ABCS)}, which uses the structure of BCS and utilizes AP as a subroutine. ABCS has a hyperparameter $r \geq 1$, which can be fixed at any value before implementing the algorithm, and represents our confidence in the ML predictions. 
If we choose $r = 1$, then the algorithm works as a purely online one and disregards all predictions. 
In this case, ABCS is $5$-competitive, as before. 
However, for any fixed $r>1$, we prove (Corollary~\ref{cor:bcr_wcr}) that the competitive ratio of ABCS is at most
\blue{
\begin{equation}
\label{eq:aca-cr}
    \textsc{CR}(\eta) \leq \min\left( ( 1 + \mathcal{O}(\eta) ) \cdot \left( 1 + r^{-1} + \mathcal{O}\left( r^{-2} \right) \right), 16 \sqrt{2} r^{7/2} \right).
\end{equation}
}
where $\eta$ is the prediction's accuracy as before.
There are a number of consequences of the above result. 
We start by emphasizing that although the competitive ratio is a function of the error $\eta$, the algorithm is completely oblivious to it.
Now, the higher we fix the value of $r$ to be, the closer the competitive ratio of ABCS gets to 1 if the predictions turn out to be accurate in hindsight.
If the predictions are completely inaccurate ($\eta = \infty$), the competitive ratio is at most \blue{$16 \sqrt{2} r^{7/2}$}, a constant that depends only on $r$ and \emph{not} on $\eta$.
ABCS is therefore robust against unpredictable surges in workload, while providing near-optimal performance if the predictions are accurate. 

Another interesting thing to note is that for $r > 1$, the competitive ratio in~\eqref{eq:aca-cr} is the minimum of two terms: the first term, which we call the \emph{Optimistic Competitive Ratio} (OCR), is smaller when the prediction is accurate and the second term, which we call the \emph{Pessimistic Competitive Ratio} (PCR), is smaller when the prediction is inaccurate.
From the algorithm designer's perspective, there is a  clear trade-off between \textsc{OCR} and \textsc{PCR}, which is conveniently controlled by the confidence hyperparameter $r$.
It is important to note that ABCS provides performance guarantees for \emph{any fixed} $r\geq 1$ irrespective of the model parameters or the accuracy of the predictions. 
However, \emph{the choice of $r$ reflects the risk that the system designer is willing to take in the pessimistic case against the gain in the optimistic case.}
See Remark~\ref{rem:ocr-pcr} for further discussion.
This trade-off, however, is not specific to our algorithm. 
In fact, we prove a negative result in Proposition~\ref{prop:lowerboundcreta} that \emph{any} algorithm which is $(1 + \delta)$-competitive in the optimistic case, has a competitive ratio of at least $1 / (4 \delta)$ in the pessimistic case. \\

\noindent
\blue{\textbf{(3) Offline algorithm for regular workloads.} Finally, we consider the scenario in which the workload $\lambda(\cdot)$ is known perfectly upfront. We propose an offline algorithm which solves a linear program and prove (Theorem \ref{th:lin-appx}) that, if the workload is sufficiently regular (see Assumption \ref{as:regular}), then the offline algorithm is $(1 + \mathcal{O}(\delta))$-competitive with respect to the offline optimal algorithm. Here, $\delta$ is a hyperparameter of the algorithm that measures the desired accuracy. As $\delta$ decreases, the accuracy of the solution increases, however, the dimension of the linear program (in terms of the number of decision variables and constraints) increases at rate $1 / \delta$ as well. The offline algorithm may be used as a subroutine in AP, even if the predictions are unreliable.} \\

To test the performance of our algorithms in practice, we implemented them on both a real-world dataset of DNS requests observed at a campus network~\cite{singh2016data} and a set of artificial datasets, and the performance turns out to be excellent. See Section~\ref{sec:numericalexperiments} for details.

\subsection{Related works.}
Over the past two decades, the rapid growth of data centers and its immense power consumption have inspired a vibrant line of research in optimizing the energy efficiency of such systems~\cite{urgaonkar2005dynamic, barroso2007case, dayarathna2015data, rong2016optimizing}. 
Below, we provide an overview of a few influential works relevant to the current paper. \\

% Capacity scaling
\noindent
The capacity scaling problem was introduced in a seminal paper by Lin et al.~\cite{lin2012dynamic}, who analyze a discrete time model of a data center. At each time step $t$, the cost of operating $m(t)$ servers is determined by the switching cost and an arbitrary convex function $g_t(m(t))$, which, for example, specifies the cost of increased power consumption versus response time. 
At time step $t$, the system reveals the function $g_t$ and accurate functions $g_{t+1}, g_{t+2}, \dots, g_{t+w}$ in a prediction window of $w$ future time steps.
Lin et al.~\cite{lin2012dynamic} propose an algorithm, called the LCP algorithm, and prove that it is $3$-competitive. 
Surprisingly, the performance of the LCP algorithm does not improve if $w > 0$, i.e., if predictions are available. 
\blue{We consider a modified model in continuous time, where predictions are not necessarily accurate. Moreover, the performance of our algorithm increases provably in the presence of predictions.}

Lu et al.~\cite{lu2012simple} consider a scenario where tasks cannot wait in queue and must be served immediately upon arrival.
They discover that in this case, the capacity scaling problem reduces to solving a number of independent ski-rental problems. 
The authors then propose an algorithm and prove that it is $2$-competitive. 
Our model, in addition, includes the response time, which directly generalizes the framework of~\cite{lu2012simple}. 
This flexibility introduces a whole new dimension in the space of possible decisions. 
For example, since the results of Lu et al.~\cite{lu2012simple} lack any form of delay, tasks are processed at the same time by any algorithm. 
Our model allows an algorithm-dependent delay of serving tasks, which \emph{desynchronizes} the time at which tasks are processed at a server across different algorithms and hence, significantly complicates the analysis.
Mazzucco and Dyachuk~\cite{mazzucco2012optimizing} analyze a related problem in which the number of servers is periodically updated and a task is lost if a server is not immediately available to serve it. The goal of their algorithm is to balance the power consumption and the cost of losing tasks.
Galloway et al.~\cite{galloway2012empirical} and later Gandhi et al.~\cite{gandhi2012autoscale, gandhi2014adaptive} perform an empirical study of data centers. Their results show that significant power savings are possible, while maintaining much of the latency of the network.\\

% Speed scaling
A well-studied problem that is somewhat related to our setup is \emph{speed scaling}. 
Here, the goal is to optimize the processing speed of a single server and to minimize the weighted sum of the flow-time and power consumption, while the switching cost is zero. The power consumption is typically cubic in the processing speed. In contrast to our model, the scheduling of jobs also play a crucial role here.
A seminal paper in this area is by Bansal et al.~\cite{bansal2009speed}, which proposes an algorithm that schedules the task with the shortest remaining processing time (SRPT) first and processes it at a speed such that the power consumption is equal to the number of waiting tasks plus one.
The authors prove that this algorithm is $(3 + \varepsilon)$-competitive. 
Later papers have extended the case of the single server to processor sharing systems \cite{wierman2009power} and parallel processors with deadline constraints \cite{albers2014speed}. The problem of speed scaling has also been analyzed in the case the inter-arrival times and required processing times are exponentially distributed \cite{ata2006dynamic}.

% Ski-rental problem
Any algorithm for the capacity scaling problem consists of two components: first, to activate servers and second, to deactivate servers. 
For a single server, a natural abstraction of the latter problem is the famous ski-rental problem, as first introduced by Karlin et al.~\cite{karlin1988competitive}.
The ski-rental problem has been applied to cases of capital investment~\cite{azar1999capital, damaschke2003nearly}, TCP acknowledgement~\cite{karlin2003dynamic} and cache coherence~\cite{anderson1996two}. 
Irani et al.~\cite{irani2002competitive} analyze the ski-rental problem when multiple power-down states are available, such as active, sleeping, hibernating, and inactive. 
The power consumption in each state is different and moving between the states incurs a switching cost. 
Augustine et al.~\cite{augustine2004optimal} generalize these results when the transition costs between the different states are not additive. Although the current work focuses on only two states, i.e., active and inactive, we expect that the algorithm and proofs are general enough to accommodate multiple power-down states, which we leave as interesting future work. 
Khanafer et al.~\cite{khanafer2013constrained} analyze the ski-rental problem in a stochastic context.\\

% Predictions
\blue{Two papers are often independently credited for initiating the study of online algorithms augmented by ML predictions: Lykouris and Vassilvtiskii \cite{lykouris2018competitive} in the context of caching and Mahdian et al.~\cite{mahdian2012online} in the context of allocation of online advertisement space, load balancing and facility location.
Lykouris and Vassilvtiskii \cite{lykouris2018competitive} show how to adapt the marker algorithm for the caching problem to obtain a competitive ratio of $2$ if the predictions are perfectly accurate, and a bounded competitive ratio if the predictions are inaccurate. 
Mahdian et al.~\cite{mahdian2012online} propose an algorithm which naively switches between an optimistic and a pessimistic scheduling algorithm to minimize the makespan when routing tasks to multiple machines. We here mostly follow the terminology of~\cite{lykouris2018competitive}.
Since then, the ideas have} been applied to bipartite matching \cite{kumar2018semi}, ski-rental and scheduling on a single machine \cite{purohit2018improving}, bloom-filters \cite{mitzenmacher2018model}, and frequency estimation \cite{hsu2018learning}.
Lee et al.~\cite{lee2019learning} propose an algorithm which operates onsite generators to reduce the peak energy usage of data centers. Although related to the current work, their algorithm works independent to the capacity scaling happening inside the data center. Bamas et al.~\cite{bamas2020learning} discuss an algorithm augmented by predictions for the related problem of speed scaling discussed above, in the case of parallel processors with deadline constraints. Similar to our results, Bamas et al.~\cite{bamas2020learning} identify a trade-off between what we call an Optimistic Competitive Ratio and a Pessimistic Competitive Ratio. We prove, for the capacity scaling problem considered in the current work, that \emph{any} algorithm must exhibit such a trade-off (see Proposition \ref{prop:lowerboundcreta} for details).~\blue{Antoniadis et al.~\cite{antoniadis2020online} discuss algorithms augmented with predictions for the general framework of metrical task systems. Note that the problem in the current paper cannot be described in the form of a metrical task system. For example, if one casts the problem in a metric space which contains both the number of servers and the workload in the queue, then the possible transitions depend on a non-trivial combination of the arrival function, number of servers and the workload in the queue. The metrical task system allows the possible transitions to depend on either the metric between two points or the arrival function, but not a combination thereof. However, one cannot omit the number of servers or the workload in the queue from the metric space either since the cost depends on both. Therefore, the workload in the queue adds a completely new way in which decisions between rounds are coupled that cannot be captured by a metric space.}

Recently, the notion of a predictor has also emerged in stochastic scheduling. Mitzenmacher \cite{mitzenmacher2019scheduling} introduces the predictor as a probability density function $g(x, y)$ for a task with actual service time $x$ and predicted service time $y$. 
Here, the author analyzes the shortest predicted job first (SPJF) and shortest predicted remaining processing time (SPRPT) queueing disciplines for a single queue and determines the price of misprediction, i.e. the ratio of the cost if perfect information of the service time distribution is known and the cost if only predictions are available. For multiple queues, Mitzenmacher \cite{mitzenmacher2019supermarket} has simulated the supermarket model or the `power-of-$d$' model, to show empirically that the availability of predictions greatly improves performance.

A different line of work called online algorithms with advice questions how many bits of \emph{perfect} future information are necessary to reproduce the optimal offline algorithm (see \cite{boyar2017online} for a survey). The difference with the current work is that we do not assume that the predictions are perfect but instead have arbitrary accuracy.\\

% Stochastic scheduling
When the arrival process and service times are stochastic, there are several major works that consider energy efficiency of the system.
Gandhi et al.~\cite{gandhi2013exact} provide an exact analysis of the M/M/k/setup system. 
The system is similar to the M/M/k class of Markov chains, i.e., tasks arrive according to a Poisson process and require an exponentially distributed processing time. 
To process the tasks, the system has access to a maximum of $k$ servers. 
According to the algorithm in \cite{gandhi2013exact}, if a task arrives and there are no available servers, the system moves one server to its setup state, where it remains for an exponentially random time before the server becomes active. 
The authors provide a sophisticated method to analyze the system exactly.
Maccio and Down~\cite{maccio2015optimal} analyze a similar system for a broader class of cost functions. 
When each server has a dedicated separate queue, Mukherjee et al.~\cite{mukherjee2017optimal} and Mukherjee and Stolyar \cite{mukherjee2019join}  analyze the case where the setup times and standby times (the time a server remains idle before it is deactivated) are independent exponentially distributed. 
In this case they propose an algorithm that achieves asymptotic optimality for both the response time and power consumption in the large-system limit.
Earlier research has also modeled the response time as a constraint rather than charging a cost for the response-time \cite{goldman2000online}. 
Here, each task is presented with a deadline and the task should be served before this deadline or it is irrevocably lost. The earliest deadline first (EDF) queueing discipline has been proven to be effective in this case \cite{doytchinov2001real}.

\subsection{Notation and organization.}
The remainder of the paper is organized as follows. Section~\ref{sec:modeldescription} describes the model. Section~\ref{sec:prelim} introduces some preliminary concepts and definitions related to the ML predictions, such as the error. 
Section~\ref{sec:mainresults} introduces our algorithms and the main results, of which the high-level proof ideas are provided in Section~\ref{sec:proofs}. Most of the technical proofs are given in the appendix.
Section~\ref{sec:numericalexperiments} presents extensive numerical experiments, including the performance of our algorithms on a real-world dataset. 
Finally, Section~\ref{sec:conclusion} concludes our work and presents directions for future research.

\section{Model description.}
\label{sec:modeldescription}

We now introduce a general model for capacity scaling. 
Let $\omega, \beta$, and $\theta$ be fixed non-negative parameters of the model.
We will assume that the tasks waiting in the buffer accumulate a waiting cost at rate $\omega > 0$, the cost of activating a server is $\beta > 0$, and each active server accumulates a power consumption cost at rate $\theta \geq 0$. \blue{We denote the workload in the buffer at time $t$ by $q(t)$.}

\blue{An instance consists of a known finite time horizon $T > 0$ and an unknown and arbitrary function $\lambda:[0,T] \to \mathbbm{R}_+$ representing the arrival process.
The model is:
\begin{equation}
\begin{aligned}
\label{eq:model}
\underset{m: [0, T] \to \mathbbm{R}_+}{\text{minimize}}\hspace{3mm} &\omega \cdot \int_0^T q(s) \diff s + \beta \cdot \limsup\limits_{\delta \downarrow 0} \sum_{i = 0}^{\lfloor T / \delta \rfloor} [m(i \delta + \delta) - m(i \delta)]^+ + \theta \cdot \int_0^T m(s) \diff s \\
\text{subject to}\hspace{3mm} &\begin{aligned}[t]
    &q(t) = \int_0^t (\lambda(s) - m(s)) \mathbbm{1}\{ q(s) > 0 \text{ or } \lambda(s) \geq m(s) \} \diff s \text{ for all } t \in [0, T] \\
    &m(0) = 0, \; m(t) \geq 0 \text{ for all } t \in (0, T]
\end{aligned}
\end{aligned}
\end{equation}
where $[x]^+ = \max(x, 0)$.} 
To solve the optimization problem above, an algorithm needs to determine the function $m(\cdot)$, given the parameters $\omega, \beta, \theta$.
Note that our goal is to investigate \emph{online} algorithms, meaning that $\lambda(\cdot)$ is revealed to the algorithm in an online fashion. In other words, at time $t$, the algorithm must determine $m(t)$ depending only on $\lambda(s)$ for $s \in [0, t]$. \blue{Note that the system may equivalently reveal the total workload received before time $s \in [0, t]$, as $\lambda$ is simply its rate of increase.} For an algorithm that runs $m(t)$ servers at time $t$, the cost accumulated until time $t$ is defined as
\begin{equation}
\label{eq:cost}
    \textsc{Cost}^{\lambda}(m, t) := \omega \cdot \int_0^t q(s) \diff s + \beta \cdot \limsup\limits_{\delta \downarrow 0} \sum_{i = 0}^{\lfloor t / \delta \rfloor} [m(i \delta + \delta) - m(i \delta)]^+ + \theta \cdot \int_0^t m(s) \diff s 
\end{equation}
We will compare the total cost $\textsc{Cost}^{\lambda}(m, T)$ for an online algorithm to that of the \emph{offline} minimum defined as
\begin{equation}
\label{eq:opt}
    \textsc{Opt} := \inf_{m: (0, T] \to \mathbbm{R}_+} \textsc{Cost}^\lambda(m, T),
\end{equation}
and without loss of generality, we will assume $\textsc{Opt} < \infty$ throughout the paper.

\begin{remark}
\label{rem:minexists}
\blue{The minimizer of} \eqref{eq:opt} exists, as stated by the next proposition. The proof of Proposition \ref{prop:existence} 
is given in Appendix~\ref{app:existence}.
The difficulty in the proof lies in dealing with the second term in~\eqref{eq:cost}, which makes the function $\textsc{Cost}^\lambda(m, T)$ discontinuous in $m(\cdot)$ w.r.t.~the $L_1$ norm.
\end{remark}

\begin{proposition}
\label{prop:existence}
There exists $m^*: [0, T] \to \mathbbm{R}_+$ such that $\textsc{Cost}^\lambda(m^*, T) = \textsc{Opt}$.
\end{proposition}

\begin{remark}
\blue{The model in \eqref{eq:model} assumes that $m(0) = 0$ for the sake of clarity of exposition. The results in this paper extend to any $m(0)$ by adding an additive constant of $\mathcal{O}(\beta \cdot m(0))$ to each of the performance bounds. The proofs in the Appendix are presented for this more general case.}
\end{remark}

\begin{figure}
\centering
\includegraphics[width=0.5\textwidth]{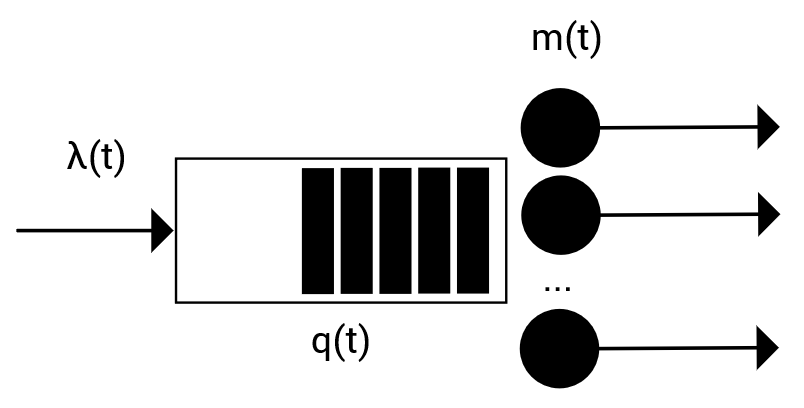}
\caption{The system receives task at rate $\lambda(t)$ and operates $m(t)$ servers. The workload is $q(t)$.}
\label{fig:capacityscaling}
\end{figure}

The model in~\eqref{eq:model} actually combines some well-studied state-of-the-art models~\cite{lin2012dynamic, lu2012simple, maccio2015optimal, bansal2009speed, Albers2007}.
To see how it relates to the problem of capacity scaling, note that the objective function in~\eqref{eq:model} is a weighted sum of three metrics.
Below, we clarify each of them.
These three metrics are common performance measures of the system, such as the response time or the power consumption. 
The parameters $\omega$, $\beta$ and $\theta$ represent the weights assigned to each of these metrics. 
The three metrics are as follows:
\begin{enumerate}[(i)]
\item
\textbf{The flow-time.} The flow-time is defined as the total time a task spends in the system and captures the response time of the system. Note that the average response time per unit of workload is $\frac{\int_0^T q(s) \diff s}{\int_0^T \lambda(s) \diff s}$; see also~\cite{bansal2009speed, Albers2007}. 
The weight $\omega$ is the cost attributed to the response time (e.g., in dollars per second). The weight $\omega$ could, for example, be determined based on loss of revenue or user dissatisfaction as a result of increased response time.

\item
\textbf{The switching cost.}
As in~\cite{lu2012simple, lin2012dynamic, rzadca2020autopilot}, the parameter $\beta$ can be viewed as the cost to increase the number of active servers (e.g., in dollars per server).
This may include for example, the cost to terminate a lower priority service and related migration costs. 
In practice, these costs are usually equivalent to the cost of running the server for multiple hours~\cite{lin2012dynamic}.
The total switching cost is $\beta$ times the number of times a server is made active. 

\item
\textbf{The power consumption.} The power consumption is proportional to the total time servers are in the active state~\cite{lu2012simple}. The weight $\theta$ represents the cost of power (e.g., in dollars per server per second).
\end{enumerate}
Also, the constraints in \eqref{eq:model} model the dynamics of capacity scaling and $q(\cdot)$ can be viewed as the queue length process or the remaining workload process.
Note that \eqref{eq:model} does not require $q(t)$ or $m(t)$ to be integer-valued. 
This is a fairly standard relaxation, since a service may typically request a fraction of the server's capacity \cite{tirmazi2020borg, rzadca2020autopilot} and a single task is tiny; see for example, \cite{lin2012dynamic, mukherjee2017optimal}. 
The system in~\eqref{eq:model} can also be interpreted as a \emph{fluid counterpart} of a discrete system.
Figure~\ref{fig:capacityscaling} depicts the model schematically.

\begin{remark}
The model in \eqref{eq:model} assumes that the service capacity can be increased nearly instantaneously. Hence, it does not include the so-called setup time.
Besides being a common assumption in competitive analysis (see for example \cite{lin2012dynamic, lu2012simple}), this is also not completely unreasonable in practice. 
This is mainly because servers are not usually physically turned off in reality. Instead, when a server becomes ``inactive'', the server's capacity will be used by other low-priority services. 
Then, activating a server means quickly terminating such low-priority services, see~\cite{tirmazi2020borg, rzadca2020autopilot} for a more detailed account.
From a theoretical standpoint, the assumption of a zero setup time is also necessary to get a uniformly bounded competitive ratio, as stated in the next lemma. 
For the sake of Lemma \ref{lemma:warmuptime}, let us assume that in the capacity scaling problem in \eqref{eq:model} there is an additional setup time $t_0 > 0$ before the number of servers can be increased. In other words, if the online algorithm decides to turn on a server at time $t$, then the number of servers is increased at time $t + t_0$.
The proof of Lemma \ref{lemma:warmuptime} is provided in Appendix \ref{app:proofwarmuptime}.
\end{remark}

\begin{lemma}
\label{lemma:warmuptime}
\blue{

Let $\mathcal{A}$ be any deterministic algorithm for the capacity scaling problem in \eqref{eq:model} and assume that there is an additional setup time $t_0 > 0$ before the number of servers can be increased. Also, let $\textsc{CR}$ denote the competitive ratio of $\mathcal{A}$ (see Definition \ref{def:competitiveratio} for a formal definition of the competitive ratio). Then, there exists $\theta$ such that
$
     \textsc{CR} \geq \frac{\omega t_0^2}{2 \beta}.
$
In short, there does not exist any deterministic online algorithm with uniformly bounded competitive ratio.
}
\end{lemma}

Formulation \eqref{eq:model} is fairly easy to solve as an \emph{offline} optimization problem. Section \ref{sec:offlinelp} presents a linear program that solves the offline problem. 
However, as mentioned earlier, we are interested in an \emph{online} algorithm. Specifically, we distinguish two scenarios:
\begin{enumerate}
    \item
\textbf{Purely online scenario.} \blue{The system reveals $\omega$, $\beta$, $\theta$ and, at time $t$, also $\lambda(s)$ for $s \in [0, t]$ to the online algorithm, but not $\lambda(s)$ for any $s > t$.} The purely online scenario corresponds to the setting where predictions may not be available and provides a natural starting point of our investigation. 
We discuss a competitive algorithm for the purely online scenario in Section~\ref{sec:purelyonline}. Additionally, in this purely online scenario, our algorithm does not require the system to reveal the finite time horizon $T$ upfront.

    \item
\textbf{Machine learning scenario.} In addition to the assumptions in the purely online scenario above, at time $t = 0$, an ML predictor predicts the arrival rate function of the entire interval, that is, it predicts the arrival rate function to be $\tlambda: [0, T] \to \mathbbm{R}_+$. The ML predictor may, for example, be trained on the past observed workload on a day. For the purpose of the current work, we treat the predictor as a black box. We discuss a consistent algorithm for the machine learning scenario in Section \ref{sec:ftp}, for which the competitive ratio degrades gracefully with the prediction's accuracy. However, the algorithm is not competitive in the worst-case.
Finally, in Section \ref{sec:robustml} we discuss an algorithm for the machine learning scenario, which is simultaneously competitive and consistent, by combining the algorithms from Sections \ref{sec:purelyonline} and \ref{sec:ftp}.
\end{enumerate}
The idea of using online algorithms with unreliable machine-learned advice was first introduced \blue{in~\cite{mahdian2012online} in the context of allocation of online advertisement space, load balancing and facility location}, and in \cite{lykouris2018competitive} in the context of competitive caching. The next section provides the necessary details of the framework of~\cite{lykouris2018competitive}.

\section{Preliminary concepts.}
\label{sec:prelim}
This section briefly presents the competitive analysis framework for algorithms that have access to ML predictions.
The mostly follow the setup as introduced in~\cite{lykouris2018competitive} and adapt it here for the current scenario.
We measure the errors in predictions by the mean absolute error (MAE) between the true and the predicted label, which is commonly used in state-of-the-art machine learning algorithms \cite{gao2014machine, qi2020mean}.
\begin{definition}
\label{def:mae}
The error in the prediction $\tlambda(\cdot)$ with respect to the actual arrival rate $\lambda(\cdot)$ is
\begin{equation}
\label{eq:error}
    \MAE{\tlambda}{\lambda} = \frac{1}{T} \int_0^T \left\lvert \tlambda(t) - \lambda(t) \right\rvert \diff t.
\end{equation}
\end{definition}
To measure the performance of an algorithm augmented by an ML predictor, we will define the competitive ratio as a function of the prediction's accuracy.
However, before stating the definition of competitive ratio, we introduce the level of accuracy of a prediction.

\begin{definition}
\label{def:error}
\blue{

Fix a finite time horizon $T$ and arrival rate function $\lambda(\cdot)$. 
Let $\textsc{Opt}$ be as defined in~\eqref{eq:opt}. For $\eta>0$, we say that a prediction $\tlambda$ is $\eta$-accurate for the instance $(T, \lambda)$ if
\begin{equation}
    \MAE{\tlambda}{\lambda} \leq \eta \cdot \frac{\textsc{Opt}}{T}.
\end{equation}

}
\end{definition}

The definition of the prediction's accuracy is intimately tied to the cost of the optimal solution. 
As already argued in \cite{lykouris2018competitive}, since $\textsc{Opt}$ is a linear functional of $\lambda$,
normalizing the error by the cost of the optimal solution is necessary.
This is because the definition should be invariant to scaling and padding arguments.
For example, if we double both $\lambda(\cdot)$ and $\tlambda(\cdot)$, then the prediction's accuracy should still be the same.

Let $\mathcal{A}$ be any algorithm for~\eqref{eq:model}. The performance of $\mathcal{A}$ is measured by the competitive ratio $\textsc{CR}(\eta)$, which itself is a function of the accuracy $\eta$.
The following definition is an adaptation of \cite[Definition 3]{lykouris2018competitive} for the current setup.
\begin{definition}
\label{def:competitiveratio}
\blue{

Fix a finite time horizon $T$ and arrival rate function $\lambda(\cdot)$. 
Let $\mathcal{A}$ be any algorithm for \eqref{eq:model} and $m(t)$ denote its number of servers when it has access to a prediction $\tlambda$, and $\textsc{Opt}$ be as defined in~\eqref{eq:opt}. 
We say that $\mathcal{A}$ has a competitive ratio at most $\textsc{CR}$ for the instance $(T, \lambda)$ and prediction $\tlambda$ if
\begin{equation}
\label{eq:defcr}
    \textsc{Cost}^\lambda(m, T) \leq \textsc{CR} \cdot \textsc{Opt},
\end{equation}
We say that the competitive ratio of $\mathcal{A}$ is at most $\textsc{CR}(\eta)$, if the competitive ratio is at most $\textsc{CR}(\eta)$ for all instances $(T, \lambda)$ and any $\eta$-accurate prediction $\tlambda$. By convention, we say that $\textsc{CR}(\eta) = \infty$ if such a finite $\textsc{CR}(\eta)$ does not exist.
}
\end{definition}

Note that, although the competitive ratio depends on the prediction's accuracy, the algorithm is oblivious to this accuracy. 
We desire three properties of an online algorithm that has access to a prediction. The algorithm's performance should 
(i) be close to the optimal solution if the prediction is perfect, 
(ii) degrade gracefully with the prediction's error, and 
(iii) be bounded regardless of the prediction's accuracy. 
The definitions of consistency, robustness, and competitiveness summarize these desiderata; see also \cite[Definition 4, 5, and 6]{lykouris2018competitive}.

\begin{definition}
\label{def:desiderata}
Let $\mathcal{A}$ be any algorithm for~\eqref{eq:model}, and $\textsc{CR}(\eta)$ denote its competitive ratio when it has access to an $\eta$-accurate prediction. Then, we say that algorithm $\mathcal{A}$ is
\begin{enumerate}[{\normalfont (i)}]
    \item 
$\rho$-consistent if $\textsc{CR}(0) = \rho$;

    \item
$\alpha$-robust if $\textsc{CR}(\eta) = \mathcal{O}(\alpha(\eta))$;

    \item
$\gamma$-competitive if $\textsc{CR}(\eta) \leq \gamma$ for all $\eta \in [0, \infty]$.
\end{enumerate}
\end{definition}

\section{Main results.}
\label{sec:mainresults}

\subsection{Balanced Capacity Scaling algorithm.}
\label{sec:purelyonline}

We first discuss a competitive algorithm in the purely online scenario. \blue{Recall that in this case the system reveals $\omega$, $\beta$, $\theta$ and, at time $t$, also $\lambda(s)$ for $s \in [0, t]$ to the algorithm, but not $\lambda(s)$ for any $s > t$.}
Moreover, as mentioned earlier, the results in this section also hold when the finite time horizon $T$ is not revealed upfront. 
The Balanced Capacity Scaling (BCS) algorithm that we propose is parameterized by two non-negative numbers $r_1$ and $r_2$. Algorithm~\ref{alg:bcsalg} below describes BCS for any fixed choices of $r_1$ and $r_2$.

\begin{algorithm}
\DontPrintSemicolon
Choose $m(\cdot)$ such that at each time $t \geq 0$:
\begin{equation}
\label{eq:diffcapacity}
    \frac{\diff m(t)}{\diff t} = \frac{r_1 \omega \cdot q(t) - r_2 \theta \cdot m(t)}{\beta}.
\end{equation}
\caption{ BCS ($r_1, r_2$)}
\label{alg:bcsalg}
\end{algorithm}

We start by briefly discussing the intuition behind BCS. 
At each time $t \geq 0$, BCS computes the derivative of the number of servers, 
i.e., how fast the system should increase or decrease the service capacity. 
Note that if we solve equation \eqref{eq:diffcapacity}, then we obtain the number of servers $m(t)$, which is differentiable for all $t \geq 0$. 
The two parameters $r_1$ and $r_2$ control how fast the algorithm reacts, by increasing or decreasing the number of servers respectively. 
If the workload $q(t)$ is non-zero, then the first term in the right-hand side of equation \eqref{eq:diffcapacity} increases the number of servers at rate $r_1$.
The second term is an ``inertia term'' which decreases the number of servers at rate $r_2$. Note that if we integrate equation \eqref{eq:diffcapacity}, we obtain
\begin{equation}
    \int_0^t r_1 \omega \cdot q(s)\diff s = \int_0^t \beta \cdot \frac{\diff m(s)}{\diff s} \diff s + \int_0^t r_2 \theta \cdot m(s) \diff s.
\end{equation}
This means that \emph{BCS aims to carefully balance the flow-time with the switching cost plus the power consumption}.
\blue{BCS} is memoryless and computationally cheap. 
The derivative of the number of servers depends only on the current workload and number of servers, without requiring knowledge about the past workload, number of servers or arrival rate. 
BCS can therefore be implemented without any memory requirements. \\

We are able to characterize the performance of BCS analytically, for any fixed choices of $r_1$ and $r_2$. Theorem \ref{th:cr1bcs} below characterizes the competitive ratio of \blue{BCS}. The proof of Theorem \ref{th:cr1bcs} is provided in Section \ref{sec:proofbcs1}.

\begin{theorem}
\label{th:cr1bcs}
Let \textsc{CR} denote the competitive ratio of BCS (Algorithm \ref{alg:bcsalg}). Then,
\begin{equation}
    \textsc{CR} \leq \left(1 + \frac{1}{r_1} + \frac{1}{r_2} \right) \max\left( 2, r_1, 2 r_2 \right).
\end{equation}
\end{theorem}

The optimal choice of the parameters is $r_1 = 2$ and $r_2 = 1$. Corollary \ref{cor:5comp} states that BCS is $5$-competitive in this case.

\begin{corollary}
\label{cor:5comp}
Let \textsc{CR} denote the competitive ratio of BCS (Algorithm \ref{alg:bcsalg}).
If $r_1 = 2$ and $r_2 = 1$, then $\textsc{CR} \leq 5$.
\end{corollary}

Moreover, in the special case when tasks are not allowed to wait and must be served immediately upon arrival ($\omega = \infty$), BCS turns out to be $2$-competitive, as stated in Theorem \ref{th:crnowait}. \blue{Note that the algorithm introduced by Lu et al.~\cite{lu2012simple} in the special case $\omega = \infty$ is also $2$-competitive and the authors prove that this is in fact optimal.} The proof of Theorem \ref{th:crnowait} is given in Appendix \ref{app:proofnowait}.

\begin{theorem}
\label{th:crnowait}
Let \textsc{CR} denote the competitive ratio of BCS (Algorithm \ref{alg:bcsalg}).
If $r_1 = 2$, $r_2 = 1$ and $\omega = \infty$, then $\textsc{CR} \leq 2$.
\end{theorem}

Note that the capacity scaling problem has previously been related to the classical ski-rental problem~\cite{lu2012simple, augustine2004optimal, irani2002competitive} which is 2-competitive. As it turns out, when tasks are allowed to wait, the formulation in \eqref{eq:model} of the capacity scaling problem is strictly harder than the ski-rental problem, as Proposition \ref{prop:lowerbound} states below. Proposition \ref{prop:lowerbound} is proved in Appendix \ref{app:proofpurelyonline-lower-bd}.

\begin{proposition}
\label{prop:lowerbound}
Let $\mathcal{A}$ be any deterministic algorithm for the capacity scaling problem in \eqref{eq:model} in the purely online scenario, and $\textsc{CR}$ denote its competitive ratio. There exist choices for $\omega$, $\beta$, and $\theta$ such that $\textsc{CR} \geq 2.549$.
In other words, any deterministic algorithm is at least $2.549$-competitive.
\end{proposition}

\begin{remark}
We should note that  the proof of Proposition \ref{prop:lowerbound} assumes that the finite time horizon $T$ is not revealed upfront. 
We leave it to future work to identify a (possibly weaker) lower bound if $T$ is known to the algorithm.
\end{remark}

\subsection{Augmenting unreliable ML predictions.}
\label{sec:machinelearning}

To augment \blue{BCS} with machine learning predictions, we proceed in two steps. First, in Section \ref{sec:ftp}, we introduce Adapt to the Prediction (AP). We prove that the competitive ratio of AP degrades gracefully with the prediction's accuracy, although AP is not competitive. Second, in Section \ref{sec:robustml}, we discuss how to combine \blue{BCS and AP} to \blue{obtain ABCS, which} follows the predictions but is robust against inaccurate predictions and therefore competitive.

\subsubsection{Adapt to the Prediction algorithm.}
\label{sec:ftp}

We will now turn our attention to the machine learning scenario. Recall that in this case, at time $t = 0$, the algorithm receives a predicted arrival rate function $\tlambda: [0, T] \to \mathbbm{R}_+$. Note that a trivial way to implement the predictions is to blindly trust the predictions, i.e., to let
\begin{equation}
    m \in \underset{m: (0, T] \to \mathbbm{R}_+}{\arg\min} \textsc{Cost}^{\tlambda}(m, T).
\end{equation}
The above minimum exists (see Remark \ref{rem:minexists}). 
However, in this case, the performance decays drastically even for relatively small prediction errors. 
Indeed, if the actual arrival rate $\lambda(\cdot)$ is higher than the predicted arrival rate $\tlambda(\cdot)$ at the start, then the associated workload could stay in the queue until the end of the time horizon $[0,T]$ and incur a significant waiting cost.
We instead propose Adapt to the Prediction (AP), which consists of an offline and an online component. The offline component computes an estimate for the number of servers upfront based on $\tlambda(\cdot)$. The online component follows the offline estimate, but dynamically adapts the number of servers based on discrepancies between the predicted and actual arrival rates. Let us define 
$$\Delta\lambda(t) :=
\begin{cases}
\left( \lambda(t) - \tlambda(t) \right)^+ & \text{for}\quad t \geq 0\\
0 & \text{for}\quad t < 0
\end{cases}$$ 
Algorithm \ref{alg:ftpalg} below describes AP.

\begin{algorithm}
\DontPrintSemicolon
Choose $m(\cdot)$ such that at each time $t \geq 0$:
\begin{equation}
    m(t) = m_1(t) + m_2(t),
\end{equation}
where
\begin{align}
    m_1 &\in \underset{m: (0, T] \to \mathbbm{R}_+}{\arg\min} \textsc{Cost}^{\tlambda}(m, T), \\
    \frac{\diff m_2(t)}{\diff t} &= \sqrt{\frac{\omega}{2 \beta}} \cdot \left( \Delta\lambda(t) - \Delta\lambda\left( t - \sqrt{2 \beta/\omega} \right) \right). \label{eq:seccomftp}
\end{align}
\caption{ AP }
\label{alg:ftpalg}
\end{algorithm}

The number of servers under AP consists of two components, an offline component $m_1$ and an online component $m_2$. The offline component $m_1$ is determined upfront by the optimal number of servers to handle the predicted arrival rate $\tlambda$. The online component $m_2$ is determined in an online manner and it reacts if the actual arrival rate turns out to be higher than the predicted arrival rate. Note that if we solve equation \eqref{eq:seccomftp}, then we obtain the number of servers $m_2(t)$, which is differentiable for all $t \geq 0$. The online component works as follows.
If $\Delta \lambda(t) > 0$, then the online component increases the service capacity at rate $\sqrt{\omega / (2 \beta)}$. In other words, for each additional unit of workload received, the number of servers is increased by $\sqrt{\omega / (2 \beta)}$. After a fixed time of $\sqrt{2 \beta / \omega}$ the number of servers is decreased again. Intuitively, if $\omega \gg \beta$, then the online component turns on many servers for a short period of time, whereas if $\beta \gg \omega$, then the online component turns on a few servers for a longer period of time.

\begin{remark}
\blue{The constants $\sqrt{\omega / (2 \beta)}$ and $\sqrt{2 \beta / \omega}$ in~\eqref{eq:seccomftp} are chosen to minimize the competitive ratio of AP. More specifically, one can prove a similar performance guarantee as in Theorem \ref{th:crftp} below, but for any arbitrary choice of these constants. The choice of $\sqrt{\omega / (2 \beta)}$ and $\sqrt{2 \beta / \omega}$ is the unique minimizer of the competitive ratio. Hence, AP in its current form outperforms any other choice of constants in the worst-case.}
\end{remark}

Although the optimization in the offline component might be expensive, it has to be performed only once at the start. \blue{To solve the minimization problem, one could, for example, use the offline approximation technique which we describe in Section \ref{sec:offlinelp}.} Moreover, if the predictions are based on historical data, the offline component $m_1$ might even be precomputed and retrieved from memory at the start. The online component in contrast is computationally cheap. \\

The competitive ratio of AP, of course, depends on the accuracy of the predictions. Theorem~\ref{th:crftp} characterizes the performance of AP. Recall the definition of the competitive ratio $\textsc{CR}(\eta)$ from Section~\ref{sec:prelim}. 

\begin{theorem}
\label{th:crftp}
\blue{

Fix any finite time horizon $T$, arrival rate function $\lambda(\cdot)$, and prediction $\tlambda(\cdot)$. 
Let $m(t)$ be the number of servers of AP (Algorithm \ref{alg:ftpalg}) and $\textsc{Opt}$ be as defined in~\eqref{eq:opt}. Then,
\begin{equation}
    \textsc{Cost}^\lambda(m, T) \leq \textsc{Opt} + \left( \sqrt{2 \omega \beta} + \theta \right) T \cdot \lVert \tlambda - \lambda \rVert_{MAE}.
\end{equation}
Let $\textsc{CR}(\eta)$ denote the competitive ratio of AP (Algorithm \ref{alg:ftpalg}) when it has access to an $\eta$-accurate prediction. Then, as a result of the above,
\begin{equation}
    \textsc{CR}(\eta) \leq 1 + \left( \sqrt{2 \omega \beta} + \theta \right) \eta.
\end{equation}

}
\end{theorem}

The proof of Theorem \ref{th:crftp} is provided in Appendix \ref{app:proofftp}. If $\eta$ is small, then the competitive ratio is close to one. In fact, AP replicates the optimal solution exactly if the predictions turn out to be accurate and hence, is $1$-consistent. Moreover, the competitive ratio also degrades gracefully in the prediction's accuracy, which, as discussed earlier, is not achieved by the offline component $m_1$ alone.

\begin{remark}
\label{rem:blindlyfollow}
Although AP does not follow the predictions blindly, AP is not competitive, since it is not hard to verify that $\textsc{CR}(\eta) \to \infty$ as $\eta \to \infty$ (e.g. let $\tlambda(t) \to \infty$ uniformly for all $t \in [0, T]$). 
\blue{Note that earlier algorithms proposed in the literature, such as the RHC and LCP algorithms from~\cite{lin2012dynamic}, are proven to be competitive only if predictions are accurate, i.e., consistent in the terminology of the current paper. As these algorithms follow the predictions blindly, these algorithms are therefore not competitive if predictions are inaccurate.} Hence, the goal in the next subsection is to combine the above approaches of BCS and AP to obtain an algorithm which follows the predictions most of the time, but ignores the predictions when appropriate.
\end{remark}

\begin{remark}
\blue{The competitive ratio bound is scale-invariant in the weights $\omega$, $\beta$, $\theta$. For example, if each of the weights $\omega$, $\beta$, $\theta$ is doubled, then the factor $\sqrt{2 \omega \beta} + \theta$ doubles as well. However, since $\textsc{Opt}$ doubles, the accuracy $\eta$ is halved (recall Definition \ref{def:error}).}
\end{remark}

\subsubsection{Adaptive Balanced Capacity Scaling.}
\label{sec:robustml}

We now answer the question of how to follow the predictions most of the time without trusting them blindly. The Adaptive Balanced Capacity Scaling (ABCS) algorithm we propose, strategically combines BCS and AP introduced earlier. 
Let $\tm(\cdot)$ be the number of servers as turned on by AP (Algorithm \ref{alg:ftpalg}). Let $\tq(\cdot)$ be the queue length process under AP, that is,
\begin{equation}
    \tq(t) = \int_0^t (\lambda(s) - \tm(s)) \mathbbm{1}\{ \tq(s) > 0 \text{ or } \lambda(s) \geq \tm(s) \} \diff s \qquad\text{ for all } t \geq 0.
\end{equation}
\blue{ABCS is} parameterized by four non-negative numbers $R_1 \geq r_1 \geq 0$ and $R_2 \geq r_2 \geq 0$. Algorithm \ref{alg:bcsmlalg} below describes ABCS for any fixed choices of $R_1, r_1, R_2, r_2$. \\

\begin{algorithm}
\DontPrintSemicolon
Choose $m(\cdot)$ such that at each time $t \geq 0$:
\begin{equation}
\label{eq:diffcapacityml}
    \frac{\diff m(t)}{\diff t} = \frac{\hat{r}_1(t) \omega \cdot q(t) - \hat{r}_2(t) \theta \cdot m(t)}{\beta},
\end{equation}
where
\begin{equation}
\begin{aligned}
    \hat{r}_1(t) &= \begin{cases}
        r_1 &\text{ if } m(t) - \tm(t) > [q(t) - \tq(t)]^+ \cdot \sqrt{\frac{\omega}{2 \beta}}, \\
        R_1 &\text{ if } m(t) - \tm(t) \leq [q(t) - \tq(t)]^+ \cdot \sqrt{\frac{\omega}{2 \beta}},
    \end{cases} \\
    \hat{r}_2(t) &= \begin{cases}
        R_2 &\text{ if } m(t) > \tm(t) \text{ and } q(t) \leq \tq(t), \\
        r_2 &\text{ if } m(t) \leq \tm(t) \text{ or } q(t) > \tq(t).
    \end{cases}
\end{aligned}
\end{equation}
\caption{ ABCS ($r_1, r_2, R_1, R_2$)}
\label{alg:bcsmlalg}
\end{algorithm}

\begin{remark}
\label{rem:othererrormeasure}
\blue{It is worthwhile to highlight that ABCS is oblivious to the choice of AP as the source of the advised number of servers $\tm(\cdot)$. Therefore, if there exists an algorithm similar to AP but with a better error dependence, then it is straightforward to extend ABCS to use this algorithm as the source for the advised number of servers instead. The improved error dependence carries over immediately into the competitive ratio of ABCS (see also Proposition \ref{prop:predcr}).}
\end{remark}

In spirit, \blue{ABCS} works similarly to \blue{BCS}. In fact, if $R_1 = r_1$ and $R_2 = r_2$, then \blue{ABCS} is equivalent to \blue{BCS} and disregards predictions altogether. 
However, in contrast to the constant rates $r_1$ and $r_2$ of BCS, the rates at which ABCS reacts, is captured by the state-dependent rate functions $\hat{r}_1(t)$ and $\hat{r}_2(t)$. The reason behind the precise choices of $\hat{r}_1(t)$ and $\hat{r}_2(t)$ will be clear later from the performance of the algorithm.
From a high-level perspective, these are chosen to approach the behavior of the advised number of servers $\tm(t)$ of AP. 
Indeed, if ABCS has less than the advised number of servers $\tm(t)$, then it increases $m(t)$ at the higher rate $R_1$ and decreases it at the lower rate $r_2$. 
Similarly, if ABCS has ``sufficiently more'' servers than the advised number $\tm(t)$, then it increases $m(t)$ at the lower rate $r_1$ and decreases it at the higher rate $R_2$. 
The number of servers of ABCS therefore judiciously approaches the number of advised servers. 
However, it does not blindly follow $\tm(t)$ to protect against inaccurate predictions.
For example, if the workload $q(t)$ is significantly higher than the current number of servers $m(t)$, then ABCS will always increase the number of servers at a non-zero rate.

Our main result characterizes the performance of ABCS analytically, which is presented in Theorem~\ref{th:crbcs} below. 
The proof of Theorem~\ref{th:crbcs} is provided in Section \ref{sec:proofbcs}. 
Recall the definition of the competitive ratio $\textsc{CR}(\eta)$ from Section~\ref{sec:prelim}.

\begin{theorem}
\label{th:crbcs}
Let $\textsc{CR}(\eta)$ denote the competitive ratio of ABCS (Algorithm \ref{alg:bcsmlalg}) when it has access to an $\eta$-accurate prediction. Then,
\begin{equation}
    \textsc{CR}(\eta) \leq \min\left((1 + (\sqrt{2 \omega \beta} + \theta) \eta) \cdot \textsc{OCR}, \textsc{PCR} \right),
\end{equation}
where
\begin{equation}
\begin{gathered}
\label{eq:defrhogamma}
    \textsc{OCR} = \max\left( c_1 r_1, \frac{c_2 R_1}{\sqrt{1 + 2 R_1}}, c_2 + c_3, c_4 \right), \quad \textsc{PCR} = \max\left(c_5 R_1, 2 c_6, 2 c_6 R_2 + 1 - \frac{R_2}{r_2} \right), \\
    \begin{aligned}[t]
        c_1 &= 1 + \frac{1}{r_1} + \frac{1}{R_2}, &
        c_2 &= \frac{c_1 \sqrt{1 + 2 r_1} - c_1 + c_3}{\sqrt{1 + 2 R_1}}, &
        c_3 &= 1 + \frac{1}{R_1} + \frac{1}{R_2}, \\
        c_4 &= 1 + r_2 + \frac{r_2}{R_1} + c_2 r_2, &
        c_5 &= 1 + \frac{1}{r_1} + \frac{1}{r_2}, &
        c_6 &= c_5 \sqrt{\frac{R_1}{r_1}}.
    \end{aligned}
\end{gathered}
\end{equation}
\end{theorem}

Theorem~\ref{th:crbcs} characterizes the competitive ratio of ABCS explicitly for any choices of the parameters. 
Note that for any value of $\eta$, the competitive ratio is at most \textsc{PCR}. 
Moreover, if $\eta$ is small, then the competitive ratio is close to \textsc{OCR}. 
It is straightforward to check from Theorem \ref{th:crbcs} that ABCS satisfies the three desiderata of Definition~\ref{def:desiderata}. 
In particular, ABCS is \textsc{OCR}-consistent and \textsc{PCR}-competitive. 
The constants \textsc{OCR} and \textsc{PCR}, of course, depend on the parameters $R_1$, $r_1$, $R_2$ and $r_2$.
\blue{Corollary~\ref{cor:bcr_wcr} provides guidance on how to choose these parameters asymptotically optimally.}

\begin{figure}
\centerline{\begin{subfigure}[t]{.5\textwidth}
    \centering
    \includegraphics[width=\textwidth]{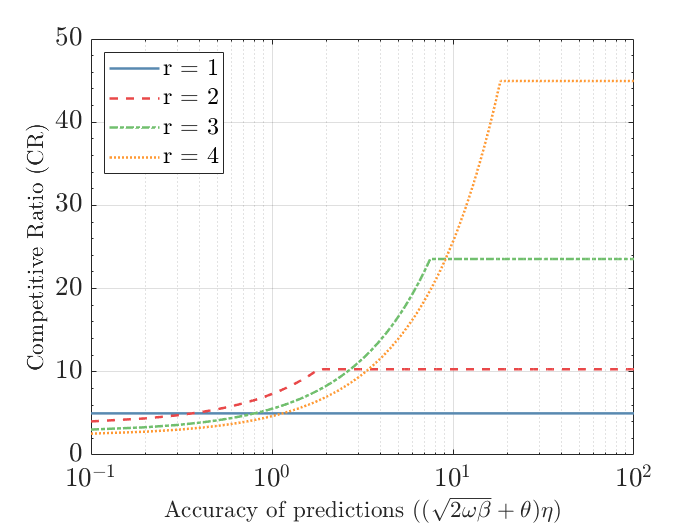}
    \caption{\blue{The competitive ratio as a function of the normalized accuracy of the predictions $(\sqrt{2 \omega \beta} + \theta) \eta$ for varying values of the confidence $r$. The competitive ratio increases as predictions are less accurate, but remains bounded.}}
    \label{fig:cr_eta}
\end{subfigure}\hspace{5mm}%
\begin{subfigure}[t]{.5\textwidth}
    \centering
    \includegraphics[width=\textwidth]{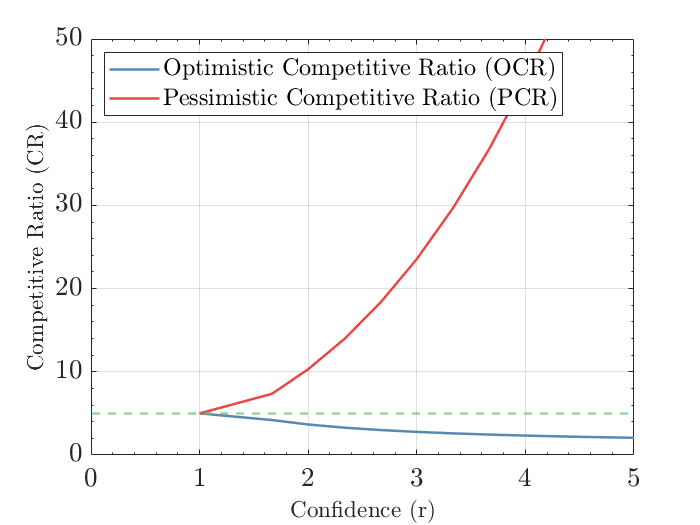}
    \caption{\blue{The Pessimistic Competitive Ratio (PCR) and the Optimistic Competitive Ratio (OCR) as a function of the confidence $r$. The figure interpolates between the purely online scenario ($\textsc{OCR} = \textsc{PCR} = 5$) and the machine learning scenario ($\textsc{OCR} = 1$ and $\textsc{PCR} = \infty$).}}
    \label{fig:bcr_wcr}
\end{subfigure}}
\caption{The analytical performance of ABCS (Algorithm \ref{alg:bcsmlalg}).}
\end{figure}

\begin{corollary}
\label{cor:bcr_wcr}
Let $r \geq 1.1$ be a hyperparameter, representing the confidence in the predictions. Let $\textsc{CR}(\eta)$ denote the competitive ratio of ABCS (Algorithm \ref{alg:bcsmlalg}) when it has access to an $\eta$-accurate prediction.
\blue{If $R_1 = 8 r (r - 1)$, $r_1 = r^{-1}$, $R_2 = 2 r$ and $r_2 = r^{-1}$, then
\begin{equation}
    \textsc{CR}(\eta) \leq \min\left( ( 1 + (\sqrt{2 \omega \beta} + \theta) \eta ) \cdot \left( 1 + \frac{1}{r} + \frac{5}{8 r^2} + \mathcal{O}\left( \frac{1}{r^3} \right) \right), 16 \sqrt{2} r^{7/2} \right).
\end{equation}}
\end{corollary}
Corollary \ref{cor:bcr_wcr} characterizes the trade-off between the \textsc{OCR} and the \textsc{PCR}. 
If the confidence in the predictions $r$ is set at a high value, then the \textsc{OCR} tends to $1$. 
However, the value of \textsc{PCR} tends to become large in this case, even though, importantly, it remains uniformly bounded as $\eta \to \infty$.
\blue{Figure \ref{fig:cr_eta} plots the competitive ratio as a function of $\eta$ and the confidence hyperparameter $r$. For fixed $r$, the competitive ratio increases linearly in $\eta$ (note that the x-axis is on log-scale). However, if $\eta$ is large, the competitive ratio remains constant in $\eta$ at a value of \textsc{PCR}. Note that, for any $\eta$, the competitive ratio is always~$5$ in the case of zero confidence ($R_1 = r_1$ and $R_2 = r_2$).}

\begin{remark}
\label{rem:cons1}
\blue{The fact that ABCS achieves a consistency close to one, albeit in trade-off with robustness, is quite unique. For example, the seminal paper by  Lykouris and Vassilvtiskii \cite{lykouris2018competitive} provides an algorithm in the context of caching that is at most $2$-consistent. Also, Antoniadis et al. \cite{antoniadis2020online} provide a $9$-consistent deterministic algorithm for the problem of metrical task systems. The randomized algorithm introduced in the same paper is $(1 + \varepsilon)$-consistent but has a large additive factor. Corollary \ref{cor:bcr_wcr} shows that ABCS achieves $(1 + \varepsilon)$-consistency without any additive factor.}
\end{remark}

\begin{remark}
\label{rem:ocr-pcr}
\blue{Figure \ref{fig:bcr_wcr} plots the values of \textsc{PCR} and \textsc{OCR} as a function of the confidence hyperparameter $r$.}
It depicts the interpolation between the purely online scenario ($\textsc{OCR} = \textsc{PCR} = 5$) and the machine learning scenario ($\textsc{OCR} = 1$ and $\textsc{PCR} = \infty$). The current work generalizes these two extremes to any scenario in between.
As mentioned in the introduction, we provide performance guarantees for ABCS for any value of the confidence hyperparameter $r \geq 1$. 
However, the specific choice of $r$ would depend on where the system designer wants to place the system on the red \blue{and blue curves} in Figure~\ref{fig:bcr_wcr}; view it as a risk-vs-gain curve. 
\blue{For example, the figure shows that if one chooses a value of $r$, so that if the predictions turn out to be accurate, ABCS would be 3-competitive, then that would put the system at the risk of being up to about 18-competitive, if the predictions turn out to be completely wrong.
Later, in Proposition~\ref{prop:lowerboundcreta} we show that the trade-off between \textsc{OCR}-vs-\textsc{PCR} that we obtain for ABCS is necessary in the sense that any algorithm which is $(1 + \delta)$-competitive in the optimistic case must be at least $1 / (4 \delta)$-competitive in the pessimistic case.}
\end{remark}

\begin{remark}
\label{rem:predacc}
Recently, there has been some interest in understanding the performance of algorithms when an estimate of the prediction's accuracy $\eta$ is available in terms of some probability distribution~\cite{mitzenmacher2019scheduling, mitzenmacher2019supermarket}.
In such cases, Theorem \ref{th:crbcs} allows one to calculate the optimal choice of confidence hyperparameter $r$ that minimizes the expected competitive ratio.
Assume that the prediction's accuracy $\eta$ follows some distribution $\mu(\cdot)$. The distribution $\mu(\cdot)$ might, for example, be estimated based on historically observed data. 
For a fixed $r$, note that \textsc{OCR} and \textsc{PCR} are functions of $r$. 
Denote
$$\zeta(r) := \frac{\textsc{PCR} - \textsc{OCR}}{2 \textsc{OCR}}.$$
The expected value of the \emph{random} competitive ratio of ABCS is then
\begin{equation}
\begin{aligned}
    \E_{\eta \sim \mu}[\textsc{CR}(\eta)]
    &= \int_0^\infty \min\left( (1 + 2 \eta) \cdot \textsc{OCR}, \textsc{PCR} \right) \diff \mu(\eta) \\
    &= 2 \textsc{OCR} \cdot \int_0^{\zeta(r)} \eta \diff \mu(\eta) + \textsc{OCR} \cdot \int_0^{\zeta(r)} \diff \mu(\eta) + \textsc{PCR} \cdot \int_{\zeta(r)}^\infty \diff \mu(\eta) \\
    &= \begin{gathered}[t]
    2 \textsc{OCR} \cdot \E\left[ \eta \mathbbm{1}\left\{ \eta \leq \zeta(r) \right\} \right] + \textsc{OCR} \cdot \pro\left( \eta \leq \zeta(r) \right) 
    + \textsc{PCR} \cdot \pro\left( \eta > \zeta(r) \right).
    \end{gathered}
\end{aligned}
\end{equation}
Therefore, if either the distribution or an estimate thereof is known, then the parameters of ABCS can be chosen to minimize the expected competitive ratio.
\end{remark}

Theorem \ref{th:crbcs} and Corollary~\ref{cor:bcr_wcr} demonstrate that there is a trade-off between the \textsc{OCR} and the \textsc{PCR}. The following proposition shows that this trade-off is in fact, inherent to the problem and is not an artifact of the algorithm.

\begin{proposition}
\label{prop:lowerboundcreta}
Let $\mathcal{A}$ be any deterministic algorithm for the capacity scaling problem in \eqref{eq:model}, and $\textsc{CR}(\eta)$ denote its competitive ratio when it is has access to an $\eta$-accurate prediction. \blue{There exist choices of $\omega$, $\beta$, and $\theta$ such that, for any $\delta > 0$, if $\textsc{CR}(0) \leq 1 + \delta$, then
\begin{equation}
\label{eq:lowerboundcreta}
    \textsc{CR}\left( \frac{1}{\delta} \right) \geq \frac{1}{4 \delta}.
\end{equation}
In short, any deterministic algorithm that is $(1 + \delta)$-consistent must be $\Omega( 1 / \delta)$-competitive.}
\end{proposition}

Proposition \ref{prop:lowerboundcreta} is proved in Appendix \ref{app:proofcreta-lower-bd}. \blue{In comparing Corollary~\ref{cor:bcr_wcr} and Proposition~\ref{prop:lowerboundcreta}, one may notice that there is a gap between the consistency-competitiveness trade-off achieved by ABCS and the provable lower bound on this trade-off.  Improving the lower bound result in Proposition \ref{prop:lowerboundcreta} or designing an algorithm with a better trade-off is left as an interesting future research direction.}

\blue{As mentioned earlier, an algorithm for capacity scaling must consist of two components: one component decides when to activate a server and the other component decides when to deactivate a server.}
For the latter problem, a popular state-of-the-art approach is to implement a power-down timer \cite{karlin1988competitive, irani2002competitive, augustine2004optimal, lu2012simple, gandhi2013exact, mukherjee2017optimal}. 
The power-down timer works as follows: each time a server becomes idle, the system starts a timer corresponding to that server. 
If the server remains idle after the timer expires, then the server is deactivated. 
\blue{Algorithm \ref{alg:timeralg} shows the Timer algorithm for any choice of power-down timer $\tau: \mathbb{R}_+^3 \to (0, \infty)$.}

\begin{algorithm}
\DontPrintSemicolon

At each time $t \geq 0$: \\
\hspace{5mm} \blue{Turn off a server if the server has been idle for more than $\tau(\omega, \beta, \theta)$ time.}

\caption{The Timer algorithm ($\tau$)}
\label{alg:timeralg}
\end{algorithm}
We end this section by pointing out that, 
although the Timer algorithm above has proven to be successful under specific (especially stochastic) scenarios, 
the worst-case performance of the algorithm in the current context is poor as the following proposition shows. 
\blue{In fact, Proposition~\ref{prop:timer-lower-bd} shows that there do not exist any choices of $\omega$, $\beta$, and $\theta$ such that the Timer algorithm has a bounded competitive ratio}. 
To the best of our knowledge, there does not exist \blue{any competitive algorithm} for the capacity scaling problem where $\omega$ is finite, until in the current work. Proposition~\ref{prop:timer-lower-bd} is proved in Appendix~\ref{app:prooftimer-lower-bd}.
\begin{proposition}
\label{prop:timer-lower-bd}
\blue{
Let $\tau: \mathbb{R}_+^3 \to (0, \infty)$ be any arbitrary function.
For any fixed $\omega, \beta, \theta > 0$, let $\textsc{CR}$ be the competitive ratio of the Timer algorithm (Algorithm \ref{alg:timeralg}) with power-down timer $\tau(\omega, \beta, \theta)$. Then, $\textsc{CR} = \infty$.
}
\end{proposition}

\subsection{Offline algorithm.}
\label{sec:offlinelp}

\blue{We end the main results by providing an approximation algorithm for the offline problem. 
%, i.e., in the case that the weights $\beta$, $\omega$, $\theta$, the arrival rate function $\lambda(\cdot)$ and the finite time horizon $T$ are known upfront. 
Although finding an efficient solution to the offline problem is not the main focus of the current paper, the results in this section will be used to run the numerical experiments in Section \ref{sec:numericalexperiments}. Also, the algorithm may be used by AP to compute the optimal number of servers given the \emph{predicted} arrival function (see Section \ref{sec:ftp}).
Moreover, the approximation algorithm raises a question about the trade-off between numerical complexity and accuracy of the solution, which might be of independent interest.
%Note that, in general, it is non-trivial to compute an optimal solution to \eqref{eq:model}, even if the arrival function is known, since the minimization is taken over a function space of infinite dimension. Therefore, we here introduce an offline algorithm that computes an approximation of the offline optimal solution. Also, 

As a measure of numerical complexity, let us introduce the following regularity assumption on the arrival rate function.
\begin{assumption}[Regular]
\label{as:regular}
We say that a function $\lambda: [0, T] \to \mathbb{R}_+$ is $\delta$-regular if $\lambda(i \delta + s) = \lambda(i \delta)$ for all $s \in [0, \delta)$ and $i = 0, 1, \dots, \lfloor T / \delta \rfloor$.
\end{assumption}
The regularity assumption is a reasonable approximation for any arrival rate function occurring in practice for $\delta$ sufficiently small. 
Now, we state the proposed offline approximation algorithm. For the sake of notation, assume that $T$ is divisible by $\delta$. Let $n = T / \delta$, $q_1 = 0$ and $m_0 = 0$. The offline algorithm solves the following linear program:
\begin{equation}
\begin{aligned}
\label{eq:offlinelp}
\underset{m, d \in \mathbbm{R}^n, q \in \mathbbm{R}^{n+1}}{\text{minimize}}\hspace{3mm} &\omega \delta \cdot \sum_{i = 1}^n \frac{q_i + q_{i+1}}{2} + \beta \cdot \sum_{i = 1}^n d_i + \theta \delta \cdot \sum_{i = 1}^n m_i \\
\text{subject to}\hspace{3mm} &\begin{aligned}[t]
    &q_{i+1} \geq q_i + \int_{(i-1) \delta}^{i \delta} \lambda(t) \diff t - \delta m_i &\text{ for all } i = 1, \dots, n \\
    &d_i \geq m_i - m_{i-1} &\text{ for all } i = 1, \dots, n \\
    &q_{i+1}, d_i, m_i \geq 0 &\text{ for all } i = 1, \dots, n
\end{aligned}
\end{aligned}
\end{equation}
The linear program is a discretization of the problem in \eqref{eq:model}. The vectors $q$ and $m$ represent the workload in the buffer and the number of servers, respectively. To obtain a solution from \eqref{eq:offlinelp} for the original problem in \eqref{eq:model}, set $m(i \delta + s) = m_{i+1}$ for all $s \in [0, \delta)$ and $i = 0, 1, \dots, T / \delta - 1$. Hence, the linear program in \eqref{eq:model} computes the number of servers that would minimize the cost but where the number of servers is restricted to be $\delta$-regular. The constraints and objective value in \eqref{eq:model} then directly reduce to the constraints and objective value in \eqref{eq:offlinelp}.
Theorem~\ref{th:lin-appx} below characterizes the competitive ratio of the solution to the above linear program w.r.t.~the offline optimum \textsc{Opt} in~\eqref{eq:opt}, for any fixed choice of $\delta > 0$. The proof of Theorem~\ref{th:lin-appx} is given in Appendix~\ref{app:prooflinx-appx}. 

\begin{theorem}
\label{th:lin-appx}
Let \textsc{CR} denote the competitive ratio of the solution to the linear program \eqref{eq:offlinelp}. If $\lambda$ is $\delta$-regular then,
\begin{equation}
    \textsc{CR} \leq \left( 1 + \frac{\omega \delta}{2 \theta} \right) \left( 1 + \frac{\omega \delta^2}{\beta} \right).
\end{equation}
\end{theorem}
%We use the offline approximation algorithm in Section \ref{sec:numericalexperiments} to compute the optimal offline policy. Also, the offline approximation algorithm may be used by AP to compute the optimal number of servers given the \emph{predicted} arrival function (see Section \ref{sec:ftp}).
}

\section{Numerical experiments.}
\label{sec:numericalexperiments}

\blue{

We implemented the algorithms proposed in the current paper and evaluated their performance on both a real-world dataset of internet traffic and two artificially generated workloads. The real-world dataset consists of DNS requests observed at a campus network across four consecutive days in April 2016 \cite{singh2016data}. We let $\lambda(t)$ to be the number of requests per second according to this dataset. To empirically verify the performance of our algorithms against extreme cases, we also tested the algorithms on two artificially generated arrival rate functions. The arrival rate functions present highly stylized versions of particular patterns which may occur in real-world traffic. Here, we let $\lambda(t)$ to be either sinusoidal or a step-function. The weights $\beta$, $\theta$, $\omega$ are modeled after realistic parameters \cite{Shebabi2016, lin2012dynamic}. In particular, we assume that each server consumes 850W at a price of 0.15 cents/kWh, $\beta$ is equal to the power-cost of running a server for four hours and $\omega =$ 0.1 cents.

\subsection{Purely online scenario.}

We first test the algorithms AP, BCS, and the Timer algorithm that do not require predictions (for the AP algorithm, we let $\tlambda(t) = 0$ for all $t$). The Timer algorithm has a threshold of $\tau(\omega, \beta, \theta) = \beta / \theta$ and turns servers on whenever $\lambda(t) > m(t)$, which is typically used in the literature \cite{karlin1988competitive, irani2002competitive, augustine2004optimal, lu2012simple, gandhi2013exact, mukherjee2017optimal}. Figure \ref{fig:time_plot} shows the number of servers on the real-world dataset and the artificial patterns. Table \ref{tab:crnopred} summarizes the competitive ratios in each scenario.

The performance of BCS on the real-world dataset is excellent, only 20\% more than the offline optimal solution, even without predictions. On the artificial patterns, BCS is outperformed by both AP and the Timer algorithm. AP and the Timer algorithm seem to work well if the data does not contain any sudden spikes of workload. Also, note that the competitive ratio of BCS in both cases is significantly lower than the worst-case competitive ratio of 5.

\begin{figure}
\centerline{\begin{subfigure}[t]{.33\textwidth}
    \centering
    \includegraphics[width=\textwidth]{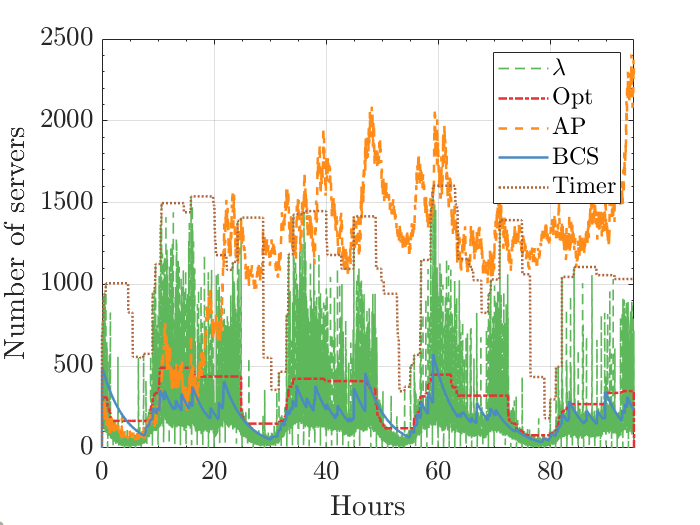}
    \caption{Real-world}
\end{subfigure}\hspace{5mm}%
\begin{subfigure}[t]{.33\textwidth}
    \centering
    \includegraphics[width=\textwidth]{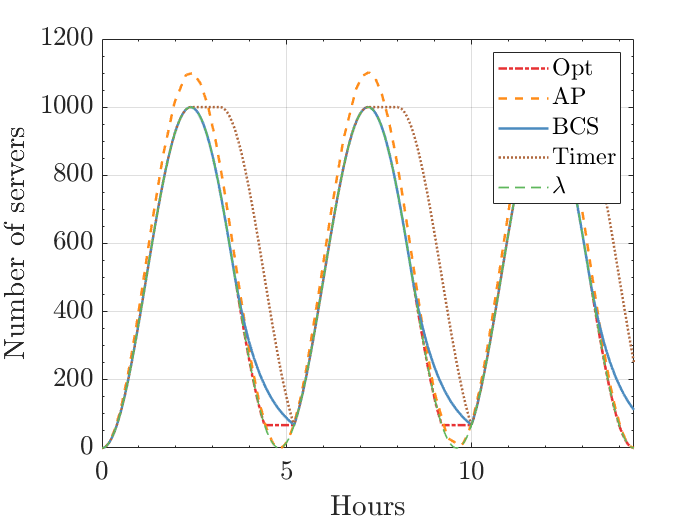}
    \caption{Sinusoidal}
\end{subfigure}\hspace{5mm}%
\begin{subfigure}[t]{.33\textwidth}
    \centering
    \includegraphics[width=\textwidth]{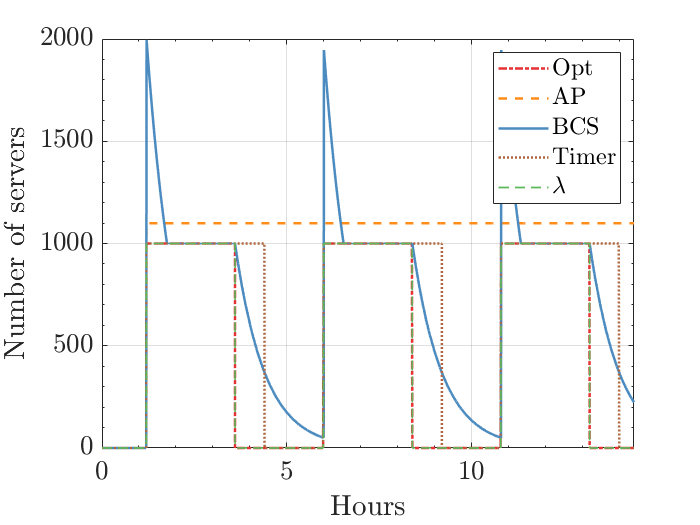}
    \caption{Step-function}
\end{subfigure}}
\caption{The real-world arrival pattern and two artificial arrival patterns considered and the number of servers of Opt, AP, BCS and the Timer algorithm.}
\label{fig:time_plot}
\end{figure}

\begin{table}
\centering
\begin{tabular}{|c|ccc|}
\hline
\textsc{CR} & AP & BCS & Timer \\
\hline
Real-world & 27.7 & 1.2 & 3.1 \\ 
\hline
Sinusoidal & 1.1 & 1.4 & 1.2 \\
\hline
Step-function & 1.8 & 2.2 & 1.3 \\
\hline
\end{tabular}
\caption{The competitive ratio of AP, BCS and the Timer algorithm without predictions.}
\label{tab:crnopred}
\end{table}

\subsection{Machine learning scenario.}

Next, we test the algorithms AP and ABCS in the case that predictions are provided. For the real-world dataset, we evaluate three types of predictions:
\begin{itemize}
    \item \textit{Type 1.} The system does not reveal any predictions, i.e., $\tlambda(t) = 0$ for all $t \in [0, T]$.
    \item \textit{Type 2.} The system reports the moving average (MA) across three hours, i.e., $\tlambda(t) = (\min( t, 1.5 ) + \min( T - t, 1.5 ))^{-1} \int_{\max(t - 1.5, 0)}^{\min(t + 1.5, T)} \lambda(t) \diff t$ for all $t \in [0, T]$.
    \item \textit{Type 3.} The system provides perfect predictions, i.e., $\tlambda(t) = \lambda(t)$ for all $t \in [0, T]$.
\end{itemize}
For the artificial patterns, we evaluate four types of predictions:
\begin{itemize}
    \item \textit{Type 1.} The system does not reveal any predictions, i.e., $\tlambda(t) = 0$ for all $t \in [0, T]$.
    \item \textit{Type 2.} The system predicts only the average of the arrival rate, i.e., $\tlambda(t) = 500$ for all $t \in [0, T]$.
    \item \textit{Type 3.} The system predicts the opposite of the arrival rate, i.e., $\tlambda(t) = 1000 - \lambda(t)$ for all $t \in [0, T]$.
    \item \textit{Type 4.} The system provides perfect predictions, i.e., $\tlambda(t) = \lambda(t)$ for all $t \in [0, T]$.
\end{itemize}
Table \ref{tab:crpred} summarizes the competitive ratios in each scenario, where ABCS is evaluated for three choices of the confidence hyperparameter (low confidence ($r = 1$), medium confidence ($r = 3$) and high confidence ($r = 5$)).

The performance of ABCS on the real-world dataset is excellent. ABCS even reproduces an optimal solution in the case that perfect predictions are available and the confidence is medium or larger. The performance generally improves as more accurate predictions are available. Moreover, in contrast to AP, ABCS is robust against inaccurate predictions. The competitive ratio of ABCS is close to one for type 1 predictions (and type 3 predictions for the artificial patterns), even in the case of high confidence. Hence, in practice, ABCS is able to reproduce the optimal solution if sufficiently accurate predictions are provided while maintaining a competitive ratio close to one, even if the predictions are completely inaccurate.

\begin{table}
\centering
\begin{tabular}{|cc|cccc|}
\hline
$\textsc{CR}(\eta)$ && AP & ABCS (low conf.) & ABCS (medium conf.) & ABCS (high conf.)  \\
\hline
Real-world & type 1 & 27.7 & 1.19 & 1.00 & 1.00 \\
& type 2 & 14.2 & 1.19 & 1.07 & 1.38 \\
& type 3 & 1.00 & 1.19 & 1.00 & 1.00 \\
\hline
Sinusoidal & type 1 & 1.10 & 1.43 & 1.16 & 1.12 \\
& type 2 & 1.21 & 1.43 & 1.18 & 1.14 \\
& type 3 & 1.46 & 1.43 & 1.17 & 1.17 \\
& type 4 & 1.00 & 1.43 & 1.11 & 1.15 \\
\hline
Step-function & type 1 & 1.85 & 2.17 & 1.68 & 1.64 \\
& type 2 & 1.72 & 2.17 & 1.61 & 1.56 \\
& type 3 & 1.90 & 2.17 & 1.68 & 1.63 \\
& type 4 & 1.00 & 2.17 & 1.37 & 1.15 \\
\hline
\end{tabular}
\caption{The competitive ratio of AP and ABCS for different values of the confidence hyperparameter in the presence of predictions.}
\label{tab:crpred}
\end{table}

}

\section{Proofs.}
\label{sec:proofs}

\subsection{Proof of Theorem \ref{th:cr1bcs}.}
\label{sec:proofbcs1}

We will provide a high-level overview of the proof of Theorem \ref{th:cr1bcs} and refer to the appendix for the details.
Recall that \blue{BCS} is a special case of \blue{ABCS} (let $R_1 = r_1$, $R_2 = r_2$).
To prove Theorem \ref{th:cr1bcs}, we will, in fact, establish a  more general result in Proposition \ref{prop:worstcasecr} below, where the rates $r_1$ and $r_2$ may vary as rate functions over time. 
Proposition \ref{prop:worstcasecr} states that the competitive ratio of ABCS never exceeds \textsc{PCR} irrespective of the magnitude of the error in prediction. Theorem \ref{th:cr1bcs} thus follows immediately by letting $R_1 = r_1$ and $R_2 = r_2$. 

\begin{proposition}
\label{prop:worstcasecr}
\blue{

Fix a finite time horizon $T$ and arrival rate function $\lambda(\cdot)$. Let $\textsc{Opt}$ be as defined in~\eqref{eq:opt} and  $m(t)$ be the number of servers of ABCS (Algorithm \ref{alg:bcsmlalg}) when it has access to a prediction $\tlambda$. Then
\begin{equation}
    \textsc{Cost}^\lambda(m, T) \leq \textsc{PCR} \cdot \textsc{Opt},
\end{equation}
for all instances $(T, \lambda)$ and predictions $\tlambda$, where \textsc{PCR} is as defined in \eqref{eq:defrhogamma}.

}
\end{proposition}

The proof of Proposition \ref{prop:worstcasecr} is based on a potential function argument and is provided in Appendix \ref{app:proofworstcasecr}. We end this section by giving a proof-sketch of Proposition \ref{prop:worstcasecr}.

\begin{proof}{Proof sketch of Proposition \ref{prop:worstcasecr}.}
Let $m(\cdot)$ be the number of servers of ABCS and $m^*(\cdot)$ be a differentiable optimal solution to the offline optimization problem \eqref{eq:model}.
Appendix \ref{app:proofworstcasecr} shows how to extend this to arbitrary non-differentiable solutions. 
\blue{Let $\Phi(\cdot)$ be a non-negative potential function such that
\begin{equation}
\label{eq:overview_potentialidea}
    \frac{\diff \Phi(t)}{\diff t} + \frac{\partial \textsc{Cost}^\lambda(m, t)}{\partial t} \leq \textsc{PCR} \cdot \frac{\partial \textsc{Cost}^\lambda(m^*, t)}{\partial t},
\end{equation}
and $\Phi(0) = 0$, assuming, for now, that such a $\Phi(\cdot)$ exists. We integrate equation \eqref{eq:overview_potentialidea} from time $t = 0$ to $t = T$ to obtain
\begin{equation}
    \textsc{Cost}^\lambda(m, T) \leq \textsc{PCR} \cdot \textsc{Cost}^\lambda(m^*, T) + \Phi(0) - \Phi(T) \leq \textsc{PCR} \cdot \textsc{Cost}^\lambda(m^*, T),
\end{equation}
where the last step follows because $\Phi(T)$ is non-negative and $\Phi(0) = 0$.} The proof of Proposition \ref{prop:worstcasecr} is therefore completed if we manage to find a potential function $\Phi(t)$ satisfying equation \eqref{eq:overview_potentialidea} and \blue{$\Phi(0) = 0$}. Define the potential function $\Phi(t)$ such that
\begin{equation}
\begin{aligned}
\label{eq:potential}
    \Phi(t) = &\begin{cases}
        c_5 \beta \cdot \left( d_{R_1}(t) - m(t) + m^*(t) \right), &\text{ if } m(t) > m^*(t) \\
        c_6 \beta \cdot \left( d_{r_1}(t) - m(t) + m^*(t) \right) &\text{ if } m(t) \leq m^*(t) \\
    \end{cases} \\
     &+ \frac{\beta \cdot m(t)}{r_2}
    + c_6 R_2 \theta \cdot [q(t) - q^*(t)]^+,
\end{aligned}
\end{equation}
where $c_5$ and $c_6$ are as defined in equation \eqref{eq:defrhogamma} and
\begin{equation}
    d_r(t) = \sqrt{\frac{r \omega \cdot \left([q(t) - q^*(t)]^+\right)^2}{\beta} + (m(t) - m^*(t))^2}.
\end{equation}
Note that $\Phi(t)$ is non-negative and \blue{$\Phi(0) = 0$}.
\begin{remark}
\blue{Let us, at least quantitatively, discuss the intuition behind the potential function~$\Phi(t)$. Note that the higher the value of $\Phi(t)$, the worse ABCS is doing in comparison to the offline optimum. The most important term in $\Phi(t)$ is the term involving $d_r(t) - m(t) + m^*(t)$, which is responsible for bounding the switching and waiting cost of ABCS. The term consists of the difference $[q(t) - q^*(t)]^+$, which is zero if $q(t) \leq q^*(t)$ and strictly positive otherwise. Clearly, if $q(t) > q^*(t)$, then ABCS needs to spend more in the future to reduce the additional workload. Now, the workload difference is enclosed in an $\ell_2$-metric together with the difference in the number of servers. This is the crux of the potential function and leads to a diminishing marginal penalty of the queue length difference. The larger the difference $m(t) - m^*(t)$, the less influence a unit increase of $[q(t) - q^*(t)]^+$ has on the potential function. Hence, the penalty of the queue length difference is measured relative to the difference in the number of servers, which is natural and essential in the proof. For example, both the derivative of $q(t)$ is a function of $m(t)$ and, for ABCS, the derivative of $m(t)$ is a function of $q(t)$. The rest of the proof sketch is also instructional to understand the intuition behind the choice of the potential function.}
\end{remark}
It remains to show that $\Phi(t)$ satisfies equation \eqref{eq:overview_potentialidea}, which profoundly relies on $d_r(t)$. The full argument involves a case distinction and is provided in Appendix \ref{app:proofworstcasecr}. For this proof sketch, let us only consider the case that $m(t) > m^*(t)$, $q(t) > q^*(t)$, $\frac{\diff m}{\diff t} \geq 0$ and $\frac{\diff m^*}{\diff t} \geq 0$. Recall that, by the definition of ABCS,
\begin{equation}
\begin{aligned}
    \frac{\diff q(t)}{\diff t} &= \lambda(t) - m(t), &
    \frac{\diff m(t)}{\diff t} &= \frac{\hat{r}_1(t) \omega \cdot q(t) - \hat{r}_2(t) \theta \cdot m(t)}{\beta} \leq \frac{R_1 \omega \cdot q(t)}{\beta}.
\end{aligned}
\end{equation}
The derivative of $d_{R_1}(t)$ is at most
\begin{equation}
\begin{aligned}
    \beta \cdot \frac{\diff d_{R_1}}{\diff t}
    &\leq d_{R_1}(t)^{-1} \cdot
    \left( \begin{aligned}
        &R_1 \omega \cdot (q(t) - q^*(t)) (\lambda(t) - m(t)) \\
        + &R_1 \omega \cdot (q^*(t) - q(t)) (\lambda(t) - m^*(t)) \\
        + &R_1 \omega \cdot q(t) \cdot (m(t) - m^*(t)) \\
        + &\beta \cdot \frac{\diff m^*}{\diff t} \cdot (m^*(t) - m(t))
    \end{aligned} \right) \\
    &= d_{R_1}(t)^{-1} \cdot (m(t) - m^*(t)) \left( R_1 \omega \cdot q^*(t) - \beta \cdot \frac{\diff m^*}{\diff t} \right) 
    \leq R_1 \omega \cdot q^*(t).
\end{aligned}
\end{equation}
Crucially, the derivative does not contain any terms involving $m(t)$ or $q(t)$, but only $q^*(t)$, which is easy to bound by the cost of the optimal solution. Thus, the derivative of $\Phi(t)$ can be upper bounded as follows:
\begin{equation}
\label{eq:eq:exdiffphi}
    \frac{\diff \Phi(t)}{\diff t}
    \leq c_5 R_1 \omega \cdot q^*(t)
    - \left( 1 + \frac{1}{r_1} \right) \beta \cdot \frac{\diff m}{\diff t}
    + c_5 \beta \cdot \frac{\diff m^*}{\diff t}
    + \left( 1 + \frac{R_2}{r_1} \right) \theta \cdot (m^*(t) - m(t)),
\end{equation}
where \blue{the constants} $c_5$ and $c_6$ have been expanded according to their definitions in \eqref{eq:defrhogamma}. Observe that the derivative of the potential function $\Phi(t)$ exactly cancels the cost of ABCS, since $\omega \cdot q(t) \leq \frac{\beta}{r_1} \cdot \frac{\diff m(t)}{\diff t} + \frac{R_2 \theta m(t)}{r_1}$. We therefore obtain
\begin{equation}
\begin{aligned}
    \frac{\diff \Phi(t)}{\diff t} + \frac{\partial \textsc{Cost}^\lambda(m, t)}{\partial t}
    &\leq c_5 R_1 \omega \cdot q^*(t)
    + c_5 \beta \cdot \frac{\diff m^*}{\diff t}
    + \left( 1 + \frac{R_2}{r_1} \right) \theta \cdot m^*(t) \\
    &\leq \max\left( c_5 R_1, c_5, 1 + \frac{R_2}{r_1} \right) \cdot \frac{\partial \textsc{Cost}^\lambda(m^*, t)}{\partial t}.
\end{aligned}
\end{equation}
Note that removing the term $d_r(t)$ from $\Phi(t)$ would yield a similar form as equation \eqref{eq:eq:exdiffphi}. However, the resulting potential function is not non-negative, hence the need for the term $d_r(t)$. \hfill$\blacksquare$
\end{proof}

\subsection{Proof of Theorem \ref{th:crbcs}.}
\label{sec:proofbcs}

Theorem \ref{th:crbcs} states that the competitive ratio is at most the minimum of the Optimistic Competitive Ratio (\textsc{OCR}) and the Pessimistic Competitive Ratio (\textsc{PCR}). Proposition \ref{prop:worstcasecr} in the previous section showed that the competitive ratio of ABCS is at most \textsc{PCR}. To prove the bound on the competitive ratio by \textsc{OCR}, we will relate the performance of ABCS to the cost achieved by the subroutine AP. Proposition \ref{prop:predcr} below states that the cost of ABCS differs by at most a factor of \textsc{OCR} from the cost of AP.

\begin{proposition}
\label{prop:predcr}
\blue{

Fix a finite time horizon $T$ and arrival rate function $\lambda(\cdot)$. Let $\textsc{Opt}$ be as defined in~\eqref{eq:opt} and $m(t)$ be the number of servers of BCS (Algorithm \ref{alg:bcsalg}) when it has access to a prediction $\tlambda$. Then
\begin{equation}
    \textsc{Cost}^\lambda(m, T) \leq \textsc{OCR} \cdot \textsc{Cost}^\lambda(\tm, T),
\end{equation}
for all instances $(T, \lambda)$ and predictions $\tlambda$, where $\tm(t)$ represents the advised number of servers of AP and $\textsc{OCR}$ is as defined in \eqref{eq:defrhogamma}.

}
\end{proposition}

\blue{As already hinted at in Remark \ref{rem:predacc}, Proposition \ref{prop:predcr} is independent of the source of the advice and holds for any function $\tm(\cdot)$. Therefore, if there exists another algorithm that provides advice, besides AP, then this advice may be readily used in ABCS and the competitive ratio of ABCS with respect to the new advice is again at most \textsc{OCR}.}
The proof of Proposition \ref{prop:predcr} is based on a potential function argument and is provided in Appendix \ref{app:proofpredcr}. The proof follows a similar structure as the proof of Proposition \ref{prop:worstcasecr}. We now have all the ingredients to prove Theorem~\ref{th:crbcs}. 

\begin{proof}{Proof of Theorem \ref{th:crbcs}.}
\blue{

The proof follows almost immediately from Proposition \ref{prop:worstcasecr} and \ref{prop:predcr}. Note that $\textsc{CR}(\eta) \leq \textsc{PCR}$ because
\begin{equation}
    \textsc{Cost}^\lambda(m, T) \leq \textsc{PCR} \cdot \textsc{Opt},
\end{equation}
by Proposition \ref{prop:worstcasecr}. Next, $\textsc{CR}(\eta) \leq (1 + (\sqrt{2 \omega \beta} + \theta) \eta) \cdot \textsc{OCR}$ because
\begin{equation}
    \textsc{Cost}^\lambda(m, T)
    \leq \textsc{OCR} \cdot \textsc{Cost}^\lambda(\tm, T)
    \leq (1 + (\sqrt{2 \omega \beta} + \theta) \eta) \cdot \textsc{OCR} \cdot \textsc{Opt},
\end{equation}
by Proposition \ref{prop:predcr} and Theorem \ref{th:crftp}. \hfill$\blacksquare$

}
\end{proof}

\section{Conclusion.}
\label{sec:conclusion}

In this paper, we explored how ML predictions can be used to improve the performance of capacity scaling solutions without sacrificing robustness. We extend the state of the art in capacity scaling by analyzing a more general model in continuous time where tasks are allowed to wait, in which case popular earlier proposed algorithms are not competitive. The Balanced Capacity Scaling (BCS) algorithm we proposed is $5$-competitive in the general case. We also introduced Adapt to the Prediction (AP) which is $1$-competitive if the ML predictions are accurate. Finally, we proposed \blue{Adaptive Balanced Capacity Scaling (ABCS)}, which combines the ideas behind BCS and AP.
We proved that, in the presence of accurate ML predictions, ABCS is $(1 + \varepsilon)$-competitive and its performance degrades gracefully in the prediction's accuracy. Moreover, the competitive ratio of ABCS is at most $\mathcal{O}\left( \varepsilon^{-7/2} \right)$ when ML predictions are inaccurate. Although the competitive ratio of ABCS depends on the accuracy, the algorithm is oblivious to it. In the context of data centers, since real-world internet traffic is erratic, any implemented capacity scaling solution must be robust against sudden unpredictable surges in workload. Our results yield significant reductions in power consumption while maintaining robustness against these sudden spikes. %The theoretical results are supported by extensive numerical experiments on a real-world dataset.

\blue{An interesting, yet challenging direction for future work is to restrict the number of servers to be integer valued. A common approach is to randomly round a fractional solution, such as the one returned by ABCS, and prove that the increase in competitive ratio due to the random rounding is at most a constant factor in expectation. Apart from the fact that one cannot expect $(1 + \varepsilon)$-consistency by this procedure, the integrality gap is in fact unbounded for this problem, as shown by Lemma \ref{lemma:integralgap} below. The proof of Lemma \ref{lemma:integralgap} is provided in Appendix \ref{app:integralgap}.
\begin{lemma}
\label{lemma:integralgap}
Let $\textsc{Opt}_{int}$ be the offline minimum cost where $m(\cdot)$ is restricted to be integer-valued, i.e.,
\begin{equation}
    \textsc{Opt}_{int} := \inf_{m: (0, T] \to \mathbb{N}} \textsc{Cost}^\lambda(m, T).
\end{equation}
Also, let $CR$ denote the competitive ratio of $\textsc{Opt}_{int}$ w.r.t.~\textsc{Opt} in \eqref{eq:opt}. Then, $\textsc{CR} = \infty$.
\end{lemma}
Lemma \ref{lemma:integralgap} forbids any rounding procedure from being competitive and we therefore leave the integral case as an open question.} Finally, in an ongoing work, we are exploring how the confidence hyperparameter of ABCS can be learned over time if there are statistical guarantees on the prediction's accuracy.

% Acknowledgments here
\section*{Acknowledgements.}
The authors thank Ravi Kumar (Google Research) for inspiring discussions that started this project.
The work is partially supported by the National Science Foundation under Grant No.~2113027.
We also gratefully acknowledge the financial support for this project from the ARC-TRIAD fellowship at Georgia Tech.

\def\UrlBreaks{\do\/\do-}
\bibliographystyle{apalike}
\bibliography{main}

\begin{thebibliography}{}

\bibitem[Albers and Fujiwara, 2007]{Albers2007}
Albers, S. and Fujiwara, H. (2007).
\newblock Energy-efficient algorithms for flow time minimization.
\newblock {\em ACM Transactions on Algorithms (TALG)}, 3(4):49--es.

\bibitem[Albers et~al., 2014]{albers2014speed}
Albers, S., M{\"u}ller, F., and Schmelzer, S. (2014).
\newblock Speed scaling on parallel processors.
\newblock {\em Algorithmica}, 68(2):404--425.

\bibitem[Anderson and Karlin, 1996]{anderson1996two}
Anderson, C. and Karlin, A.~R. (1996).
\newblock Two adaptive hybrid cache coherency protocols.
\newblock In {\em Proceedings. Second International Symposium on
  High-Performance Computer Architecture}, pages 303--313. IEEE.

\bibitem[Antoniadis et~al., 2020]{antoniadis2020online}
Antoniadis, A., Coester, C., Elias, M., Polak, A., and Simon, B. (2020).
\newblock Online metric algorithms with untrusted predictions.
\newblock In {\em International Conference on Machine Learning}, pages
  345--355. PMLR.

\bibitem[Ata and Shneorson, 2006]{ata2006dynamic}
Ata, B. and Shneorson, S. (2006).
\newblock {Dynamic control of an M/M/1 service system with adjustable arrival
  and service rates}.
\newblock {\em Management Science}, 52(11):1778--1791.

\bibitem[Augustine et~al., 2004]{augustine2004optimal}
Augustine, J., Irani, S., and Swamy, C. (2004).
\newblock Optimal power-down strategies.
\newblock In {\em 45th Annual IEEE Symposium on Foundations of Computer
  Science}, pages 530--539. IEEE.

\bibitem[Azar et~al., 1999]{azar1999capital}
Azar, Y., Bartal, Y., Feuerstein, E., Fiat, A., Leonardi, S., and Ros{\'e}n, A.
  (1999).
\newblock On capital investment.
\newblock {\em Algorithmica}, 25(1):22--36.

\bibitem[Bamas et~al., 2020]{bamas2020learning}
Bamas, E., Maggiori, A., Rohwedder, L., and Svensson, O. (2020).
\newblock Learning augmented energy minimization via speed scaling.
\newblock {\em arXiv preprint arXiv:2010.11629}.

\bibitem[Bansal et~al., 2009]{bansal2009speed}
Bansal, N., Chan, H.-L., and Pruhs, K. (2009).
\newblock Speed scaling with an arbitrary power function.
\newblock In {\em Proceedings of the twentieth annual ACM-SIAM symposium on
  Discrete algorithms}, pages 693--701. SIAM.

\bibitem[Barbu and Precupanu, 2012]{barbu2012convexity}
Barbu, V. and Precupanu, T. (2012).
\newblock {\em Convexity and optimization in Banach spaces}.
\newblock Springer Science \& Business Media.

\bibitem[Barroso and H{\"o}lzle, 2007]{barroso2007case}
Barroso, L.~A. and H{\"o}lzle, U. (2007).
\newblock The case for energy-proportional computing.
\newblock {\em Computer}, 40(12):33--37.

\bibitem[Bodik, 2010]{bodik2010automating}
Bodik, P. (2010).
\newblock {\em Automating datacenter operations using machine learning}.
\newblock PhD thesis, UC Berkeley.

\bibitem[Boyar et~al., 2017]{boyar2017online}
Boyar, J., Favrholdt, L.~M., Kudahl, C., Larsen, K.~S., and Mikkelsen, J.~W.
  (2017).
\newblock Online algorithms with advice: A survey.
\newblock {\em ACM Computing Surveys (CSUR)}, 50(2):1--34.

\bibitem[Cortez et~al., 2017]{cortez2017resource}
Cortez, E., Bonde, A., Muzio, A., Russinovich, M., Fontoura, M., and Bianchini,
  R. (2017).
\newblock Resource central: Understanding and predicting workloads for improved
  resource management in large cloud platforms.
\newblock In {\em Proceedings of the 26th Symposium on Operating Systems
  Principles}, pages 153--167.

\bibitem[Damaschke, 2003]{damaschke2003nearly}
Damaschke, P. (2003).
\newblock Nearly optimal strategies for special cases of on-line capital
  investment.
\newblock {\em Theoretical Computer Science}, 302(1-3):35--44.

\bibitem[Dayarathna et~al., 2015]{dayarathna2015data}
Dayarathna, M., Wen, Y., and Fan, R. (2015).
\newblock Data center energy consumption modeling: A survey.
\newblock {\em IEEE Communications Surveys \& Tutorials}, 18(1):732--794.

\bibitem[Doytchinov et~al., 2001]{doytchinov2001real}
Doytchinov, B., Lehoczky, J., and Shreve, S. (2001).
\newblock Real-time queues in heavy traffic with earliest-deadline-first queue
  discipline.
\newblock {\em Annals of Applied Probability}, pages 332--378.

\bibitem[Facebook, 2014]{FacebookScaling2014}
Facebook (2014).
\newblock {Making Facebook’s software infrastructure more energy efficient
  with Autoscale}.
\newblock
  \url{https://engineering.fb.com/production-engineering/making-facebook-s-software-infrastructure-more-energy-efficient-with-autoscale/}.

\bibitem[Galloway et~al., 2012]{galloway2012empirical}
Galloway, J., Smith, K., and Carver, J. (2012).
\newblock An empirical study of power aware load balancing in local cloud
  architectures.
\newblock In {\em 2012 Ninth International Conference on Information
  Technology-New Generations}, pages 232--236. IEEE.

\bibitem[Gandhi et~al., 2013]{gandhi2013exact}
Gandhi, A., Doroudi, S., Harchol-Balter, M., and Scheller-Wolf, A. (2013).
\newblock {Exact analysis of the M/M/k/setup class of Markov chains via
  recursive renewal reward}.
\newblock In {\em Proceedings of the ACM SIGMETRICS/International Conference on
  Measurement and Modeling of Computer Systems}, pages 153--166.

\bibitem[Gandhi et~al., 2014]{gandhi2014adaptive}
Gandhi, A., Dube, P., Karve, A., Kochut, A., and Zhang, L. (2014).
\newblock Adaptive, model-driven autoscaling for cloud applications.
\newblock In {\em 11th International Conference on Autonomic Computing (ICAC
  14)}, pages 57--64.

\bibitem[Gandhi et~al., 2010]{Gandhi2010}
Gandhi, A., Gupta, V., Harchol-Balter, M., and Kozuch, M.~A. (2010).
\newblock Optimality analysis of energy-performance trade-off for server farm
  management.
\newblock {\em Performance Evaluation}, 67(11):1155--1171.

\bibitem[Gandhi et~al., 2012]{gandhi2012autoscale}
Gandhi, A., Harchol-Balter, M., Raghunathan, R., and Kozuch, M.~A. (2012).
\newblock Autoscale: Dynamic, robust capacity management for multi-tier data
  centers.
\newblock {\em ACM Transactions on Computer Systems (TOCS)}, 30(4):1--26.

\bibitem[Gao, 2014]{gao2014machine}
Gao, J. (2014).
\newblock Machine learning applications for data center optimization.

\bibitem[Goldman et~al., 2000]{goldman2000online}
Goldman, S.~A., Parwatikar, J., and Suri, S. (2000).
\newblock Online scheduling with hard deadlines.
\newblock {\em Journal of Algorithms}, 34(2):370--389.

\bibitem[Google, 2016]{GoogleScaling2016}
Google (2016).
\newblock {Data centers get fit on efficiency}.
\newblock
  \url{https://blog.google/outreach-initiatives/environment/data-centers-get-fit-on-efficiency/}.

\bibitem[Hsu et~al., 2019]{hsu2018learning}
Hsu, C.-Y., Indyk, P., Katabi, D., and Vakilian, A. (2019).
\newblock Learning-based frequency estimation algorithms.
\newblock In {\em International Conference on Learning Representations}.

\bibitem[Irani et~al., 2002]{irani2002competitive}
Irani, S., Shukla, S., and Gupta, R. (2002).
\newblock Competitive analysis of dynamic power management strategies for
  systems with multiple power saving states.
\newblock In {\em Proceedings 2002 Design, Automation and Test in Europe
  Conference and Exhibition}, pages 117--123. IEEE.

\bibitem[Karlin et~al., 2003]{karlin2003dynamic}
Karlin, A.~R., Kenyon, C., and Randall, D. (2003).
\newblock {Dynamic TCP acknowledgment and other stories about e/(e- 1)}.
\newblock {\em Algorithmica}, 36:209--224.

\bibitem[Karlin et~al., 1988]{karlin1988competitive}
Karlin, A.~R., Manasse, M.~S., Rudolph, L., and Sleator, D.~D. (1988).
\newblock Competitive snoopy caching.
\newblock {\em Algorithmica}, 3(1-4):79--119.

\bibitem[Khanafer et~al., 2013]{khanafer2013constrained}
Khanafer, A., Kodialam, M., and Puttaswamy, K.~P. (2013).
\newblock The constrained ski-rental problem and its application to online
  cloud cost optimization.
\newblock In {\em 2013 Proceedings IEEE INFOCOM}, pages 1492--1500. IEEE.

\bibitem[Kumar et~al., 2018]{kumar2018semi}
Kumar, R., Purohit, M., Schild, A., Svitkina, Z., and Vee, E. (2018).
\newblock Semi-online bipartite matching.
\newblock {\em arXiv preprint arXiv:1812.00134}.

\bibitem[Lassettre et~al., 2003]{lassettre2003dynamic}
Lassettre, E., Coleman, D.~W., Diao, Y., Froehlich, S., Hellerstein, J.~L.,
  Hsiung, L., Mummert, T., Raghavachari, M., Parker, G., Russell, L., et~al.
  (2003).
\newblock Dynamic surge protection: An approach to handling unexpected workload
  surges with resource actions that have lead times.
\newblock In {\em International Workshop on Distributed Systems: Operations and
  Management}, pages 82--92. Springer.

\bibitem[Lee et~al., 2019]{lee2019learning}
Lee, R., Hajiesmaili, M.~H., and Li, J. (2019).
\newblock Learning-assisted competitive algorithms for peak-aware energy
  scheduling.
\newblock {\em arXiv preprint arXiv:1911.07972}.

\bibitem[Lin et~al., 2012]{lin2012dynamic}
Lin, M., Wierman, A., Andrew, L.~L., and Thereska, E. (2012).
\newblock Dynamic right-sizing for power-proportional data centers.
\newblock {\em IEEE/ACM Transactions on Networking}, 21(5):1378--1391.

\bibitem[Lu et~al., 2012]{lu2012simple}
Lu, T., Chen, M., and Andrew, L.~L. (2012).
\newblock Simple and effective dynamic provisioning for power-proportional data
  centers.
\newblock {\em IEEE Transactions on Parallel and Distributed Systems},
  24(6):1161--1171.

\bibitem[Lykouris and Vassilvtiskii, 2018]{lykouris2018competitive}
Lykouris, T. and Vassilvtiskii, S. (2018).
\newblock Competitive caching with machine learned advice.
\newblock In {\em International Conference on Machine Learning}, pages
  3296--3305.

\bibitem[Maccio and Down, 2015]{maccio2015optimal}
Maccio, V.~J. and Down, D.~G. (2015).
\newblock On optimal policies for energy-aware servers.
\newblock {\em Performance Evaluation}, 90:36--52.

\bibitem[Mahdian et~al., 2012]{mahdian2012online}
Mahdian, M., Nazerzadeh, H., and Saberi, A. (2012).
\newblock Online optimization with uncertain information.
\newblock {\em ACM Transactions on Algorithms (TALG)}, 8(1):1--29.

\bibitem[Manmeet et~al., 2019]{singh2016data}
Manmeet, S., Maninder, S., and Sanmeet, K. (2019).
\newblock {TI-2016 DNS dataset}.
\newblock {\em IEEE Dataport}.

\bibitem[Mazzucco and Dyachuk, 2012]{mazzucco2012optimizing}
Mazzucco, M. and Dyachuk, D. (2012).
\newblock Optimizing cloud providers revenues via energy efficient server
  allocation.
\newblock {\em Sustainable Computing: Informatics and Systems}, 2(1):1--12.

\bibitem[Mitzenmacher, 2018]{mitzenmacher2018model}
Mitzenmacher, M. (2018).
\newblock A model for learned bloom filters and optimizing by sandwiching.
\newblock In {\em Advances in Neural Information Processing Systems}, pages
  464--473.

\bibitem[Mitzenmacher, 2019a]{mitzenmacher2019scheduling}
Mitzenmacher, M. (2019a).
\newblock Scheduling with predictions and the price of misprediction.
\newblock {\em arXiv preprint arXiv:1902.00732}.

\bibitem[Mitzenmacher, 2019b]{mitzenmacher2019supermarket}
Mitzenmacher, M. (2019b).
\newblock The supermarket model with known and predicted service times.
\newblock {\em arXiv preprint arXiv:1905.12155}.

\bibitem[Mukherjee et~al., 2017]{mukherjee2017optimal}
Mukherjee, D., Dhara, S., Borst, S.~C., and van Leeuwaarden, J. S.~H. (2017).
\newblock Optimal service elasticity in large-scale distributed systems.
\newblock {\em Proceedings of the ACM on Measurement and Analysis of Computing
  Systems}, 1(1):1--28.

\bibitem[Mukherjee and Stolyar, 2019]{mukherjee2019join}
Mukherjee, D. and Stolyar, A. (2019).
\newblock Join idle queue with service elasticity: Large-scale asymptotics of a
  nonmonotone system.
\newblock {\em Stochastic Systems}, 9(4):338--358.

\bibitem[Nature, 2018]{Nature2018}
Nature (2018).
\newblock {How to stop data centres from gobbling up the world’s
  electricity}.
\newblock \url{https://www.nature.com/articles/d41586-018-06610-y}.

\bibitem[Netflix, 2013]{NetflixScryer2013}
Netflix (2013).
\newblock {Scryer: Netflix’s Predictive Auto Scaling Engine}.
\newblock
  \url{https://netflixtechblog.com/scryer-netflixs-predictive-auto-scaling-engine-a3f8fc922270}.

\bibitem[Purohit et~al., 2018]{purohit2018improving}
Purohit, M., Svitkina, Z., and Kumar, R. (2018).
\newblock Improving online algorithms via ml predictions.
\newblock In {\em Advances in Neural Information Processing Systems}, pages
  9661--9670.

\bibitem[Qi et~al., 2020]{qi2020mean}
Qi, J., Du, J., Siniscalchi, S.~M., Ma, X., and Lee, C.-H. (2020).
\newblock On mean absolute error for deep neural network based vector-to-vector
  regression.
\newblock {\em IEEE Signal Processing Letters}.

\bibitem[Rong et~al., 2016]{rong2016optimizing}
Rong, H., Zhang, H., Xiao, S., Li, C., and Hu, C. (2016).
\newblock Optimizing energy consumption for data centers.
\newblock {\em Renewable and Sustainable Energy Reviews}, 58:674--691.

\bibitem[Rzadca et~al., 2020]{rzadca2020autopilot}
Rzadca, K., Findeisen, P., Swiderski, J., Zych, P., Broniek, P., Kusmierek, J.,
  Nowak, P., Strack, B., Witusowski, P., Hand, S., et~al. (2020).
\newblock Autopilot: workload autoscaling at google.
\newblock In {\em Proceedings of the Fifteenth European Conference on Computer
  Systems}, pages 1--16.

\bibitem[Shehabi et~al., 2016]{Shebabi2016}
Shehabi, A., Smith, S., Sartor, D., Brown, R., Herrlin, M., Koomey, J.,
  Masanet, E., Horner, N., Azevedo, I., and Lintner, W. (2016).
\newblock {United States Data Center Energy Usage Report}.
\newblock Technical report, {Lawrence Berkeley National Lab}.

\bibitem[Sverdlik, 2020]{ZoomCapacity2020}
Sverdlik, Y. (2020).
\newblock {How Zoom, Netflix, and Dropbox are Staying Online During the
  Pandemic}.
\newblock
  \url{https://www.datacenterknowledge.com/uptime/how-zoom-netflix-and-dropbox-are-staying-online-during-pandemic}.

\bibitem[Tirmazi et~al., 2020]{tirmazi2020borg}
Tirmazi, M., Barker, A., Deng, N., Haque, M.~E., Qin, Z.~G., Hand, S.,
  Harchol-Balter, M., and Wilkes, J. (2020).
\newblock Borg: the next generation.
\newblock In {\em Proceedings of the Fifteenth European Conference on Computer
  Systems}, pages 1--14.

\bibitem[Urgaonkar et~al., 2005]{urgaonkar2005dynamic}
Urgaonkar, B., Shenoy, P., Chandra, A., and Goyal, P. (2005).
\newblock Dynamic provisioning of multi-tier internet applications.
\newblock In {\em Second International Conference on Autonomic Computing
  (ICAC'05)}, pages 217--228. IEEE.

\bibitem[Wierman et~al., 2009]{wierman2009power}
Wierman, A., Andrew, L.~L., and Tang, A. (2009).
\newblock Power-aware speed scaling in processor sharing systems.
\newblock In {\em IEEE INFOCOM 2009}, pages 2007--2015. IEEE.

\end{thebibliography}

\newpage
\appendix

\section{Proofs.}

This section provides the proofs which have been omitted from the main text.

\subsection{Proof of Proposition \ref{prop:existence}.}
\label{app:existence}

Fix a finite time horizon $T$, arrival rate function $\lambda(\cdot)$ and initial number of servers $m(0)$. Let
\begin{equation}
    V^+(f) = \limsup\limits_{\delta \downarrow 0} \sum_{i = 0}^{\lfloor T / \delta \rfloor} \left[ f(i \delta + \delta) - f(i \delta) \right]^+
\end{equation}
for a function $f: [0, T] \to \mathbbm{R}$. The definition of $V^+(f)$ is closely related to the notion of bounded variation. The bounded variation of a function $f: [0, T] \to \mathbbm{R}$ is defined as
\begin{equation}
    V(f) = \sup\left\{ \sum_{i = 1}^n \left\lvert f(z_i) - f(z_{i-1}) \right\rvert \text{ such that } \{ z_i \}_{i=0}^n \text{ is a partition of } [0, T] \right\}.
\end{equation}
Let $m_n$ be a sequence of functions such that $\textsc{Cost}^\lambda(m_n, T) \to \textsc{Opt}$ as $n \to \infty$. There exists $N \in \mathbbm{N}$ such that $\textsc{Cost}^\lambda(m_n, T) \leq 2 \cdot \textsc{Opt}$ for all $n \geq N$. As a result, $V^+(m_n)$ is uniformly bounded for $n \geq N$. Note that, without increasing the cost, we can set $m_n(T) = 0$. The bounded variation and $V^+(m_n)$ are then related as
\begin{equation}
\label{eq:bvdiffplus}
    V\left( m_n \right) = 2 V^+\left( m_n \right) + m(0).
\end{equation}
The rest of the proof depends on the following compactness theorem \cite{barbu2012convexity}.
%\footnote{DM: Can you cite precise theorem number?}

\begin{theorem}
\label{th:helly}
(Helly's selection theorem)
Let $f_n: [0, T] \to \mathbbm{R}$ be a sequence of functions and suppose that the next two conditions hold:
\begin{enumerate}[(i)]
    \item The sequence $f_n$ has uniformly bounded variation, i.e., $\sup_{n \in \mathbbm{N}} V(f_n) < \infty$,
    \item The sequence $f_n$ is uniformly bounded at a point, i.e., there exists $t \in [0, T]$ such that $\{ f_n(t) \}_{n = 1}^\infty$ is a bounded set.
\end{enumerate}
Then, there exists a subsequence $f_{n_k}$ of $f_n$ and a function $f: [0, T] \to \mathbbm{R}$ such that
\begin{enumerate}[(i)]
    \item $f_{n_k}$ converges to $f$ pointwise as $k \to \infty$,
    \item $f_{n_k}$ converges to $f$ in $L_1$ as $k \to \infty$,
    \item $V(f) \leq \liminf_{k \to \infty} V(f_{n_k})$.
\end{enumerate}
\end{theorem}

Recall the infinite sequence $m_N, m_{N+1}, \dots$ introduced above. Condition (i) in Theorem \ref{th:helly} holds because
\begin{equation}
    V\left( m_n \right) = 2 V^+\left( m_n \right) + m(0) \leq \frac{4 \textsc{Opt}}{\beta} + m(0),
\end{equation}
for all $n \geq N$. Moreover, condition (ii) in Theorem \ref{th:helly} holds for $t = 0$, since $m_n(0) = m(0)$ for all $n \in \mathbbm{N}$. Hence, there exists a subsequence $m_{n_k}$ of $m_n$ and a function $m^*: [0, T] \to \mathbbm{R}$ such that $m_{n_k} \to m^*$ pointwise and in the $L_1$ norm, as $k \to \infty$, and
\begin{equation}
    V^+(m^*) = (V(m^*) - m(0)) / 2 \leq \liminf_{k \to \infty} (V(m_{n_k}) - m(0)) / 2 = \liminf_{k \to \infty} V^+(m_{n_k}).
\end{equation}
Therefore, since the flow-time and the power cost are continuous in $m$ with respect to the $L_1$ norm,
\begin{equation}
\begin{aligned}
    \textsc{Cost}^\lambda(m^*, T)
    &= \omega \cdot \int_0^T q^*(t) \diff t + \beta \cdot V^+(m^*) + \theta \cdot \int_0^T m^*(t) \diff t \\
    &\leq \omega \cdot \lim_{k \to \infty} \int_0^T q_{n_k}(t) \diff t + \beta \cdot \liminf_{k \to \infty} V^+(m_{n_k}) + \theta \cdot \lim_{k \to \infty} \int_0^T m_{n_k}(t) \diff t \\
    &\leq  \lim_{k \to \infty} \textsc{Cost}^\lambda(m_{n_k}, T)
    = \textsc{Opt},
\end{aligned}
\end{equation}
which completes the proof of the proposition. \hfill$\blacksquare$

\subsection{Proof of Lemma \ref{lemma:warmuptime}.}
\label{app:proofwarmuptime}

\blue{

Fix any online algorithm $\mathcal{A}$ and parameters $\omega$ and $\beta$. We will construct an instance for which $\textsc{Cost}^\lambda(m, T) \geq \frac{\omega t_0^2}{2 \beta} \cdot \textsc{Opt}$.

Let $\lambda(t) = 0$ for $t \in [0, t_0]$ and $\lambda(t) = \rho$ for $t \in (t_0, 2 t_0]$, where the value of $\rho$ will be chosen (adversarially) later. Let the finite time horizon $T = 2 t_0$ and $\theta = 0$. Let $m(t)$ be the number of servers of $\mathcal{A}$ for the instance. We distinguish two cases depending on $\bar{m} := \sup_{t \in (t_0, 2 t_0]} m(t)$. Crucially, note that any decision to increase $m(t)$ during the interval $[t_0, 2 t_0]$ must be taken during the interval $[0, t_0]$, since $t_0$ is the setup time.
\begin{enumerate}
    \item
First, consider the case when $\bar{m} > 0$. The increment in $m(t)$ must have been initiated during $[0, t_0]$. In that case, fix $\rho = 0$. One possible solution of \eqref{eq:model} is $m^*(t) = 0$ for $t \in [0, T]$ and the cost of the optimal solution is therefore at most $\textsc{Opt} = 0$. However, the cost of $\mathcal{A}$ is at least $\textsc{Cost}^\lambda(m, T) \geq \beta \bar{m} > 0$. Therefore, there does not exists a constant $\textsc{CR}$ such that $\textsc{Cost}^\lambda(m, T) \leq \textsc{CR} \cdot\textsc{Opt}$ and hence $\textsc{CR} = \infty$ by definition.

    \item
Next, consider the case when $\bar{m} = 0$, i.e., no increment in $m(t)$ has been initiated during $[0, t_0]$. In that case, fix $\rho = 1$. One possible solution of \eqref{eq:model} is $m^*(t) = 1$ for $t \in [0, T]$. This solution does not incur any waiting cost and the cost of the optimal solution is therefore at most $\textsc{Opt} \leq \beta$. However, the cost of $\mathcal{A}$ is $\textsc{Cost}^\lambda(m, T) = \omega \cdot \int_0^{t_0} \rho s \diff s = \frac{\omega t_0^2}{2} \geq \frac{\omega t_0^2}{2 \beta} \cdot \textsc{Opt}$.
\end{enumerate}

This completes the proof of the Lemma. \hfill$\blacksquare$

}
\subsection{Proof of Theorem \ref{th:lin-appx}.}
\label{app:prooflinx-appx}

%\begin{proof}[Proof of Theorem~\ref{th:lin-appx}]
Fix a finite time horizon $T$, arrival rate function $\lambda(\cdot)$ and initial number of servers $m(0)$. Assume that $T$ is divisible by $\delta$ and let $n = T / \delta$. Let
\begin{equation}
\mathcal{C} := \left\{ f: [0, T] \to \mathbbm{R}_+ \mid f(i \delta + s) = f(i \delta) \text{ for all } s \in [0, \delta) \text{ and } i = 0, 1, \dots, n - 1 \right\}
\end{equation}
be the subspace of the space of functions which are constant in each $\delta$-interval. Recall that by assumption, $\lambda \in \mathcal{C}$. We note that each $f \in \mathcal{C}$ is equivalently represented by a vector $f = \left( f(0), f(\delta), \dots, f(n - 1) \right) \in \mathbbm{R}^n$ and vice versa. We will therefore interchangeably use vector notation to denote an element from $\mathcal{C}$.

\begin{claim}
We claim that
\begin{equation}
    \underset{m \in \mathcal{C}}{\inf} \; \textsc{Cost}^\lambda(m, T) \leq \left( 1 + \frac{\omega \delta^2}{\beta} \right) \underset{m: [0, T] \to \mathbbm{R}_+}{\inf} \textsc{Cost}^\lambda(m, T) + \frac{\omega \delta^2 \cdot m(0)}{2}.
\end{equation}
\end{claim}

\begin{proof}{Proof.}
Let $m^*: [0, T] \to \mathbbm{R}_+$ be arbitrary and let $m_i = \frac{1}{\delta} \int_{(i-1) \delta}^{i \delta} m^*(t) \diff t$. We will prove that
\begin{equation}
    \textsc{Cost}^\lambda(m, T) \leq \left( 1 + \frac{\omega \delta^2}{\beta} \right) \textsc{Cost}^\lambda(m^*, T) + \frac{\omega \delta^2 \cdot m(0)}{2},
\end{equation}
which finishes the proof of the claim. Note that it follows immediately by construction that the switching cost of $m$ is at most the switching cost of $m^*$ and the power cost of $m$ is equal to the power cost of $m^*$. We will therefore focus on the flow-time cost. The queue length of $m^*$ at the endpoints of each $\delta$-interval is at least the queue length of $m$ as follows,
\begin{equation}
\begin{aligned}
\label{eq:qlengthorder}
    q^*(i \delta) &= q^*((i-1) \delta) + \int_{(i-1) \delta}^{i \delta} \left( \lambda_i - m^*(s) \right) \mathbbm{1}\{ q^*(s) > 0 \text{ or } \lambda_i \geq m^*(s) \} \diff s \\
    &\geq \left[ q^*((i-1) \delta) + \int_{(i-1) \delta}^{i \delta} \left( \lambda_i - m^*(s) \right) \diff s \right]^+ \\
    &\geq \left[ q((i-1) \delta) + \delta \lambda_i - \delta m_i \right]^+
    = q(i \delta),
\end{aligned}
\end{equation}
where the inequality $q^*((i-1) \delta) \geq q((i-1) \delta)$ follows by induction on $i$. Define
\begin{equation}
    \Delta_i = \sup_{t \in [(i-1) \delta, i \delta]} m^*(t) - \inf_{t \in [(i-1) \delta, i \delta]} m^*(t),
\end{equation}
and observe that
\begin{equation}
\label{eq:upperboudndeltai}
    \sum_{i = 1}^n \Delta_i \leq m(0) + 2 \lim_{\varepsilon \downarrow 0} \sum_{i = 0}^{\lfloor T / \varepsilon \rfloor} [m^*(i \varepsilon + \varepsilon) - m^*(i \varepsilon)]^+
    \leq m(0) + \frac{2 \textsc{Cost}^\lambda(m^*, T)}{\beta}.
\end{equation}
Then, the flow-time cost of $m^*$ in each $\delta$-interval is at least,
\begin{equation}
\begin{aligned}
    \int_{(i-1) \delta}^{i \delta} q^*(t) \diff t
    &= \int_{(i-1) \delta}^{i \delta} \left[ q^*((i-1) \delta) + \int_{(i-1) \delta}^{t} \left( \lambda_i - m(s) \right) \mathbbm{1}\{ q^*(s) > 0 \text{ or } \lambda_i \geq m(s) \} \diff s \right] \diff t \\
    &\geq \int_{(i-1) \delta}^{i \delta} \left[ q^*((i-1) \delta) + \int_{(i-1) \delta}^{t} \left( \lambda_i - m(s) \right) \diff s \right]^+ \diff t \\
    &\geq \int_{(i-1) \delta}^{i \delta} \left[ q((i-1) \delta) + \int_{(i-1) \delta}^{t} \left( \lambda_i - m_i - \Delta_i \right) \diff s \right]^+ \diff t \\
    &= \int_{(i-1) \delta}^{i \delta} \left[ q(t) - (t - (i-1) \delta) \Delta_i \right]^+ \diff t
    \geq \int_{(i-1) \delta}^{i \delta} q(t) \diff t - \frac{\delta^2 \Delta_i}{2},
\end{aligned}
\end{equation}
where the second inequality uses \eqref{eq:qlengthorder}. Therefore,
\begin{equation}
    \omega \cdot \int_0^T q(t) \diff t - \omega \cdot \int_0^T q^*(t) \diff t
    \leq \frac{\omega \delta^2}{2} \cdot \sum_{i = 1}^n \Delta^i
    \leq \frac{\omega \delta^2 \cdot m(0)}{2} + \frac{\omega \delta^2 \cdot \textsc{Cost}^\lambda(m^*, T)}{\beta},
\end{equation}
where the second inequality follows by \eqref{eq:upperboudndeltai}. This completes the proof of the claim. \hfill$\blacksquare$
\end{proof}

Let $\textsc{Obj}^\lambda(m, T)$ denote the value of the objective in \eqref{eq:offlinelp} for $m \in \mathcal{C}$.

\begin{claim}
We claim that
\begin{equation}
\begin{aligned}
    \textsc{Obj}^\lambda(m, T) &= \textsc{Cost}^\lambda(m, T) + \omega \cdot \sum_{i = 1}^n \left(\delta \cdot \frac{q_i}{2} - \frac{q_i^2}{2 (m_i - \lambda_i)} \right) \mathbbm{1}\{ q_i > 0 \text{ and } q_{i+1} = 0 \} \\
    &\leq \left( 1 + \frac{\omega \delta}{2 \theta} \right) \textsc{Cost}^\lambda(m, T),
\end{aligned}
\end{equation}
for any $m \in \mathcal{C}$.
\end{claim}

\begin{proof}{Proof.}
Let $m \in \mathcal{C}$ be arbitrary. Note that, since $m \in \mathcal{C}$, the switching cost is $\sum_{i = 1}^n \left[ m_i - m_{i-1} \right]^+$ and the power cost is $\sum_{i = 1}^n \delta m_i$, which matches the terms in $\textsc{Obj}^\lambda(m, T)$. We will therefore focus on the flow-time cost. Denote $q = \left( q(0), q(\delta), \dots, q(n) \right) \in \mathbbm{R}^{n+1}$. The flow-time is equal to
\begin{equation}
\begin{aligned}
    \int_{(i-1) \delta}^{i \delta} q(t) \diff t
    &= \int_{0}^{\delta} \left[ q_i + (\lambda_i - m_i) t \right]^+ \diff t \\
    &= \delta \cdot \frac{q_i + q_{i+1}}{2} \mathbbm{1}\{ q_i = 0 \text{ or } q_{i+1} > 0 \} + \frac{q_i^2}{2 (m_i - \lambda_i)} \mathbbm{1}\{ q_i > 0 \text{ and } q_{i+1} = 0 \},
\end{aligned}
\end{equation}
because $q(\cdot)$ increases or decreases linearly. This completes the equality in the claim. To see why the inequality holds, note that
\begin{equation}
\begin{gathered}
    \sum_{i = 1}^n \left(\delta \cdot \frac{q_i}{2} - \frac{q_i^2}{2 (m_i - \lambda_i)} \right) \mathbbm{1}\{ q_i > 0 \text{ and } q_{i+1} = 0 \}
    \leq \frac{\delta}{2} \cdot \sum_{i = 1}^n q_i \mathbbm{1}\{ q_i > 0 \text{ and } q_{i+1} = 0 \} \\
    \leq \frac{\delta}{2} \cdot \sum_{i = 1}^n \delta m_i,
    \leq \frac{\delta \cdot \textsc{Cost}^\lambda(m, T)}{2 \theta},
\end{gathered}
\end{equation}
where the second inequality follows because $\delta (m_i - \lambda_i) \geq q_i$. This completes the proof of the claim.

\hfill$\blacksquare$
\end{proof}

We now finish the proof of Theorem \ref{th:lin-appx}. Let $m \in \mathcal{C}$ be an optimal solution to \eqref{eq:offlinelp}. Moreover, define
\begin{equation}
    m^* = \underset{m \in \mathcal{C}}{\arg\min} \; \textsc{Cost}^\lambda(m, T).
\end{equation}
Then, the cost of $m$ is at most,
\begin{equation}
\begin{aligned}
    \textsc{Cost}^\lambda(m, T)
    &\leq \textsc{Obj}^\lambda(m, T)
    \leq \textsc{Obj}^\lambda(m^*, T) \\
    &\leq \left( 1 + \frac{\omega \delta}{2 \theta} \right) \textsc{Cost}^\lambda(m^*, T) \\
    &\leq \left( 1 + \frac{\omega \delta}{2 \theta} \right) \left( \left( 1 + \frac{\omega \delta^2}{\beta} \right) \textsc{Opt} + \frac{\omega \delta^2 m(0)}{2} \right),
\end{aligned}
\end{equation}
which completes the proof of the theorem. \hfill$\blacksquare$
%\end{proof}
\subsection{Proof of Theorem \ref{th:crnowait}.}
\label{app:proofnowait}

Fix a finite time horizon $T$, arrival rate function $\lambda(\cdot)$ and initial number of servers $m(0)$.
Let $m^*(t)$ be a solution of the offline optimization problem \eqref{eq:model} and $m(t)$ be the number of servers of BCS (Algorithm \ref{alg:bcsalg}). We will prove that
\begin{equation}
    \textsc{Cost}(m, T) \leq 2 \cdot \textsc{Cost}(m^*, T) + \beta \cdot m(0),
\end{equation}
where we have omitted $\lambda$ from the notation $\textsc{Cost}^\lambda(m, T)$.

\paragraph{Overview of the proof.} Let $t_1 \leq t_2 \leq \dots $ be a partitioning of the interval $[0, T]$ such that (i) $m(t)$ is monotone in $[t_k, t_{k+1}]$ and (ii) either $m(t) > m^*(t)$ or $m(t) \leq m^*(t)$ in $[t_k, t_{k+1}]$ for all $k \in \mathbbm{N}$. The goal of the proof will be to find a non-negative potential function $\Phi(t)$ such that
\begin{equation}
\label{eq:phidelta}
    \Phi(t_{k+1}) - \Phi(t_k) + \textsc{Cost}(m, t_{k+1}) - \textsc{Cost}(m, t_k) \leq 2 \cdot (\textsc{Cost}(m^*, t_{k+1}) - \textsc{Cost}(m^*, t_k)),
\end{equation}
for all $k \in \mathbbm{N}$. We sum equation \eqref{eq:phidelta} over $k \in \mathbbm{N}$ to obtain
\begin{equation}
    \textsc{Cost}(m, T) \leq 2 \cdot \textsc{Cost}(m^*, T) + \Phi(0) - \Phi(T) \leq 2 \cdot \textsc{Cost}(m^*, T) + \Phi(0),
\end{equation}
where the last step follows because $\Phi(T)$ is non-negative. The proof of Theorem \ref{th:crnowait} is therefore completed if we manage to find a non-negative potential function $\Phi(t)$ satisfying equation \eqref{eq:phidelta} and $\Phi(0) = \beta \cdot m(0)$.

\paragraph{Choice of $\Phi(t)$.} Define the potential function $\Phi(t)$ such that
\begin{equation}
    \Phi(t) = \begin{cases}
        \beta \cdot m(t), &\text{ if } m(t) > m^*(t) \\
        2 \beta \cdot m^*(t) - \beta \cdot m(t) &\text{ if } m(t) \leq m^*(t)
    \end{cases}
\end{equation}
Note that $\Phi(t)$ is non-negative and $\Phi(0) = \beta \cdot m(0)$.

\paragraph{Verification of \eqref{eq:phidelta}.} We continue by verifying equation \eqref{eq:phidelta}. Fix $k \in \mathbbm{N}$. We distinguish two cases, depending on whether $m(t)$ is decreasing or non-decreasing in $[t_k, t_{k+1}]$.
\begin{enumerate}[(i)]
    \item
Assume that $m(s)$ is decreasing for $s \in [t_k, t_{k+1}]$. Recall that, by definition,
\begin{equation}
    \frac{\diff m(t)}{\diff t} = - \frac{\theta \cdot m(t)}{\beta},
\end{equation}
for $t \in [t_k, t_{k+1}]$ and therefore
\begin{equation}
    m(t_k + s) = m(t_k) \cdot \exp\left( - \frac{\theta \cdot s}{\beta} \right),
\end{equation}
for $s \in [0, t_{k+1} - t_k]$ and hence
\begin{equation}
    \theta \cdot \int_{t_k}^{t_{k+1}} m(s) \diff s
    = \beta \cdot m(t_k) \left( 1 - \exp\left( - \frac{\theta \cdot (t_{k+1} - t_k)}{\beta} \right) \right)
    = \beta \cdot (m(t_k) - m(t_{k+1})).
\end{equation}
We further distinguish two cases depending on whether $m(t) > m^*(t)$ or $m(t) \leq m^*(t)$ in $[t_k, t_{k+1}]$. First, consider the case that $m(s) > m^*(s)$ for $s \in [t_k, t_{k+1}]$. Then,
\begin{equation}
\begin{aligned}
    &\Phi(t_{k+1}) - \Phi(t_k) + \textsc{Cost}(m, t_{k+1}) - \textsc{Cost}(m, t_k) \\
    &= \beta \cdot (m(t_{k+1}) - m(t_k)) + \beta \cdot (m(t_k) - m(t_{k+1})) \\
    &= 0
    \leq 2 (\textsc{Cost}(m^*, t_{k+1}) - \textsc{Cost}(m^*, t_k)).
\end{aligned}
\end{equation}
Next, consider the case that that $m(s) \leq m^*(s)$ for $s \in [t_k, t_{k+1}]$. Then,
\begin{equation}
\begin{aligned}
    &\Phi(t_{k+1}) - \Phi(t_k) + \textsc{Cost}(m, t_{k+1}) - \textsc{Cost}(m, t_k) \\
    &= 2 \beta \cdot (m^*(t_{k+1}) - m^*(t_k)) - \beta \cdot (m(t_{k+1}) - m(t_k)) + \beta \cdot (m(t_k) - m(t_{k+1})) \\
    &= 2 \beta \cdot (m^*(t_{k+1}) - m^*(t_k)) + 2 \theta \cdot \int_{t_k}^{t_{k+1}} m(s) \diff s \\
    &\leq 2 \beta \cdot (m^*(t_{k+1}) - m^*(t_k)) + 2 \theta \cdot \int_{t_k}^{t_{k+1}} m^*(s) \diff s \\
    &\leq 2 (\textsc{Cost}(m^*, t_{k+1}) - \textsc{Cost}(m^*, t_k)).
\end{aligned}
\end{equation}

    \item
Assume that $m(s)$ is non-decreasing for $s \in [t_k, t_{k+1}]$. Note that, since tasks are not allowed to wait, $m^*(t) \geq \lambda(t)$ for all $t \in [0, T]$. Recall that, by definition, if the arrival rate $\lambda(t)$ is higher than the number of servers $m(t)$ then BCS increases the number of servers to match the arrival rate. Therefore, $m(s) = \lambda(s)$ for $s \in [t_k, t_{k+1}]$ because $m(t)$ is non-decreasing in $[t_k, t_{k+1}]$. Hence, $m^*(s) \geq m(s) = \lambda(s)$ for $s \in [t_k, t_{k+1}]$ and
\begin{equation}
\begin{aligned}
    &\Phi(t_{k+1}) - \Phi(t_k) + \textsc{Cost}(m, t_{k+1}) - \textsc{Cost}(m, t_k) \\
    &= 2 \beta \cdot (m^*(t_{k+1}) - m^*(t_k)) - \beta \cdot (m(t_{k+1}) - m(t_k)) \\
    &+ \beta \cdot (m(t_{k+1}) - m(t_k)) + \theta \cdot \int_{t_k}^{t_{k+1}} m(s) \diff s \\
    &\leq 2 \beta \cdot (m^*(t_{k+1}) - m^*(t_k)) + \theta \cdot \int_{t_k}^{t_{k+1}} m^*(s) \diff s \\
    &\leq 2 (\textsc{Cost}(m^*, t_{k+1}) - \textsc{Cost}(m^*, t_k)).
\end{aligned}
\end{equation}
\end{enumerate}

\hfill$\blacksquare$

\subsection{Proof of Proposition \ref{prop:lowerbound}.}
\label{app:proofpurelyonline-lower-bd}

%\begin{proof}[Proof of Theorem~\ref{th:lowerbound}]
Fix any algorithm $\mathcal{A}$, and let $m(t)$ denote its number of servers.
\blue{We will construct an instance $(T, \lambda)$ for which $\textsc{Cost}^\lambda(m, T) \geq 2.549 \cdot \textsc{Opt}$.}
%If $\Phi(0)$ is not zero, then we simply scale $\lambda \to \infty$ in the example below such that $\Phi(0)$ becomes negligible.

Let $\lambda(t) = 1$ for $t \in [0, T]$. The time horizon $T$ will be specified later. Fix $\beta = \omega = 1$ and $\theta = 0$. Let the prediction $\tlambda(t) = 0$ for all $t$ and as a result the advised number of servers $\tm(t) = 0$ for all $t$. Let $m(t)$ be the number of servers of $\mathcal{A}$ for the instance.
Define $\tau := \inf \{ t \mid m(t) > 0.885 t^2 \}$ or $\tau = \infty$, if the infimum does not exist. 
We distinguish two cases depending on the value of $\tau$.
\begin{enumerate}
    \item
First, consider the case when $\tau \leq 1.225$. Fix $T = \tau$. The optimal solution to \eqref{eq:model} is $m^*(t) = 0$ for $t \in [0, T]$. The value of the optimal solution is purely due to flow-time and is equal to $\textsc{Opt} = \tau^2 / 2$.

At time $t = \tau$, algorithm $\mathcal{A}$ has at least $m(\tau) > 0.885 \tau^2$ servers. The flow-time is at least $\int_0^\tau q(t) \diff t \geq \int_0^\tau \int_0^t 1 - 0.885 s^2 \diff s \diff t \geq \tau^2 / 2 - 0.885 \tau^4 / 12$, because $m(t) \leq 0.885 t^2$ for $t \in [0, \tau)$. The cost of $\mathcal{A}$ is therefore at least $\textsc{Cost}^\lambda(m, T) \geq 0.885 \tau^2 + \tau^2 / 2 - 0.885 \tau^4 / 12 \geq 2.549 \cdot \tau^2 / 2 = 2.549 \cdot \textsc{Opt}$, where the second inequality follows because $\tau \leq 1.225$.

    \item
Next, consider the case when $\tau > 1.225$. Fix $T = 3$. The optimal solution to \eqref{eq:model} is $m^*(t) = 1$ for $t \in [0, T]$. The value of the optimal solution is purely due to switching cost and is equal to $\textsc{Opt} = 1$.

At time $t = 1.225$, the queue length of $\mathcal{A}$ is at least $q(1.225) \geq \int_0^{1.225} 1 - 0.885 t^2 \diff t = 0.682$. The optimal solution starting from time $t = 1.225$ is $m(t) = 1 + q(1.225) / \sqrt{2} \geq 1.483$ for $t \in (1.225, T]$. The flow-time is therefore at least $\int_0^T q(t) \diff t \geq \int_0^{1.225} \int_0^s 1 - 0.885 s^2 \diff s \diff t + q(1.225) / \sqrt{2} \geq 1.067$, again because $m(t) \leq 0.885 t^2$ for $t \in [0, \tau)$. The cost of $\mathcal{A}$ is therefore at least $\textsc{Cost}^\lambda(m, T) \geq 2.549 \geq 2.549 \cdot \textsc{Opt}$.
\end{enumerate}
Hence, the statement follows. \hfill$\blacksquare$
%\end{proof}
\subsection{Proof of Theorem \ref{th:crftp}.}
\label{app:proofftp}

%\begin{proof}
Fix a finite time horizon $T$, arrival rate function $\lambda(\cdot)$ and initial number of servers $m(0)$. Let $\tlambda(\cdot)$ be the predicted arrival rate. The idea to the proof is to separate the cost into the cost of the offline and the online component. More specifically, we claim that
\begin{equation}
\label{eq:costsplitclaim}
    \textsc{Cost}^\lambda(m, T) \leq \textsc{Cost}^{\tlambda}(m_1, T) + \left( \sqrt{2 \omega \beta} + \theta \right) T \cdot \lVert \Delta\lambda \rVert_{MAE}.
\end{equation}
To see why, note that
\begin{equation}
\begin{aligned}
    q(t) &= \int_0^t (\lambda(s) - m(s)) \mathbbm{1}\{ q(s) > 0 \text{ or } \lambda(s) \geq m(s) \} \diff s \\
    &\leq \int_0^t (\tlambda(s) - m_1(s)) \mathbbm{1}\{ q(s) > 0 \text{ or } \lambda(s) \geq m(s) \} \diff s \\
    &+ \int_0^t (\Delta\lambda(s) - m_2(s)) \mathbbm{1}\{ q(s) > 0 \text{ or } \lambda(s) \geq m(s) \} \diff s \\
    &\leq \int_0^t (\tlambda(s) - m_1(s)) \mathbbm{1}\{ q_1(s) > 0 \text{ or } \lambda(s) \geq m_1(s) \} \diff s \\
    &+ \int_0^t (\Delta\lambda(s) - m_2(s)) \mathbbm{1}\{ q_2(s) > 0 \text{ or } \lambda(s) \geq m_2(s) \} \diff s \\
    &= q_1(t) + q_2(t).
\end{aligned}
\end{equation}
Therefore, the flow-time of the algorithm is at most the sum of the flow-time of the offline component on $\tlambda$ and the online component on $\Delta\lambda$. Similarly, since the switching cost and the power cost are linear in the number of servers $m(\cdot)$, the cost of the algorithm is at most
\begin{equation}
\label{eq:costclaim1a}
    \textsc{Cost}^\lambda(m, T) \leq \textsc{Cost}^{\tlambda}(m_1, T) + \textsc{Cost}^{\Delta\lambda}(m_2, T).
\end{equation}
We will further bound the cost of the online component $m_2$. Let $[t, t + \delta) \subseteq [0, T]$ be an arbitrary time interval for $\delta > 0$ small and let $\Delta q(t) = \int_t^{t+\delta} \Delta\lambda(s) \diff s$. We will bound the cost due to the $\Delta q(t)$ workload received in this time interval. The number of servers $m_2(t)$ increases by $\sqrt{\omega / (2 \beta)} \cdot \Delta q(t)$ in the interval. Moreover, after a time of $\sqrt{2 \beta / \omega}$, the number of servers $m_2(t)$ decreases again by $\sqrt{\omega / (2 \beta)} \cdot \Delta q(t)$. Throughout $\left[ t, t + \sqrt{2 \beta / \omega} \right)$, the queue length due to this fraction of the workload decreases linearly as $q(t + s) = \Delta q(t) - \sqrt{\omega / (2 \beta)} \cdot \Delta q(t) \cdot s$ until the workload is completely handled. The cost due to waiting is therefore
\begin{equation}
    \omega \cdot \int_0^{\sqrt{\frac{2 \beta}{\omega}}} \left( \Delta q(t) - \sqrt{\frac{\omega}{2 \beta}} \cdot \Delta q(t) \cdot s \right) \diff s
    = \omega \cdot \sqrt{\frac{\beta}{2 \omega}} \cdot \Delta q(t).
\end{equation}
Note that, since $\delta$ can be chosen arbitrarily small, the waiting cost in the interval $[t, t + \delta)$ is negligible. The switching cost is $\beta \cdot \sqrt{\omega / (2 \beta)} \cdot \Delta q(t)$ and the power cost is $\theta \cdot \sqrt{2 \beta / \omega} \cdot \sqrt{\omega / (2 \beta)} \cdot \Delta q(t) = \theta \cdot \Delta q(t)$. The cost of the online component is therefore
\begin{equation}
\label{eq:costclaim1b}
    \textsc{Cost}^{\Delta\lambda}(m_2, T) \leq \lim_{\delta \downarrow 0} \sum_{i = 0}^{\lfloor T / \delta \rfloor} \left( \sqrt{2 \omega \beta} + \theta \right) \cdot \Delta q(i \delta) = \left( \sqrt{2 \omega \beta} + \theta \right) T \cdot \lVert \Delta \lambda \rVert_{MAE},
\end{equation}
which proves equation \eqref{eq:costsplitclaim} by combining \eqref{eq:costclaim1a} and \eqref{eq:costclaim1b}. \blue{Similarly, let $\Delta \lambda^*(t) = \left( \tlambda(t) - \lambda(t) \right)^+$}. Then, by interchanging the actual arrival rate $\lambda$ and the predicted arrival rate $\tlambda$ in equation \eqref{eq:costsplitclaim}, we find that
\begin{equation}
\label{eq:costsplitclaim2}
\blue{
    \textsc{Cost}^{\tlambda}(m^*, T) \leq \textsc{Cost}^\lambda(m^{*1}, T) + \left( \sqrt{2 \omega \beta} + \theta \right) T \cdot \lVert \Delta \lambda \rVert_{MAE},
}
\end{equation}
where $m^*(t) = m^{*1}(t) + m^{*2}(t)$ and
\begin{align}
    m^{*1} &\in \underset{m: (0, T] \to \mathbbm{R}_+}{\arg\min} \textsc{Cost}^\lambda(m, T), \\
    \frac{\diff m^{*2}(t)}{\diff t} &= \sqrt{\frac{\omega}{2 \beta}} \cdot \left( \Delta \lambda^*(t) - \Delta \lambda^*\left( t - \sqrt{2 \beta/\omega} \right) \right).
\end{align}
Finally, we combine \eqref{eq:costsplitclaim} and \eqref{eq:costsplitclaim2} to find that
\begin{equation}
\begin{aligned}
    \textsc{Cost}^\lambda(m, T)
    &\leq \textsc{Cost}^{\tlambda}(m_1, T) + \left( \sqrt{2 \omega \beta} + \theta \right) T \cdot \lVert \Delta \lambda \rVert_{MAE} \\
    &\leq \textsc{Cost}^{\tlambda}(m^*, T) + \left( \sqrt{2 \omega \beta} + \theta \right) T \cdot \lVert \Delta \lambda \rVert_{MAE} \\
    &\leq \textsc{Cost}^\lambda(m^{*1}, T) + \left( \sqrt{2 \omega \beta} + \theta \right) T \cdot \lVert \tlambda - \lambda \rVert_{MAE},
\end{aligned}
\end{equation}
where the first inequality follows by \eqref{eq:costsplitclaim}, the second inequality follows because $m_1$ achieves the minimum cost on $\tlambda$ and the third inequality follows by \eqref{eq:costsplitclaim2}. This completes the proof because $m^{*1}$ is the optimal offline solution on $\lambda$. \hfill$\blacksquare$
%\end{proof}
\subsection{Proof of Proposition \ref{prop:lowerboundcreta}.}
\label{app:proofcreta-lower-bd}

\blue{

Fix any algorithm $\mathcal{A}$, and let $\textsc{CR}(\eta)$ denote its competitive ratio when it has access to an $\eta$-accurate prediction. Fix $\delta > 0$ and assume that $\textsc{CR}(0) \leq 1 + \delta$.
We will construct an instance for which $\textsc{Cost}^\lambda(m, T) \geq \textsc{Opt} / (4 \delta)$.

Let $T = 2 + \sqrt{2} \delta$ and $\lambda(t) = 2$ for $t \in [0, \sqrt{2} \delta)$. The value of $\lambda(t)$ for $t \in [\sqrt{2} \delta, T]$ will be specified later. Fix $\beta = \omega = 1$ and $\theta = 0$. Let the prediction $\tlambda(t) = 2$ for $t \in [0, T]$ and let $m(t)$ be the number of servers of $\mathcal{A}$ for the instance. We distinguish two cases depending on the value of $\bar{m} := \sup_{t \in [0, \sqrt{2} \delta)} m(t)$.
\begin{enumerate}
    \item
First, consider the case when $\bar{m} < 1$. Fix $\lambda(t) = 2$ for $t \in [\sqrt{2} \delta, T]$. One feasible solution of \eqref{eq:model} is $m^*(t) = 2$ for $t \in (0, T]$. The cost of this solution is purely due to switching cost and therefore $\textsc{Opt} \leq 2$. As $\tlambda(t) = \lambda(t)$, the prediction $\tlambda$ is perfect. Hence, by the assumption that $\mathcal{A}$ is $(1 + \delta)$-consistent, we must have $\textsc{Cost}^\lambda(m, T) \leq (1 + \delta) \cdot \textsc{Opt}$ for this instance. We will see that this cannot be achieved under the case $\bar{m} < 1$.

Note, at time $t = \sqrt{2} \delta$, the queue length of $\mathcal{A}$ is at least $q(\sqrt{2} \delta) \geq \int_0^{\sqrt{2} \delta} (2 - \bar{m}) \diff t > \sqrt{2} \delta$. The optimal solution starting from time $t = \sqrt{2} \delta$ is $m(t) = 2 + q(\sqrt{2} \delta) / \sqrt{2} > 4 + \delta$ for $t \in (\sqrt{2} \delta, T]$. As a result, the flow-time is at least $\int_0^T q(t) \diff t \geq q(\sqrt{2} \delta) / \sqrt{2} > \delta$. The cost of $\mathcal{A}$ is therefore at least $\textsc{Cost}^\lambda(m, T) > 2 + 2 \delta \geq (1 + \delta) \cdot \textsc{Opt}$. This is a contradiction with our assumption that $\mathcal{A}$ is $(1 + \delta)$-consistent and hence, the next case must occur.

    \item
Next, consider the case when $\bar{m} \geq 1$. Fix $\lambda(t) = 0$ for $t \in [\sqrt{2} \delta, T]$. One feasible solution of~\eqref{eq:model} is $m^*(t) = 2 \delta$ for $t \in [0, T]$. The cost of the optimal solution is therefore at most $\textsc{Opt} \leq 4 \delta - 2 \delta^2 \leq 4 \delta$.
Now, the cost of $\mathcal{A}$ is at least $\textsc{Cost}^\lambda(m, T) \geq 1 \geq \textsc{Opt} / (4 \delta)$ due to switching cost. Moreover, the mean absolute error (MAE) is $T \cdot \MAE{\tlambda}{\lambda} = \int_0^T \left\lvert \tlambda(t) - \lambda(t) \right\rvert \diff t = 4$ and hence the prediction is $1 / \delta$-accurate.
\end{enumerate}
Hence, equation \eqref{eq:lowerboundcreta} follows. \hfill$\blacksquare$

}
\subsection{Proof of Proposition \ref{prop:timer-lower-bd}.}
\label{app:prooftimer-lower-bd}

\blue{

Fix any $\omega, \beta, \theta > 0$ and any function $\tau: \mathbb{R}_+^3 \to (0, \infty)$. Let $m(t)$ be the number of servers of the Timer algorithm for the current instance. We will construct a sequence of instances for which $\textsc{Cost}^\lambda(m, T) / \textsc{Opt} \to \infty$.

Let $\tau = \tau(\omega, \beta, \theta)$ and fix $0 < \varepsilon < \tau$. Let $\lambda(t) = 2$ for $t \in [0, t_0]$ and $\lambda(t_0 + i \tau + s) = \mathbbm{1}\{ s \in (0, \varepsilon]\}$ for $s \in (0, \tau]$ and $i \in \mathbb{N}\cap [0, T],$ where $t_0 = \inf \{ t \in \mathbb{R} \mid m(t) \geq 1 \}$ or $t_0 = \infty$ if the infimum does not exist. We let $T$ to be sufficiently large. Let us distinguish two cases depending on the value of $t_0$.

\begin{enumerate}
    \item
First, consider the case when $t_0 = \infty$. One feasible solution of \eqref{eq:model} is $m^*(t) = 2$. This solution does not incur any waiting cost and the cost of the optimal solution is therefore at most $\textsc{Opt} \leq 2 \beta + \theta T$. However, since $m(t) < 1$, the cost of the Timer algorithm is at least $\textsc{Cost}^\lambda(m, T) = \omega \cdot \int_0^T s \diff s = \frac{\omega T^2}{2}$. But then, $\textsc{Cost}^\lambda(m, T) / \textsc{Opt} \geq \frac{\omega T^2}{4 \beta + 2 \theta T} \to \infty$ as $T \to \infty$.

    \item
Next, consider the case when $t_0 < \infty$. One possible solution of $\eqref{eq:model}$ is $m^*(t) = 2$ for $t \in [0, t_0]$ and $m^*(t) = \varepsilon / \tau$ for $t \in (t_0, T]$. The cost of the optimal solution is therefore at most $\textsc{Opt} \leq 2 \beta + 2 \theta t_0 + \varepsilon \theta T / \tau + \omega T / \tau \cdot \int_0^\varepsilon (1 - \varepsilon / \tau) s \diff s + \omega T / \tau \cdot \int_\varepsilon^\tau (\varepsilon - \varepsilon s / \tau) \diff s \leq 2 \beta + 2 \theta t_0 + \varepsilon \theta T / \tau + \varepsilon \omega T / 2$. 
However, note that after time $t \geq t_0$, a server idles for at most $\tau - \varepsilon < \tau$ time, which means that the Timer algorithm maintains at least one server throughout $[t_0, T]$. Therefore, the cost of the Timer algorithm is at least $\textsc{Cost}^\lambda(m, T) = \beta + \theta (T - t_0)$. But then, $\textsc{Cost}^\lambda(m, T) / \textsc{Opt} \geq \frac{\beta + \theta (T - t_0)}{2 \beta + 2 \theta t_0 + \varepsilon \theta T / \tau + \varepsilon \omega T / 2} \to \infty$ as $T \to \infty$ and $\varepsilon \to 0$.
\end{enumerate}

\hfill$\blacksquare$

}
\subsection{Proof of Proposition \ref{prop:worstcasecr}.}
\label{app:proofworstcasecr}

%\begin{proof}[Proof of Theorem \ref{th:worstcasecr}]
Fix a finite time horizon $T$, arrival rate function $\lambda(\cdot)$ and initial number of servers $m(0)$.
Let $m^*(t)$ be a solution of the offline optimization problem \eqref{eq:model} and $q^*(t)$ the corresponding workload. Assume that the solution achieves a finite cost. If there does not exist a solution which achieves finite cost, then Proposition \ref{prop:worstcasecr} follows immediately. Without loss of generality, assume that $m^*(t)$ is differentiable. To see why this is possible, assume that $m^*(t)$ is not differentiable. Define the interpolation $m^*_\delta(t)$ of $m^*(t)$ such that
\begin{equation}
    m^*_\delta(t) = \int_t^{t + \delta} \frac{m^*(s)}{\delta} \diff s,
\end{equation}
which is differentiable for all $\delta > 0$. Also, note that 
\begin{equation}
    \int_{t_1}^{t_2} m^*_\delta(t) \diff t
    = \int_{t_1}^{t_2} \int_t^{t + \delta} \frac{m^*(s)}{\delta} \diff s \diff t
    \to \int_{t_1}^{t_2} m^*(t) \diff t \text{ as } \delta \to 0,
\end{equation}
for any $0 \leq t_1 \leq t_2 \leq \infty$. The cost of $m^*(t)$ and $m^*_\delta(t)$ therefore coincide, asymptotically as $\delta \to 0$. As a result, each function $m^*(\cdot)$ can be written as the limit of a sequence of differentiable functions $m^*_\delta(\cdot)$ and we therefore assume that $m^*(t)$ is differentiable without loss of generality.

\paragraph{Overview of the proof.} Let $m(t)$ be the number of servers of ABCS (Algorithm \ref{alg:bcsmlalg}) and $q(t)$ be the corresponding workload. The goal of the proof will be to find a non-negative potential function $\Phi(t)$ such that
\begin{equation}
\label{eq:worstcase_potentialidea}
    \frac{\diff \Phi(t)}{\diff t} + \frac{\partial \textsc{Cost}(m, t)}{\partial t} \leq \textsc{PCR} \cdot \frac{\partial \textsc{Cost}(m^*, t)}{\partial t},
\end{equation}
where we have omitted $\lambda$ from the notation $\textsc{Cost}^\lambda(m, t)$. Note that $\textsc{Cost}(m, t)$ and $\textsc{Cost}(m^*, t)$ are differentiable because $m(t)$ and $m^*(t)$ are differentiable. We integrate equation \eqref{eq:worstcase_potentialidea} from time $t = 0$ to $t = T$ to obtain
\begin{equation}
    \textsc{Cost}(m, T) \leq \textsc{PCR} \cdot \textsc{Cost}(m^*, T) + \Phi(0) - \Phi(T) \leq \textsc{PCR} \cdot \textsc{Cost}(m^*, T) + \Phi(0),
\end{equation}
where the last step follows because $\Phi(T)$ is non-negative. \blue{The proof of Proposition \ref{prop:worstcasecr} is therefore completed if we manage to find a differentiable potential function $\Phi(t)$ satisfying equation \eqref{eq:worstcase_potentialidea} and $\Phi(0) = \frac{\beta \cdot m(0)}{r_2} = 0$. If $m(0) > 0$ instead, then a similar statement as in Proposition \ref{prop:worstcasecr} holds, but with an additive term of $\frac{\beta \cdot m(0)}{r_2}$.}

\paragraph{Choice of $\Phi(t)$.} Define the potential function $\Phi(t)$ such that
\begin{equation}
\begin{aligned}
    \Phi(t) = &\begin{cases}
        c_5 \beta \cdot \left( d_{R_1}(t) - m(t) + m^*(t) \right), &\text{ if } m(t) > m^*(t) \\
        c_6 \beta \cdot \left( d_{r_1}(t) - m(t) + m^*(t) \right) &\text{ if } m(t) \leq m^*(t) \\
    \end{cases} \\
    + &\frac{\beta \cdot m(t)}{r_2}
    + c_6 R_2 \theta \cdot [q(t) - q^*(t)]^+,
\end{aligned}
\end{equation}
where
\begin{equation}
    d_r(t) = \sqrt{\frac{r \omega \cdot \left([q(t) - q^*(t)]^+\right)^2}{\beta} + (m(t) - m^*(t))^2}.
\end{equation}
Note that $\Phi(t)$ is non-negative and $\Phi(0) = \frac{\beta \cdot m(0)}{r_2}$. The sophisticated reader might remark that there are points in the domain for which $\Phi(t)$ is not differentiable. As there can only be countably many of these points, these points do not influence the integral of equation \eqref{eq:worstcase_potentialidea} and we simply ignore these points in the analysis.

\paragraph{Verification of \eqref{eq:worstcase_potentialidea}.} We continue by verifying equation \eqref{eq:worstcase_potentialidea}. We distinguish four cases, depending on whether $q(t) > q^*(t)$ or $q(t) \leq q^*(t)$ and $m(t) > m^*(t)$ or $m(t) \leq m^*(t)$.
\begin{enumerate}[(i)]
    \item
Assume that $q(t) > q^*(t)$ and $m(t) > m^*(t)$. Recall that, by definition,
\begin{equation}
\begin{aligned}
    \frac{\diff q}{\diff t} &= \lambda(t) - m(t), &
    \frac{\diff m}{\diff t} &= \frac{\hat{r}_1(t) \omega \cdot q(t) - \hat{r}_2(t) \theta \cdot m(t)}{\beta} \leq \frac{R_1 \omega \cdot q(t)}{\beta}.
\end{aligned}
\end{equation}
The derivative of $d_{R_1}(t)$ is therefore at most
\begin{equation}
\begin{aligned}
    \beta \cdot \frac{\diff d_{R_1}(t)}{\diff t}
    &\leq d_{R_1}(t)^{-1} \cdot
    \left( \begin{aligned}
        &R_1 \omega \cdot (q(t) - q^*(t)) (\lambda(t) - m(t)) \\
        + &R_1 \omega \cdot (q^*(t) - q(t)) (\lambda(t) - m^*(t)) \\
        + &R_1 \omega \cdot q(t) \cdot (m(t) - m^*(t)) \\
        + &\beta \cdot \frac{\diff m^*}{\diff t} \cdot (m^*(t) - m(t))
    \end{aligned} \right) \\
    &= d_{R_1}(t)^{-1} \cdot (m(t) - m^*(t)) \left( R_1 \omega \cdot q^*(t) - \beta \cdot \frac{\diff m^*}{\diff t} \right) \\
    &\leq R_1 \omega \cdot q^*(t) + \beta \cdot \left[ - \frac{\diff m^*}{\diff t} \right]^+.
\end{aligned}
\end{equation}
The derivative of the potential function $\Phi(t)$ is then
\begin{equation}
\begin{aligned}
\label{eq:case1_potentialdiff}
    \frac{\diff \Phi(t)}{\diff t}
    &\leq c_5 R_1 \omega \cdot q^*(t)
    + c_5 \beta \cdot \left( \left[ - \frac{\diff m^*}{\diff t} \right]^+ + \frac{\diff m^*}{\diff t} \right)
    - \left( c_5 - \frac{1}{r_2} \right) \beta \cdot \frac{\diff m}{\diff t} \\
    &+ c_6 R_2 \theta \cdot (\lambda(t) - m(t) - \lambda(t) + m^*(t)) \\
    &\leq c_5 R_1 \omega \cdot q^*(t)
    + c_5 \beta \cdot \left[ \frac{\diff m^*}{\diff t} \right]^+
    - \left( 1 + \frac{1}{r_1} \right) \beta \cdot \frac{\diff m}{\diff t} \\
    &+ \left( 1 + R_2 + \frac{R_2}{r_1} \right) \theta \cdot (m^*(t) - m(t)).
\end{aligned}
\end{equation}
The derivative of the cumulative cost $\textsc{Cost}(m, t)$ is
\begin{equation}
\begin{aligned}
\label{eq:case1_algdiff}
    \frac{\partial \textsc{Cost}(m, t)}{\partial t}
    &= \omega \cdot q(t) + \beta \cdot \left[ \frac{\diff m}{\diff t} \right]^+ + \theta \cdot m(t) \\
    &\leq \frac{\beta}{r_1} \cdot \frac{\diff m}{\diff t} + \beta \cdot \left[ \frac{\diff m}{\diff t} \right]^+ + \left( 1 + \frac{R_2}{r_1} \right) \theta \cdot m(t).
\end{aligned}
\end{equation}
We sum equation \eqref{eq:case1_potentialdiff} and \eqref{eq:case1_algdiff} and cancel terms to obtain
\begin{equation}
\begin{aligned}
    \frac{\diff \Phi(t)}{\diff t} + \frac{\partial \textsc{Cost}(m, t)}{\partial t}
    &\leq c_5 R_1 \omega \cdot q^*(t)
    + c_5 \beta \cdot \left[ \frac{\diff m^*}{\diff t} \right]^+
    + \left( 1 + R_2 + \frac{R_2}{r_1} \right) \theta \cdot m^*(t) \\
    &\leq \textsc{PCR} \cdot \frac{\partial \textsc{Cost}(m^*, t)}{\partial t}.
\end{aligned}
\end{equation}
Note that if $\frac{\diff m}{\diff t} \geq 0$, then the sum follows immediately. If $\frac{\diff m}{\diff t} < 0$, we apply the bound
\begin{equation}
    -\beta \cdot \frac{\diff m}{\diff t} \leq R_2 \theta \cdot m(t) - r_1 \omega \cdot q(t) \leq R_2 \theta \cdot m(t).
\end{equation}

\item
Assume that $q(t) \leq q^*(t)$ and $m(t) > m^*(t)$. The potential function $\Phi(t)$ simplifies to
\begin{equation}
    \Phi(t) = \frac{\beta \cdot m(t)}{r_2}.
\end{equation}
The derivative of the potential function $\Phi(t)$ is then
\begin{equation}
\label{eq:case2_potentialdiff}
    \frac{\diff \Phi(t)}{\diff t}
    = \frac{\beta}{r_2} \cdot \frac{\diff m}{\diff t}
    \leq \frac{R_1 \omega}{r_2} \cdot q^*(t) - \theta \cdot m(t).
\end{equation}
The derivative of the cumulative cost $\textsc{Cost}(m, t)$ is
\begin{equation}
\begin{aligned}
\label{eq:case2_algdiff}
    \frac{\partial \textsc{Cost}(m, t)}{\partial t}
    &= \omega \cdot q(t) + \beta \cdot \left[ \frac{\diff m}{\diff t} \right]^+ + \theta \cdot m(t) \\
    &\leq \omega \cdot q(t) + \beta \cdot \left[ R_1 \omega \cdot q(t) - r_2 \theta \cdot m(t) \right]^+ + \theta \cdot m(t) \\
    &\leq \left( 1 + R_1 \right) \omega \cdot q^*(t) + \theta \cdot m(t)
\end{aligned}
\end{equation}
We sum equation \eqref{eq:case2_potentialdiff} and \eqref{eq:case2_algdiff} and cancel terms to obtain
\begin{equation}
    \frac{\diff \Phi(t)}{\diff t} + \frac{\partial \textsc{Cost}(m, t)}{\partial t}
    \leq \left( 1 + R_1 + \frac{R_1}{r_2} \right) \omega \cdot q^*(t)
    \leq \textsc{PCR} \cdot \frac{\partial \textsc{Cost}(m^*, t)}{\partial t}.
\end{equation}

\item
Assume that $q(t) > q^*(t)$ and $m(t) \leq m^*(t)$. Recall that, by definition,
\begin{equation}
\begin{aligned}
    \frac{\diff q}{\diff t} &= \lambda(t) - m(t), &
    \frac{\diff m}{\diff t} &= \frac{\hat{r}_1(t) \omega \cdot q(t) - \hat{r}_2(t) \theta \cdot m(t)}{\beta} \geq \frac{r_1 \omega \cdot q(t) - R_2 \theta \cdot m(t)}{\beta}.
\end{aligned}
\end{equation}
The derivative of $d_{r_1}(t)$ is therefore at most
\begin{equation}
\begin{aligned}
    \beta \cdot \frac{\diff d_{r_1}(t)}{\diff t}
    &\leq d_{r_1}(t)^{-1} \cdot
    \left( \begin{aligned}
        &r_1 \omega \cdot (q(t) - q^*(t)) (\lambda(t) - m(t)) \\
        + &r_1 \omega \cdot (q^*(t) - q(t)) (\lambda(t) - m^*(t)) \\
        + &(r_1 \omega \cdot q(t) - R_2 \theta \cdot m(t)) (m(t) - m^*(t)) \\
        + &\beta \cdot \frac{\diff m^*}{\diff t} \cdot (m^*(t) - m(t))
    \end{aligned} \right) \\
    &\leq d_{r_1}(t)^{-1} \cdot (m^*(t) - m(t)) \left( R_2 \theta \cdot m(t) + \beta \cdot \frac{\diff m^*}{\diff t} \right) \\
    &\leq R_2 \theta \cdot m(t) + \beta \cdot \left[ \frac{\diff m^*}{\diff t} \right]^+.
\end{aligned}
\end{equation}
The derivative of the potential function $\Phi(t)$ is then
\begin{equation}
\begin{aligned}
\label{eq:case3_potentialdiff}
    \frac{\diff \Phi(t)}{\diff t}
    &\leq c_6 R_2 \theta \cdot m(t)
    + c_6 \beta \cdot \left( \left[ \frac{\diff m^*}{\diff t} \right]^+ + \frac{\diff m^*}{\diff t} \right)
    - \left( c_6 - \frac{1}{r_2} \right) \beta \cdot \frac{\diff m}{\diff t} \\
    &+ c_6 R_2 \theta \cdot (\lambda(t) - m(t) - \lambda(t) + m^*(t)) \\
    &\leq c_6 R_2 \theta \cdot m^*(t)
    + 2 c_6 \beta \cdot \left[ \frac{\diff m^*}{\diff t} \right]^+
    - \left( c_6 - \frac{1}{r_2} \right) \beta \cdot \frac{\diff m}{\diff t}.
\end{aligned}
\end{equation}
Similar to before, the derivative of the cumulative cost $\textsc{Cost}(m, t)$ is
\begin{equation}
\label{eq:case3_algdiff}
    \frac{\partial \textsc{Cost}(m, t)}{\partial t}
    \leq \frac{\beta}{r_1} \cdot \frac{\diff m}{\diff t} + \beta \cdot \left[ \frac{\diff m}{\diff t} \right]^+ + \left( 1 + \frac{R_2}{r_1} \right) \theta \cdot m(t).
\end{equation}
We sum equation \eqref{eq:case3_potentialdiff} and \eqref{eq:case3_algdiff} and cancel terms to obtain
\begin{equation}
\begin{aligned}
    \frac{\diff \Phi(t)}{\diff t} + \frac{\partial \textsc{Cost}(m, t)}{\partial t}
    &\leq 2 c_6 \beta \cdot \left[ \frac{\diff m^*}{\diff t} \right]^+
    + \left( 2 c_6 R_2 + 1 - \frac{R_2}{r_2} \right) \theta \cdot m^*(t) \\
    &\leq \textsc{PCR} \cdot \frac{\partial \textsc{Cost}(m^*, t)}{\partial t}.
\end{aligned}
\end{equation}

\item
Assume that $q(t) \leq q^*(t)$ and $m(t) \leq m^*(t)$. The potential function $\Phi(t)$ simplifies to
\begin{equation}
    \Phi(t) = 2 c_6 \beta \cdot (m^*(t) - m(t)) + \frac{\beta \cdot m(t)}{r_2}.
\end{equation}
The derivative of the potential function $\Phi(t)$ is then
\begin{equation}
\label{eq:case4_potentialdiff}
    \frac{\diff \Phi(t)}{\diff t} = 2 c_6 \beta \cdot \frac{\diff m^*}{\diff t} - \left( 2 c_6 - \frac{1}{r_2} \right) \beta \cdot \frac{\diff m}{\diff t}.
\end{equation}
Similar to before, the derivative of the cumulative cost $\textsc{Cost}(m, t)$ is
\begin{equation}
\label{eq:case4_algdiff}
    \frac{\partial \textsc{Cost}(m, t)}{\partial t}
    \leq \frac{\beta}{r_1} \cdot \frac{\diff m}{\diff t} + \beta \cdot \left[ \frac{\diff m}{\diff t} \right]^+ + \left( 1 + \frac{R_2}{r_1} \right) \theta \cdot m(t).
\end{equation}
We sum equation \eqref{eq:case4_potentialdiff} and \eqref{eq:case4_algdiff} and cancel terms to obtain
\begin{equation}
\begin{aligned}
    \frac{\diff \Phi(t)}{\diff t} + \frac{\partial \textsc{Cost}(m, t)}{\partial t}
    &\leq 2 c_6 \beta \cdot \left[ \frac{\diff m^*}{\diff t} \right]^+
    + \left( 2 c_6 R_2 + 1 - \frac{R_2}{r_2} \right) \theta \cdot m^*(t) \\
    &\leq \textsc{PCR} \cdot \frac{\partial \textsc{Cost}(m^*, t)}{\partial t}.
\end{aligned}
\end{equation}
\end{enumerate}

\hfill$\blacksquare$
%\end{proof}
\subsection{Proof of Proposition \ref{prop:predcr}.}
\label{app:proofpredcr}

%\begin{proof}[Proof of Theorem \ref{th:predcr}]
Fix a finite time horizon $T$, arrival rate function $\lambda(\cdot)$ and initial number of servers $m(0)$. Let $\tm(\cdot)$ be the number of advised servers of AP and $\tq(\cdot)$ be the corresponding workload. Assume that the advised number of servers achieves finite cost. If the advised number of servers does not achieve finite cost, then Proposition \ref{prop:predcr} follows immediately. Without loss of generality, similar to the proof of Proposition \ref{prop:worstcasecr}, assume that $\tm(t)$ is differentiable.

\paragraph{Overview of the proof.} As argued before (see the proof of Proposition \ref{prop:worstcasecr}), the proof of Proposition \ref{prop:predcr} requires us to find a non-negative potential function $\Phi(t)$ such that
\begin{equation}
\label{eq:pred_potentialidea}
    \frac{\diff \Phi(t)}{\diff t} + \frac{\partial \textsc{Cost}(m, t)}{\partial t} \leq \textsc{OCR} \cdot \frac{\partial \textsc{Cost}(\tm, t)}{\partial t},
\end{equation}
where we have omitted $\lambda$ from the notation $\textsc{Cost}^\lambda(m, t)$.

\paragraph{Choice of $\Phi(t)$.} Define the potential function $\Phi(t)$ such that
\begin{equation}
\begin{aligned}
    \Phi(t) &= \begin{cases}
        c_1 \beta \cdot (d_{r_1}(t) - m(t) + \tm(t)) &\text{ if } \hat{r}_1(t) = r_1, \\
        c_2 \beta \cdot d_{R_1}(t) - c_3 \beta \cdot (m(t) - \tm(t)) &\text{ if } \hat{r}_1(t) = R_1,
    \end{cases} \\
    &+ \frac{\beta \cdot m(t)}{R_2}
    + c_4 \theta \cdot [q(t) - \tq(t)]^+,
\end{aligned}
\end{equation}
where
\begin{equation}
    d_r(t) = \sqrt{\frac{r \omega \cdot \left([q(t) - \tq(t)]^+\right)^2}{\beta} + (m(t) - \tm(t))^2}.
\end{equation}
\blue{Note that $\Phi(0) = \frac{\beta \cdot m(0)}{R_2} = 0$. If $m(0) > 0$ instead, then a similar statement as in Proposition \ref{prop:predcr} holds, but with an additive term of $\frac{\beta \cdot m(0)}{R_2}$.} If $\hat{r}_1(t) = r_1$ or $m(t) \leq \tm(t)$ then $\Phi(t)$ is trivially non-negative. Assume that $\hat{r}_1(t) = R_1$ and $m(t) > \tm(t)$, then
\begin{equation}
    \Phi(t) \geq c_2 \beta \cdot d_{R_1}(t) - c_3 \beta \cdot (m(t) - \tm(t))
    \geq \left( c_2 \sqrt{1 + 2 R_1} - c_3 \right) \beta \cdot (m(t) - \tm(t))
    \geq 0,
\end{equation}
and hence $\Phi(t)$ is non-negative. The sophisticated reader might remark that there are points in the domain for which $\Phi(t)$ is not differentiable. As there can only be countably many of these points, these points do not influence the integral of equation \eqref{eq:pred_potentialidea} and we simply ignore these points in the analysis.

\paragraph{Verification of \eqref{eq:pred_potentialidea}.} We continue by verifying equation \eqref{eq:pred_potentialidea}. We distinguish eight cases, depending on whether $q(t) > \tq(t)$ or $q(t) \leq \tq(t)$, $m(t) > \tm(t)$ or $m(t) \leq \tm(t)$ and $\hat{r}_1(t) = r_1$ or $\hat{r}_1(t) = R_1$.
\begin{enumerate}
    \item[(i.a)]
Assume that $q(t) > \tq(t)$, $m(t) > \tm(t)$ and $\hat{r}_1(t) = r_1$. Note that $\hat{r}_2(t) = r_2$ because $q(t) > \tq(t)$. Recall that, by definition,
\begin{equation}
\begin{aligned}
    \frac{\diff q(t)}{\diff t} &= \lambda(t) - m(t), &
    \frac{\diff m(t)}{\diff t} &= \frac{r_1 \omega \cdot q(t) - r_2 \theta \cdot m(t)}{\beta}.
\end{aligned}
\end{equation}
The derivative of $d_{r_1}(t)$ is therefore at most
\begin{equation}
\begin{aligned}
    \beta \cdot \frac{\diff d_{r_1}(t)}{\diff t}
    &\leq d_{r_1}(t)^{-1} \cdot
    \left( \begin{aligned}
        &r_1 \omega \cdot (q(t) - \tq(t)) (\lambda(t) - m(t)) \\
        + &r_1 \omega \cdot (\tq(t) - q(t)) (\lambda(t) - \tm(t)) \\
        + &(r_1 \omega \cdot q(t) - r_2 \theta \cdot m(t)) (m(t) - \tm(t)) \\
        + &\beta \cdot \frac{\diff \tm}{\diff t} \cdot (\tm(t) - m(t))
    \end{aligned} \right) \\
    &= d_{r_1}(t)^{-1} \cdot (m(t) - \tm(t)) \left( r_1 \omega \cdot \tq(t) - r_2 \theta \cdot m(t) - \beta \cdot \frac{\diff \tm}{\diff t} \right) \\
    &\leq r_1 \omega \cdot \tq(t) - \frac{r_2 \theta}{\sqrt{1 + 2 r_1}} \cdot m(t)
    + \beta \cdot \left[ - \frac{\diff \tm}{\diff t} \right]^+ - \frac{\beta}{\sqrt{1 + 2 r_1}} \cdot \left[ \frac{\diff \tm}{\diff t} \right]^+.
\end{aligned}
\end{equation}
The derivative of the potential function $\Phi(t)$ is then
\begin{equation}
\begin{aligned}
\label{eq:case1a_potentialdiff}
    \frac{\diff \Phi(t)}{\diff t}
    &\leq c_1 r_1 \omega \cdot \tq(t) - \frac{c_1 r_2 \theta}{\sqrt{1 + 2 r_1}} \cdot m(t) \\
    &+ c_1 \beta \cdot \left[ - \frac{\diff \tm}{\diff t} \right]^+ - \frac{c_1 \beta}{\sqrt{1 + 2 r_1}} \cdot \left[ \frac{\diff \tm}{\diff t} \right]^+ + c_1 \beta \cdot \frac{\diff \tm}{\diff t} \\
    &- \left( c_1 - \frac{1}{R_2} \right) \beta \cdot \frac{\diff m}{\diff t}
    + c_4 \theta \cdot (\lambda(t) - m(t) - \lambda(t) + \tm(t)) \\
    &\leq c_1 r_1 \omega \cdot \tq(t) - \frac{r_2 \theta}{r_1} \cdot m(t)
    + \left( 1 + \frac{1}{R_2} \right) \beta \cdot \left[\frac{\diff \tm}{\diff t} \right]^+ \\
    &- \left( 1 + \frac{1}{r_1} \right) \beta \cdot \frac{\diff m}{\diff t} + c_4 \theta \cdot (\tm(t) - m(t)).
\end{aligned}
\end{equation}
The derivative of the cumulative cost $\textsc{Cost}(m, t)$ is
\begin{equation}
\begin{aligned}
\label{eq:case1a_algdiff}
    \frac{\partial \textsc{Cost}(m, t)}{\partial t}
    &= \omega \cdot q(t) + \beta \cdot \left[ \frac{\diff m}{\diff t} \right]^+ + \theta \cdot m(t) \\
    &= \frac{\beta}{r_1} \cdot \frac{\diff m}{\diff t} + \beta \cdot \left[ \frac{\diff m}{\diff t} \right]^+ + \left( 1 + \frac{r_2}{r_1} \right) \theta \cdot m(t).
\end{aligned}
\end{equation}
We sum equation \eqref{eq:case1a_potentialdiff} and \eqref{eq:case1a_algdiff} and cancel terms to obtain
\begin{equation}
\begin{aligned}
    \frac{\diff \Phi(t)}{\diff t} + \frac{\partial \textsc{Cost}(m, t)}{\partial t}
    &\leq \begin{aligned}[t]
    c_1 r_1 \omega \cdot \tq(t)
    + \left( 1 + \frac{1}{R_2} \right) \beta \cdot \left[ \frac{\diff \tm}{\diff t} \right]^+
    + c_4 \theta \cdot \tm(t)
    \end{aligned} \\
    &\leq \textsc{OCR} \cdot \frac{\partial \textsc{Cost}(\tm, t)}{\partial t}.
\end{aligned}
\end{equation}
Note that if $\frac{\diff m(t)}{\diff t} \geq 0$, then the sum follows immediately. If $\frac{\diff m(t)}{\diff t} < 0$, we apply the bound
\begin{equation}
    -\beta \cdot \frac{\diff m(t)}{\diff t} = r_2 \theta \cdot m(t) - r_1 \omega \cdot q(t) \leq r_2 \theta \cdot m(t).
\end{equation}

\item[(i.b)]
Assume that $q(t) > \tq(t)$, $m(t) > \tm(t)$ and $\hat{r}_1(t) = R_1$. Note that $\hat{r}_2(t) = r_2$ because $q(t) > \tq(t)$. The derivative of $d_{R_1}(t)$ is therefore at most
\begin{equation}
\begin{aligned}
    \beta \cdot \frac{\diff d_{R_1}(t)}{\diff t}
    &\leq d_{R_1}(t)^{-1} \cdot
    \left( \begin{aligned}
        &R_1 \omega \cdot (q(t) - \tq(t)) (\lambda(t) - m(t)) \\
        + &R_1 \omega \cdot (\tq(t) - q(t)) (\lambda(t) - \tm(t)) \\
        + &R_1 \omega \cdot q(t) \cdot (m(t) - \tm(t)) \\
        + &\beta \cdot \frac{\diff \tm}{\diff t} \cdot (\tm(t) - m(t))
    \end{aligned} \right) \\
    &= d_{R_1}(t)^{-1} \cdot (m(t) - \tm(t)) \left( R_1 \omega \cdot \tq(t) - \beta \cdot \frac{\diff \tm}{\diff t} \right) \\
    &\leq \frac{R_1 \omega}{\sqrt{1 + 2 R_1}} \cdot \tq(t)
    + \frac{\beta}{\sqrt{1 + 2 R_1}} \cdot \left[ - \frac{\diff \tm}{\diff t} \right]^+.
\end{aligned}
\end{equation}
The derivative of the potential function $\Phi(t)$ is then
\begin{equation}
\begin{aligned}
\label{eq:case1b_potentialdiff}
    \frac{\diff \Phi(t)}{\diff t}
    &\leq \frac{c_2 R_1 \omega}{\sqrt{1 + 2 R_1}} \cdot \tq(t)
    + \frac{c_2 \beta}{\sqrt{1 + 2 R_1}} \cdot \left[ - \frac{\diff \tm}{\diff t} \right]^+
    + c_3 \beta \cdot \frac{\diff \tm}{\diff t} \\
    &- \left( c_3 - \frac{1}{R_2} \right) \frac{\diff m}{\diff t}
    + c_4 \theta \cdot (\lambda(t) - m(t) - \lambda(t) + \tm(t)) \\
    &\leq \frac{c_2 R_1 \omega}{\sqrt{1 + 2 R_1}} \cdot \tq(t)
    + c_3 \beta \cdot \left[ \frac{\diff \tm}{\diff t} \right]^+ \\
    &- \left( 1 + \frac{1}{R_1} \right) \beta \cdot \frac{\diff m}{\diff t}
    + c_4 \theta \cdot (\tm(t) - m(t)).
\end{aligned}
\end{equation}
The derivative of the cumulative cost $\textsc{Cost}(m, t)$ is
\begin{equation}
\label{eq:case1b_algdiff}
    \frac{\partial \textsc{Cost}(m, t)}{\partial t}
    = \frac{\beta}{R_1} \cdot \frac{\diff m}{\diff t} + \beta \cdot \left[ \frac{\diff m}{\diff t} \right]^+ + \left( 1 + \frac{r_2}{R_1} \right) \theta \cdot m(t).
\end{equation}
We sum equation \eqref{eq:case1b_potentialdiff} and \eqref{eq:case1b_algdiff} and cancel terms to obtain
\begin{equation}
\begin{aligned}
    \frac{\diff \Phi(t)}{\diff t} + \frac{\partial \textsc{Cost}(m, t)}{\partial t}
    &\leq \begin{aligned}[t]
    \frac{c_2 R_1}{\sqrt{1 + 2 R_1}} \omega \cdot \tq(t)
    + c_3 \beta \cdot \left[ \frac{\diff \tm}{\diff t} \right]^+
    + c_4 \theta \cdot \tm(t)
    \end{aligned} \\
    &\leq \textsc{OCR} \cdot \frac{\partial \textsc{Cost}(\tm, t)}{\partial t}.
\end{aligned}
\end{equation}

\item[(ii.a)]
Assume that $q(t) \leq \tq(t)$, $m(t) > \tm(t)$ and $\hat{r}_1(t) = r_1$. Note that $\hat{r}_2(t) = R_2$ because $m(t) > \tm(t)$ and $q(t) \leq \tq(t)$. The potential function $\Phi(t)$ simplifies to
\begin{equation}
    \Phi(t) = \frac{\beta \cdot m(t)}{R_2}
\end{equation}
The derivative of the potential function $\Phi(t)$ is then
\begin{equation}
\label{eq:case2a_potentialdiff}
    \frac{\diff \Phi(t)}{\diff t}
    = \frac{\beta}{R_2} \cdot \frac{\diff m}{\diff t}
    = \frac{r_1 \omega}{R_2} \cdot q(t) - \theta \cdot m(t).
\end{equation}
The derivative of the cumulative cost $\textsc{Cost}(m, t)$ is
\begin{equation}
\begin{aligned}
\label{eq:case2a_algdiff}
    \frac{\partial \textsc{Cost}(m, t)}{\partial t}
    &= \omega \cdot q(t) + \beta \cdot \left[ \frac{\diff m}{\diff t} \right] + \theta \cdot m(t) \\
    &= \omega \cdot q(t) + \beta \cdot \left[ \frac{r_1 \omega \cdot q(t)}{\beta} - \frac{R_2 \theta \cdot m(t)}{\beta} \right]^+ + \theta \cdot m(t) \\
    &\leq (1 + r_1) \omega \cdot q(t) + \theta \cdot m(t).
\end{aligned}
\end{equation}
We sum equation \eqref{eq:case2a_potentialdiff} and \eqref{eq:case2a_algdiff} and cancel terms to obtain
\begin{equation}
    \frac{\diff \Phi(t)}{\diff t} + \frac{\partial \textsc{Cost}(m, t)}{\partial t}
    \leq \left( 1 + r_1 + \frac{r_1}{R_2} \right) \omega \cdot \tq(t)
    \leq \textsc{OCR} \cdot \frac{\partial \textsc{Cost}(\tm, t)}{\partial t}.
\end{equation}

\item[(ii.b)]
Assume that $q(t) \leq \tq(t)$, $m(t) > \tm(t)$ and $\hat{r}_1(t) = R_1$. However, $\hat{r}_1(t) = R_1$ implies that $m(t) - \tm(t) \leq \left[ q(t) - \tq(t) \right]^+ \cdot \sqrt{\frac{\omega}{2 \beta}} = 0$ which contradicts our assumption.

\item[(iii.a)]
Assume that $q(t) > \tq(t)$, $m(t) \leq \tm(t)$ and $\hat{r}_1(t) = r_1$. However, $\hat{r}_1(t) = r_1$ implies that $m(t) - \tm(t) > \left[ q(t) - \tq(t) \right] \cdot \sqrt{\frac{\omega}{2 \beta}} \geq 0$ which contradicts our assumption.

\item[(iii.b)]
Assume that $q(t) > \tq(t)$, $m(t) \leq \tm(t)$ and $\hat{r}_1(t) = R_1$. Note that $\hat{r}_2(t) = r_2$ because $m(t) \leq \tm(t)$. The derivative of $d_{R_1}(t)$ is therefore at most
\begin{equation}
\begin{aligned}
    \beta \cdot \frac{\diff d_{R_1}(t)}{\diff t}
    &\leq d_{R_1}(t)^{-1} \cdot
    \left( \begin{aligned}
        &R_1 \omega \cdot (q(t) - \tq(t)) (\lambda(t) - m(t)) \\
        + &R_1 \omega \cdot (\tq(t) - q(t)) (\lambda(t) - \tm(t)) \\
        + &(R_1 \omega \cdot q(t) - r_2 \theta \cdot m(t)) (m(t) - \tm(t)) \\
        + &\beta \cdot \frac{\diff \tm}{\diff t} \cdot (\tm(t) - m(t))
    \end{aligned} \right) \\
    &\leq d_{R_1}(t)^{-1} \cdot (\tm(t) - m(t)) \left( r_2 \theta \cdot m(t) + \beta \cdot \frac{\diff \tm}{\diff t} \right) \\
    &\leq r_2 \theta \cdot m(t)
    + \beta \cdot \left[ \frac{\diff \tm}{\diff t} \right]^+.
\end{aligned}
\end{equation}
The derivative of the potential function $\Phi(t)$ is then
\begin{equation}
\begin{aligned}
\label{eq:case3b_potentialdiff}
    \frac{\diff \Phi(t)}{\diff t}
    &\leq c_2 r_2 \theta \cdot m(t)
    + c_2 \beta \cdot \left[ \frac{\diff \tm}{\diff t} \right]^+
    + c_3 \beta \cdot \frac{\diff \tm}{\diff t} \\
    &- \left( c_3 - \frac{1}{R_2} \right) \beta \cdot \frac{\diff m}{\diff t}
    + c_4 \theta \cdot (\lambda(t) - m(t) - \lambda(t) + \tm(t)) \\
    &\leq c_2 r_2 \theta \cdot m(t)
    + (c_2 + c_3) \beta \cdot \left[ \frac{\diff \tm}{\diff t} \right]^+ \\
    &- \left( 1 + \frac{1}{R_1} \right) \beta \cdot \frac{\diff m}{\diff t}
    + c_4 \theta \cdot (\tm(t) - m(t)) \\
\end{aligned}
\end{equation}
The derivative of the cumulative cost $\textsc{Cost}(m, t)$ is
\begin{equation}
\label{eq:case3b_algdiff}
    \frac{\partial \textsc{Cost}(m, t)}{\partial t}
    = \frac{\beta}{R_1} \cdot \frac{\diff m}{\diff t} + \beta \cdot \left[ \frac{\diff m}{\diff t} \right]^+ + \left( 1 + \frac{r_2}{R_1} \right) \theta \cdot m(t).
\end{equation}
We sum equation \eqref{eq:case3b_potentialdiff} and \eqref{eq:case3b_algdiff} and cancel terms to obtain
\begin{equation}
\begin{aligned}
    \frac{\diff \Phi(t)}{\diff t} + \frac{\partial \textsc{Cost}(m, t)}{\partial t}
    &\leq \begin{aligned}[t]
    (c_2 + c_3) \beta \cdot \left[ \frac{\diff \tm}{\diff t} \right]^+
    + c_4 \theta \cdot \tm(t)
    \end{aligned} \\
    &\leq \textsc{OCR} \cdot \frac{\partial \textsc{Cost}(\tm, t)}{\partial t}.
\end{aligned}
\end{equation}

\item[(iv.a)]
Assume that $q(t) \leq \tq(t)$, $m(t) \leq \tm(t)$ and $\hat{r}_1(t) = r_1$. However, $\hat{r}_1(t) = r_1$ implies that $m(t) - \tm(t) > \left[ q(t) - \tq(t) \right]^+ \cdot \sqrt{\frac{\omega}{2 \beta}} = 0$ which contradicts our assumption.

\item[(iv.b)]
Assume that $q(t) \leq \tq(t)$, $m(t) \leq \tm(t)$ and $\hat{r}_1(t) = R_1$. Note that $\hat{r}_2(t) = r_2$ because $m(t) \leq \tm(t)$. The potential function $\Phi(t)$ simplifies to
\begin{equation}
    \Phi(t) = (c_2 + c_3) \beta \cdot (\tm(t) - m(t)) + \frac{\beta \cdot m(t)}{R_2}
\end{equation}
The derivative of the potential function $\Phi(t)$ is then
\begin{equation}
\label{eq:case4b_potentialdiff}
    \frac{\diff \Phi(t)}{\diff t}
    = (c_2 + c_3) \beta \cdot \frac{\diff \tm}{\diff t}
    - \left( c_2 + 1 + \frac{1}{R_1} \right) \beta \cdot \frac{\diff m}{\diff t}.
\end{equation}
The derivative of the cumulative cost $\textsc{Cost}(m, t)$ is
\begin{equation}
\label{eq:case4b_algdiff}
    \frac{\partial \textsc{Cost}(m, t)}{\partial t}
    = \frac{\beta}{R_1} \cdot \frac{\diff m}{\diff t} + \beta \cdot \left[ \frac{\diff m}{\diff t} \right]^+ + \left( 1 + \frac{r_2}{R_1} \right) \theta \cdot m(t).
\end{equation}
We sum equation \eqref{eq:case4b_potentialdiff} and \eqref{eq:case4b_algdiff} and cancel terms to obtain
\begin{equation}
\begin{aligned}
    \frac{\diff \Phi(t)}{\diff t} + \frac{\partial \textsc{Cost}(m, t)}{\partial t}
    &\leq \begin{aligned}[t]
    (c_2 + c_3) \beta \cdot \left[ \frac{\diff \tm}{\diff t} \right]^+
    + c_4 \theta \cdot \tm(t)
    \end{aligned} \\
    &\leq \textsc{OCR} \cdot \frac{\partial \textsc{Cost}(\tm, t)}{\partial t}.
\end{aligned}
\end{equation}

\end{enumerate}

\hfill$\blacksquare$
%\end{proof}
\subsection{Proof of Lemma \ref{lemma:integralgap}.}
\label{app:integralgap}

\blue{

We will construct a sequence of instances for which $\textsc{Opt}_{int} / \textsc{Opt} \to \infty$.

Let $m(t)$ be the number of servers of $\textsc{Opt}_{int}$. Fix $0 < \varepsilon < 1$ and let $\lambda(t) = \varepsilon$ for $t \in [0, T]$ where the finite time horizon $T = 1 / \varepsilon$. Let $\beta = 0$, $\omega = \infty$ and $\theta = 1$. Then, since $\omega = \infty$, $m(t) \geq 1$ for $t \in [0, T]$. Therefore, $\textsc{Opt}_{int} \geq 1 / \varepsilon$. However, one possible fractional solution turns on $m^*(t) = \varepsilon$ servers for $t \in [0, T]$ and therefore the value of the optimal solution is at most $\textsc{Opt} \leq 1$. Thus, $\textsc{Opt}_{int} / \textsc{Opt} = 1 / \varepsilon \to \infty$ as $\varepsilon \to 0$.

\hfill$\blacksquare$

}

\end{document}